\documentclass{aa}  

\usepackage{graphicx}
\usepackage{multirow}
\usepackage{lineno}
\usepackage{xcolor}
\usepackage{txfonts}
\begin{document} 

   \title{Constraining the radius and atmospheric properties of directly imaged exoplanets through multi-phase observations}

   \author{\'O. Carri\'on-Gonz\'alez\inst{1}\thanks{ \email{o.carriongonzalez@astro.physik.tu-berlin.de, oscar.carrion.gonzalez@gmail.com}} 
          \and
          A. Garc\'ia Mu\~noz\inst{2,1}
          \and
          N. C. Santos\inst{3,4}
          \and
          J. Cabrera\inst{5}
          \and
          Sz. Csizmadia\inst{5}
          \and
          H. Rauer\inst{1,5,6}
          }

   \institute{Zentrum f\"ur Astronomie und Astrophysik, Technische Universit\"at Berlin, Hardenbergstraße 36, D-10623 Berlin, Germany
         \and AIM, CEA, CNRS, Université Paris-Saclay, Université de Paris, F-91191 Gif-sur-Yvette, France
         \and
             Instituto de Astrof\'isica e Ci\^encias do Espa\c{c}o, Universidade do Porto, CAUP, Rua das Estrelas, 4150-762 Porto, Portugal
         \and
            Departamento de F\'isica e Astronomia, Faculdade de Ci\^encias, Universidade do Porto, Rua do Campo Alegre, 4169-007 Porto, Portugal
         \and
             Deutsches Zentrum f\"ur Luft- und Raumfahrt, Rutherfordstraße 2, D-12489 Berlin, Germany
         \and
             Institute of Geological Sciences, Freie Universit\"at Berlin, Malteserstraße 74-100, D-12249 Berlin, Germany
                 }

   %\date{Received September 15, 1996; accepted March 16, 1997}

% \abstract{}{}{}{}{} 
% 5 {} token are mandatory
 
  \abstract
   {
    The theory of remote sensing shows that observing a planet at multiple phase angles ($\alpha$) is a powerful strategy to characterize its atmosphere.
    Here, we study this observing strategy as applied to future disc-integrated direct imaging of exoplanets in reflected starlight.
   }
   {
    We analyse how the information contained in reflected-starlight spectra of exoplanets depends on the phase angle, and the potential of multi-phase measurements to better constrain the atmospheric properties and the planet radius ($R_p$).
   }
   {
    We simulate spectra (500$-$900 nm) at $\alpha$=37º, 85º and 123º with spectral resolution $R\sim$125$-$225 and signal-to-noise ratio $S/N$=10, consistent with the expected capabilities of future direct-imaging space telescopes.
    Assuming a H$_2$-He atmosphere, we use a seven-parameter model that includes the atmospheric methane abundance ($f_{\rm{CH_4}}$), the optical properties of a cloud layer and $R_p$.
    All these parameters are assumed unknown a priori and explored with a Markov chain Monte Carlo retrieval method. 
   }
   {
   No single-phase observation can robustly identify whether the atmosphere has clouds or not.
   A single-phase observation at $\alpha$=123º and $S/N$=10 can constrain $R_p$ with a maximum error of 35\%, regardless of the cloud coverage.
   We find that combining small (37º) and large (123º) phase angles is a generally effective strategy to break multiple parameter degeneracies.
   This enables to determine with higher confidence the presence or absence of a cloud and its main properties, $f_{\rm{CH_4}}$ and $R_p$ in all the explored scenarios.
   Other strategies, such as doubling $S/N$ to 20 for a single-phase observation or combining small (37º) and moderate (85º) phase angles, fail to achieve this.
   We show that the improvements in multi-phase retrievals are associated with the shape of the scattering phase function of the cloud aerosols and that the improvement is more modest for isotropically-scattering aerosols.
   We finally discuss that misidentifying the background gas in the retrievals of super-Earth observations leads to a systematic underestimate of the absorbing gas abundance.
   }
   {
   Exoplanets with wide ranges of observable phase angles should be prioritized for atmospheric characterization in reflected starlight.
   }


\titlerunning{Constraining $R_p$ and atmospheric properties of directly imaged exoplanets through multi-$\alpha$ observations}
\authorrunning{Carri\'on-Gonz\'alez et al.} 
   \maketitle
%
%-------------------------------------------------------------------

\section{Introduction}
\label{sec:introduction}

Directly imaging exoplanets in reflected starlight will become a reality in this decade with the launch of the Nancy Grace Roman Space Telescope (hereon, the Roman Telescope) \citep{spergeletal2013}.
This technique will enable the atmospheric characterization of a population of cold and temperate long-period planets that cannot be studied in transit.
A thorough understanding of the physical parameters of a planet that affect the reflected-starlight spectra will be required to correctly interpret such measurements.
In addition, such understanding will help define optimal observing strategies to better characterize an exoplanet.

Several works have studied the information content of reflected-starlight spectra through so-called retrieval studies that aim to infer the properties of a planet and its atmosphere from a measurement \citep[e.g.][]{lupuetal2016,nayaketal2017,fengetal2018,damiano-hu2019,carriongonzalezetal2020,damianoetal2020}.
Given the lack of direct-imaging observations of exoplanets in reflected starlight, these retrievals use simulated measurements.
For this, a synthetic spectrum is computed by solving the radiative-transfer equation for an idealized atmospheric model and some noise is added to it.

Different approaches have been explored regarding the a priori knowledge of the atmosphere in the retrieval models.
For instance, some works assume the cloud composition to be known by fixing the aerosol properties to those of a particular Solar System analogue \citep[e.g. water clouds in][]{fengetal2018}.
In other cases, the cloud composition is computed following some microphysical prescriptions for a specified atmospheric temperature-pressure ($T$-$P$) profile and information on the condensation levels of the gaseous species in the model \citep[e.g.][]{damiano-hu2019, damianoetal2020}.
In such cases, the $T$-$P$ profile can either be a reasonable guess or the result of a self-consistent computation.
The latter is the approach followed by e.g. \citet{hu2019}, which requires several planetary properties as input, such as the gravity, metallicity, and both the irradiation and intrinsic temperatures.
These self-consistent computations have as a drawback that several of the inputs, such as the gravity, metallicity and intrinsic temperature will be a priori unknown for directly imaged exoplanets and must therefore be guessed somehow.
For instance, we note that the main component of the Venus clouds (H$_2$SO$_4$-H$_2$O) is not a major constituent of its atmosphere but the result of ongoing photochemical processes. Predicting the occurrence of such clouds from first principles remains a considerable challenge.

Other works have followed a different approach, including the optical properties of the cloud as free parameters in the retrievals \citep[e.g.][]{lupuetal2016, nayaketal2017, carriongonzalezetal2020} thus omitting any prior knowledge of the clouds that could be potentially gained from micro-physical considerations.
These studies have shown important correlations between model parameters that hinder the accurate characterization of an atmosphere.
Cloud properties such as its optical thickness ($\tau_c$), its vertical position in the atmosphere or the single-scattering albedo of the aerosols ($\omega_0$) show strong degeneracies with e.g. the abundance of gaseous absorbing species.
\citet{nayaketal2017} discussed the correlations between an unknown planet radius ($R_p$) and an unknown star-planet-observer phase angle ($\alpha$) in the retrievals.
\citet{carriongonzalezetal2020} analysed the correlations between an unknown $R_p$ and the atmospheric properties of the planet by comparing retrievals in which $R_p$ was assumed known to retrievals in which $R_p$ was assumed unknown.
That work showed that if the planet radius is unknown the correlations triggered between $R_p$ and cloud properties such as $\tau_c$ degrade the quantitative findings from the atmospheric retrievals.
These correlations worsen the constraints on the absorbing gases and prevent distinguishing between cloudy and cloud-free atmospheres.
This shows that the conclusions of the retrievals are strongly dependent on very basic assumptions of the models such as the treatment of the clouds.
Most of these retrieval exercises have analysed reflected-starlight spectra at a single phase angle \citep{lupuetal2016, fengetal2018, damiano-hu2019, carriongonzalezetal2020}.

The observations of the Solar System planets, however, have shown the relevance of optical phase curves, which sample a range of phase angles, to constrain their atmospheric properties. 
The added value of phase curves over single-phase disc-integrated reflected-starlight measurements has been explored for both gas giants \citep[e.g.][]{dyudinaetal2016, mayorgaetal2016} and terrestrial planets \citep[e.g.][]{mallama2009, robinsonetal2011, garciamunozetal2014, garciamunozetal2017, mayorgaetal2020, leeetal2020, leeetal2021}.
These studies have revealed the occurrence of different atmospheric phenomena such as glories, specular reflection by liquid surfaces or differential atmospheric rotation and waves.

Our goal in this work is to determine how simultaneous retrievals of multi-phase spectra affect the atmospheric characterization of an exoplanet if neither its radius nor its cloud properties are known a priori.
This is a follow-up of \citet{carriongonzalezetal2020}, where the models and results for full-phase ($\alpha$=0$^\circ$) spectra are described in detail.
The main conclusion of that work was that if $R_p$ is unknown, as will generally happen for directly imaged exoplanets, retrievals at $\alpha$=0$^\circ$ could not determine the presence or absence of clouds.
This result was obtained for a measurement with complete coverage between $\lambda=500-900$ nm, spectral resolution $R\sim$125-225 and $S/N$=10.
Such measurements are beyond the capabilities of the Roman Telescope, but are expected to be feasible with next-generation missions such as the Habitable Exoplanet Observatory (HabEx) \citep{mennessonetal2016, gaudietal2018} or the Large UV/Optical/IR Surveyor (LUVOIR) \citep{bolcaretal2016, luvoirteam2018}.

We aim to understand in which way the observed degeneracies between model parameters (particularly, $R_p$, the cloud optical thickness and the methane abundance) can be broken.
For that, we first analyse the information content of spectra at different phase angles by means of single-phase retrievals at $\alpha$=$37^\circ$, $85^\circ$ and $123^\circ$ with $S/N$=10.
Next, we perform simultaneous retrievals combining the observation at $\alpha$=$37^\circ$ with others at larger phase angles ($85^\circ$ and $123^\circ$).
We also explore how the single-phase retrievals at $\alpha$=$37^\circ$ improve if the $S/N$ is doubled to $S/N$=20.

The paper is structured as follows.
In Sect. \ref{sec:previousworks} we compare our approach to previous approaches described in the literature and indicate our new contributions.
In Sect. \ref{sec:model} we describe the atmospheric model and the retrieval method that we use.
The results are presented in Sect. \ref{sec:results} and several hypotheses explaining these results are tested in Sect. \ref{sec:discussion}.
The final summary and conclusions are given in Sect. \ref{sec:conclusions}.

\begin{table*}[t]
\caption{Model parameters used to compute the synthetic reflected-starlight spectra. }
\label{table:model_params}
\begin{center}
 \begin{tabular}{ l  l  c } 
 \hline \hline
  Parameter   &  Description &   Range of values \\
 \hline
  $\tau_{c}$ &   Optical thickness of the cloud     &  [0.05 $-$ 50.0] \\
  $\Delta_{c}$/H$_g$    &   Geometrical vertical extension of the cloud in units of the scale height &   [1 $-$ 8] \\
  $\tau_{\rm{c\rightarrow TOA}}$  &  Optical thickness of the gas from the top of the cloud to the top  &  [1.35 $-$ 4.5$\cdot$10$^{-4}$]\\
  & of the atmosphere (TOA), at the reference wavelength $\lambda_*$=800 nm     &  \\
  {$r_{\rm{eff}}$} &  Effective radius of the cloud's aerosols [$\mu$m] &   [0.10 $-$ 10.0] \\
  $\omega_0$   &  Single-scattering albedo of the cloud's aerosols &   [0.50 $-$ 1.0] \\
  $f_{\rm{CH_4}}$    & CH$_4$ abundance relative to H$_2$-He, assumed constant over the atmosphere &   [1$\cdot$10$^{-5}$ $-$ 5$\cdot$10$^{-2}$] \\
  $R_p/R_N$ &   Planet radius normalized to that of Neptune & [0.05 $-$ 5.0] \\
 \hline
\end{tabular}
\end{center}
\end{table*}

\section{Our work in the context of precedent studies} \label{sec:previousworks}
In preparation for future phase curves of directly imaged exoplanets in reflected-starlight, \citet{nayaketal2017} and \citet{damianoetal2020} conducted atmospheric retrievals at several phase angles. 
Below we describe their methods and main results, highlighting also the differences with and contributions of our analysis.

\citet{nayaketal2017} carried out atmospheric retrievals at seven phase angles between $\alpha=0^\circ$ and $120^\circ$, with signal-to-noise ratios ($S/N$) of 5, 10 and 20.
Building upon the work by \citet{lupuetal2016}, they assumed a two-cloud model without prior knowledge of the optical properties of the cloud aerosols.
Out of the total eleven model parameters considered in their retrievals, they focused on a subset of six parameters and studied how their retrieval results vary at several values of $\alpha$.
These parameters are: the planet's surface gravity, its radius, the observed phase angle, the atmospheric methane abundance, the pressure level at the top of the upper cloud and that of the bottom.
They concluded that $R_p$ is better determined at large (crescent) phases ($\alpha > 90^\circ$) and that, for the rest of parameters, knowing the phase angle at the time of the observation would not remarkably improve the atmospheric retrievals.
A preliminary investigation was presented on the potential gains from multi-phase observations by applying an intersection criterion between the retrieval results of single-phase observations.
This criterion was applied to the marginalized posterior probability distributions of each individual parameter, such that the resulting solution when combining two observations is the subset of values that are common to both single-phase retrievals.
They suggested that observations with low $S/N$ especially benefit from the combination of phases and noted that performing simultaneous retrievals of spectra at different $\alpha$ would be needed to confirm their tentative findings.

\citet{damianoetal2020} simulated observations for a Roman-like telescope equipped with a starshade (only three wavelength bands available: $0.45-0.55$, $0.61-0.75$ and $0.82-1.0\,\mu m$, and spectral resolution $R$=50).
They also considered a future mission like HabEx, with broad spectral coverage of $0.45-1\,\mu m$ and $R$=140.
They carried out the retrievals at $\alpha=60^\circ$ and $90^\circ$ using the methods developed in \citet{hu2019} and \citet{damiano-hu2019}.
The cloud optical thickness and aerosol properties were therefore dictated through microphysical prescriptions and not included as free parameters in the retrievals.
To account for the unknown planet radius, they assumed the planet mass to be known and included the planetary gravity as a free parameter in the retrievals.
They found that a single observation at $\alpha=60^\circ$ in the green band of the Roman Telescope ($0.61-0.75\,\mu m$) with $S/N\sim18$ ($\sim$3.5 hours of integration) cannot accurately constrain the atmospheric properties due to parameter degeneracies.
Three different strategies were found to break these degeneracies with comparable efficiency.
First, combining the observation at $\alpha=60^\circ$ and $S/N\sim18$ with another one at $\alpha=90^\circ$ and $S/N$=11.5 and performing simultaneous retrievals.
Second, expanding the wavelength coverage to the red filter ($0.82-1.0\,\mu m$), with a reduced $S/N\sim2$. Third, doubling the integration time of the green-band observation at $\alpha=60^\circ$, achieving a $S/N\sim$25.
For HabEx-like observations, with higher spectral resolution and wider wavelength coverage, they concluded that an observation at $\alpha=60^\circ$ can constrain the main atmospheric properties.
In this case their retrievals would not improve significantly if the integration time is doubled or if an additional measurement at $\alpha=90^\circ$ is performed.

Our study differs from that by \citet{nayaketal2017} in that we perform simultaneous retrievals of spectra at multiple phase angles.
As shown below, the findings from our approach differ significantly from the findings of methods that combine the outcome of single-phase retrievals.
We also analyse the variations with $\alpha$ of the retrieved optical properties of the cloud and we do that for several cloud scenarios in order to derive general conclusions.
This work also differs from \citet{damianoetal2020} in that we do not assume any prior knowledge about the planet mass or the cloud properties.
Hence, we include as free parameters in our retrievals the cloud optical properties and the planet radius.
Furthermore, we also investigate the physical basis of the improvement in the simultaneous multi-phase retrievals.
Similarly to both works, we assume a H$_2$-He dominated atmosphere.

The focus of our work is thus mainly theoretical.
Nevertheless, we note that our selected values of $\alpha$ $\sim 40^\circ-120^\circ$ may be accessible for several exoplanets in an optimistic configuration of the Roman Telescope coronagraph \citep{carriongonzalezetal2021}.
These $\alpha$ will be well within the operational range of HabEx- and LUVOIR-like telescopes.

\section{Atmospheric model and retrieval} \label{sec:model}
In this section we summarize the fundamentals of our atmospheric model and retrieval method, which are thoroughly described in \citet{carriongonzalezetal2020}.
We also describe how our retrieval method is adapted to perform simultaneous retrievals of observations at multiple phase angles.

\subsection{Forward model} \label{subsec:model_atmosphere}

\begin{figure*}
   \centering
   \includegraphics[width=18.cm]{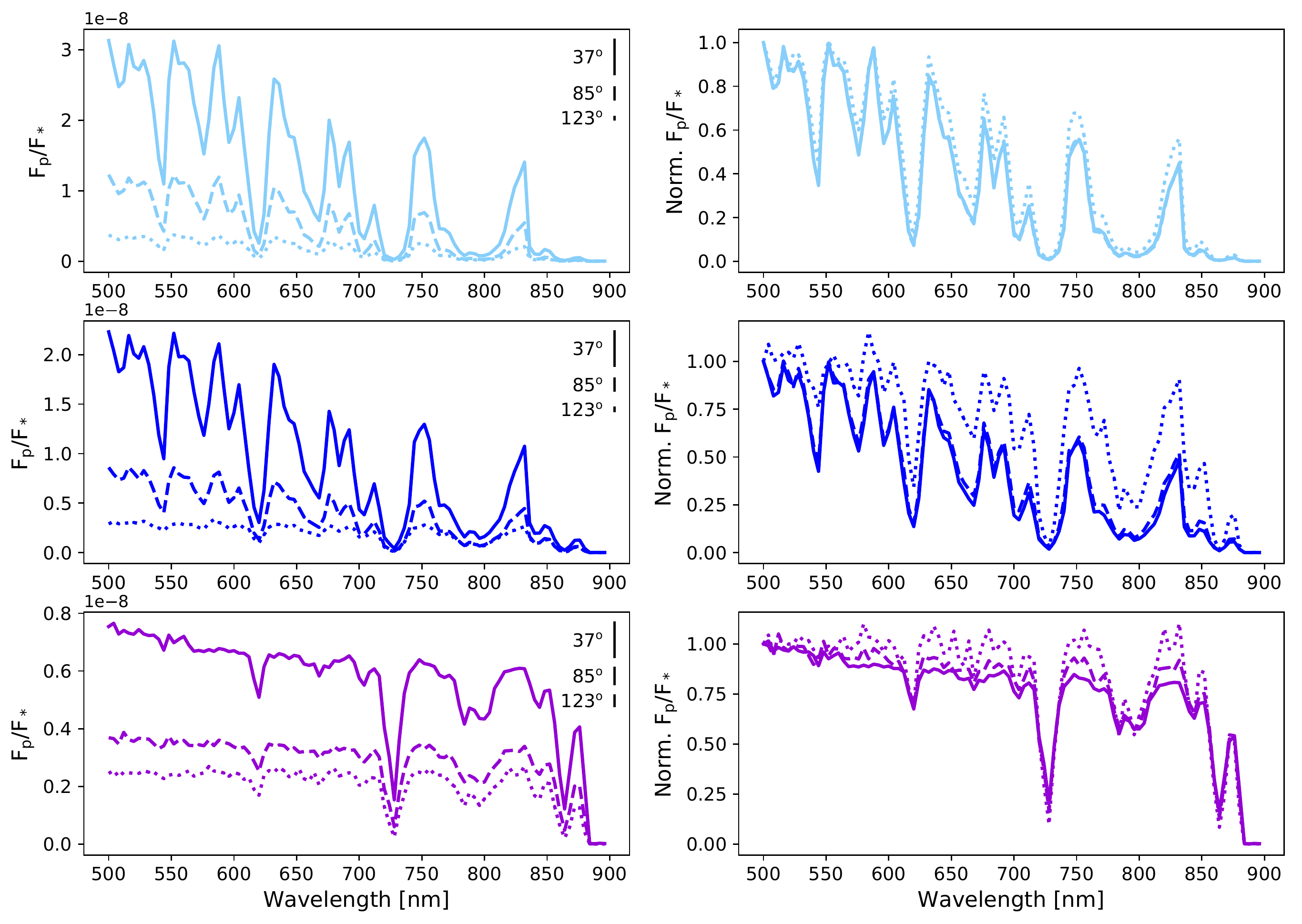}
      \caption{\label{fig:specs}
      Noiseless synthetic spectra (left column) for the no-cloud (top row and light blue), thin-cloud (middle row and darker blue) and thick-cloud (bottom row and purple) configurations.
      Solid lines in each subplot correspond to the spectrum at phase angle $\alpha=37^\circ$, dashed lines at $\alpha=85^\circ$, and dotted lines at $\alpha=123^\circ$.
      Indicated with vertical lines are the error bars corresponding to the noise at $S/N$=10 injected to create the noisy spectrum at each phase angle.
      Right column shows the same synthetic spectra but normalized with respect to the value of $F_p/F_\star(\lambda=500\,nm)$, revealing the shape changes in the spectra at different phases.
      }
   \end{figure*}

The model used here assumes an optically thick atmosphere composed of H$_2$-He with methane as the only absorbing gas and a cloud layer whose optical properties are wavelength-independent.
This configuration is generally consistent with the properties of the giant and icy planets in the Solar System.
The gas density is assumed to decrease exponentially with height and the methane abundance relative to the background gas ($f_{\rm{CH_4}}$) is assumed constant at all heights.
The cloud is parameterized by its optical thickness ($\tau_c$) and its geometrical vertical extension ($\Delta_c$).
The altitude at which the top of the cloud is located is given by the optical thickness of the gas from the top of the atmosphere (TOA) to the top of the cloud ($\tau_{\rm{c\rightarrow TOA}}$) at a reference wavelength $\lambda$=800 nm.
The aerosol particles are described by their single-scattering albedo ($\omega_0$) and the effective radius ($r_{\rm{eff}}$).
Finally, the planet radius ($R_p$) is also included as a model parameter.
Table \ref{table:model_params} summarizes our model parameters and the ranges of values defining the parameter space explored in our retrievals.

The adopted set of free parameters for the retrieval represents a fine balance between describing a realistic atmosphere and keeping a level of model complexity adequate to interpret the retrieval results. 
This balance in the choice of model parameters has long been discussed in the literature for reflected-starlight observations of Solar System objects \citep[e.g.][]{smith-tomasko1984, stephens-heidinger2000, heidinger-stephens2000, schmidetal2011} and we founded our choices on these works.

We recall that in the exponential background atmosphere adopted here the scale height of the gas $H_g$ does not appear explicitly in the radiative transfer equations \citep[see Sect. 2.3 in][]{carriongonzalezetal2020}.
In our formulation, each of the slabs in which we discretize the atmosphere is described in terms of optical thickness.
Thus, there is no need to invoke a scale height that would only be required to explicitly convert between optical thickness and altitudes.
For example, two exponential atmospheres as those utilized here with the same vertically-integrated properties but different scale heights will result in the same reflected-starlight spectrum regardless of the actual value of $H_g$. 
Other radiative transfer codes such as DISORT \citep{stamnesetal1988} also operate through the specification of optical thicknesses rather than altitudes.
For these reasons, neither $H_g$ nor the planet's gravity play a role in our calculated reflected-starlight spectra.
Indeed, although the value of $\Delta_c$ is given in units of $H_g$ (Table \ref{table:model_params}), this refers to the number of slabs in which the cloud aerosols are present \citep[Fig. 1 in][]{carriongonzalezetal2020}.

For a particular atmospheric configuration, we compute the planet-to-star contrast as:
\begin{equation}
\label{eq:Fp_Fstar}
\frac{F_p}{F_{\star}}(\alpha, \lambda)= \left(\frac{R_p}{r} \right)^2  A_g(\lambda; \;\boldsymbol{p}) 
\Phi(\alpha, \lambda; \;\boldsymbol{p})
\end{equation}
where $r$ is the planet-to-star distance.
For convenience, in our retrievals we will use the planet radius normalized to that of Neptune ($R_p/R_N$).
The geometrical albedo ($A_g$) and the normalized scattering function ($\Phi$) depend on the atmospheric properties, which are condensed in the vector of parameters $\boldsymbol{p}=\boldsymbol{p}(\tau_c; \Delta_{c}/H_g; \tau_{\rm{c\rightarrow TOA}}; r_{\rm{eff}}; \omega_0; f_{\rm{CH_4}})$ (see Table \ref{table:model_params}).
We focus on mature long-period exoplanets and hence assume no thermal-emission component at these wavelengths.

We specify our theoretical study to Barnard's Star b \citep{ribasetal2018}.
This planet candidate orbits near the snowline of its host star (0.4 AU).
Its minimum mass of $3.23 M_\oplus$ places it in the super-Earth to mini-Neptune mass regime, for which a H$_2$-He atmosphere is a physically plausible scenario \citep[see Sect. 2.2 in][]{carriongonzalezetal2020}.
Besides, this system is located at only 1.8 pc from the Sun \citep{giampapaetal1996} which makes it a promising candidate for direct-imaging observations.
\citet{carriongonzalezetal2021} found that an optimistic configuration of Roman's coronagraph could achieve a phase coverage of $\alpha \in$ [35$^{+ 23 }_{- 4 }$,120$^{+ 5 }_{- 7 }$] for this planet.
The faint magnitude of Barnard's Star ($V$=9.5 mag) will likely impede an observation with Roman. 
However, it is within the operating range of next-generation direct-imaging telescopes, which will also cover a broader range of observable phase angles.
We also note that the existence of this planet has been recently disputed by \citet{lubinetal2021}.
In our theoretical study we keep however this target as an example of a typical cool
planet around a nearby star. 
This also ensures a more straightforward comparison with \citet{carriongonzalezetal2020}.

In order to carry out a retrieval exercise, a certain atmospheric configuration has to be assumed for the observed planet. 
We refer to this as the true atmospheric configuration and in this work we analyse three possible scenarios which only differ in the cloud's optical thickness.
These are the same true atmospheric configurations as in \citet{carriongonzalezetal2020} in order to make the results comparable.
With values of $\tau_c=$ 0.05, 1.0 and 20.0, we refer to these configurations as no-, thin- and thick-cloud scenarios, respectively.
The rest of model parameters are kept the same in all three scenarios: $\Delta_{c}$/H$_g$=2, $\tau_{\rm{c\rightarrow TOA}}$=$9.1\cdot10^{-3}$, {$r_{\rm{eff}}$}=0.50 $\mu$m, $\omega_0$=0.90, and $f_{\rm{CH_4}}$=$5\cdot10^{-3}$.
We assume for this planet a radius $R_p/R_N$=0.6 and a circular orbit with $r$=0.4 AU.
We also assume that the atmosphere does not change throughout the orbit.
Cloud properties play a key role in the parameter correlations (e.g. $\tau_c - R_p$, $\tau_c - f_{\rm{CH_4}}$) that hinder the accurate atmospheric characterization with reflected-starlight spectra \citep[e.g.][]{lupuetal2016, nayaketal2017, carriongonzalezetal2020}.
By exploring different cloud scenarios, we aim to study more generally the effects of aerosols in multi-phase retrievals.

Figure \ref{fig:specs} shows (left column) the synthetic $F_p/F_\star$ for the no-, thin- and thick-cloud configurations at phase angles of 37$^\circ$ (solid lines), 85$^\circ$ (dashed lines) and 123$^\circ$ (lines).
These spectra are computed with a backward Monte Carlo radiative transfer model \citep{garciamunoz-mills2015} that ensures the accuracy of multiple-scattering for all phase angles.
The most significant trend is the reduction of the planet-to-star contrast as $\alpha$ increases and thus the apparent illuminated surface of the planet decreases.
Fig. \ref{fig:specs} also shows (right column) the spectra normalized to the value of $F_p/F_\star$ at $\lambda$=500 nm so that the relative changes in the shape of the spectra at different phase angles become evident.
We find that the normalized spectra at $\alpha=37^\circ$ and $85^\circ$ are practically identical and only at larger phases ($\alpha=123^\circ$) the relative changes in their shape are significant.

\begin{figure}
   \centering
   \includegraphics[width=9.cm]{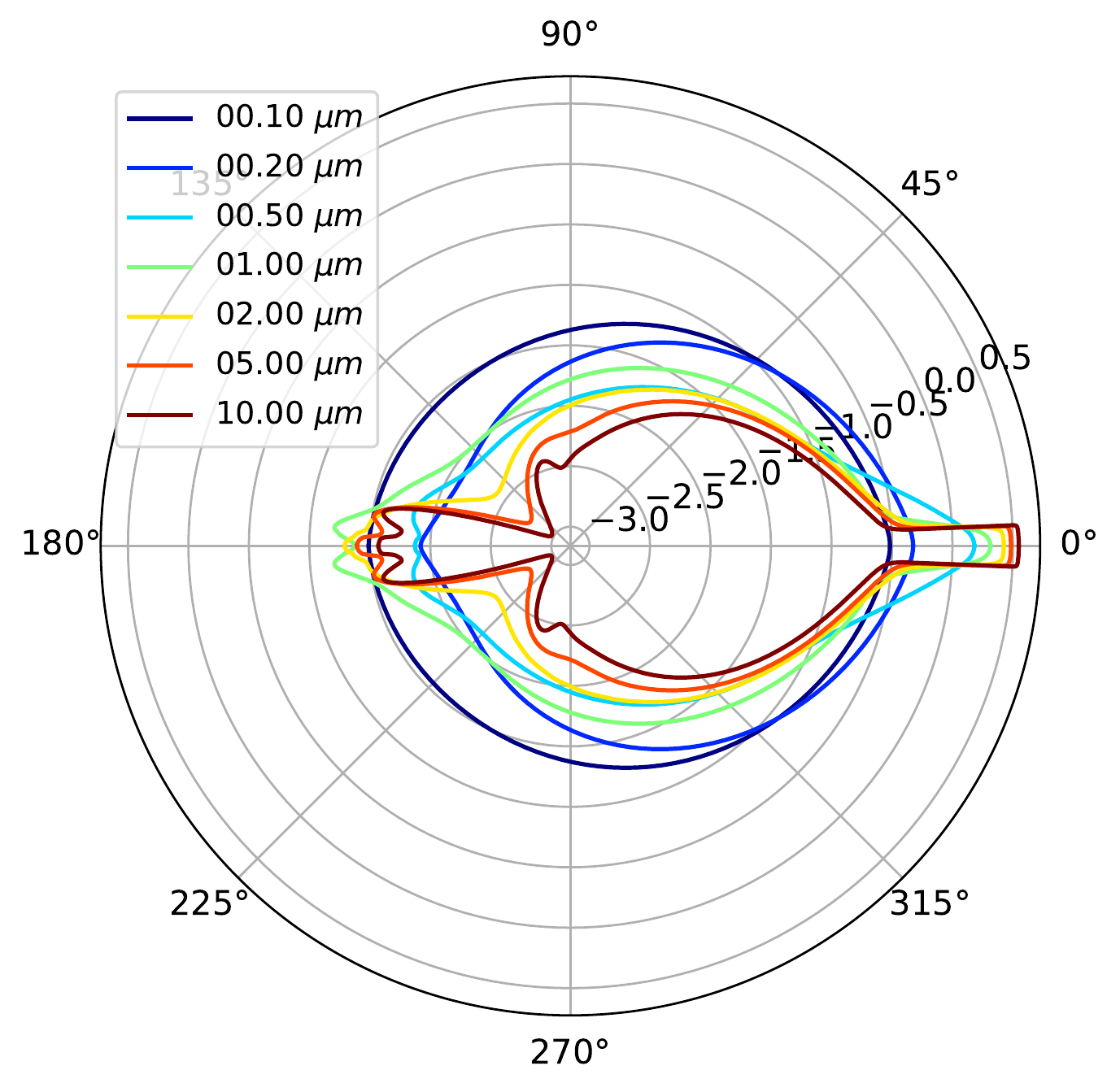}
      \caption{\label{fig:Miescattering} Dependence of the aerosol scattering phase function $p(\theta)$ with the photon scattering angle $\theta$ for the considered values of $r_{\rm{eff}}$. 
      For clarity, the plot shows $log_{10}(p(\theta))$.
      $\theta$=0$^\circ$ corresponds to forward scattering and $\theta$=180$^\circ$, to backward scattering.}
\end{figure}

In our model, $r_{\rm{eff}}$ determines the scattering phase function $p(\theta)$ of the aerosols, which is computed by means of Mie theory.
For the range of $r_{\rm{eff}}$ values considered in this work, Fig. \ref{fig:Miescattering} shows the dependence of $p(\theta)$ with the scattering angles ($\theta$).
In the single-scattering limit, $\theta$ and $\alpha$ are supplementary angles.
The forward scattering component ($\theta \sim 0^\circ$, $\alpha \sim 180^\circ$) becomes stronger for larger $r_{\rm{eff}}$.
Correspondingly, $p(\theta)$ tends to become isotropic for smaller aerosols. 
The dependence of $p(\theta)$ with $\theta$ and $r_{\rm{eff}}$ suggests that the spectra obtained at different phases contain different information and that the specific information will also depend on the macroscopic properties of the aerosols.
This consideration motivates the study of multi-phase retrievals as a strategy to constrain the atmospheric properties of the planet.

\subsection{Simultaneous retrievals at multiple phases} \label{subsec:model_sampling}
To carry out the atmospheric retrieval from an individual observation we follow the same approach as in \citet{carriongonzalezetal2020}.
We first simulate a measured spectrum by adding noise to the synthetic spectrum computed for the true atmospheric configuration. 
We assume a simplified, wavelength-independent noise model that is described by a normal distribution of mean value zero and standard deviation $\sigma_m$:
\begin{equation} \label{eq:sigma_noise}
\sigma_m= \frac{(F_p/F_{\star})_{max}}{S/N}, 
\end{equation} 
where the signal-to-noise ratio $S/N$ is given as an input and $(F_p/F_{\star})_{max}$ is the maximum value of the contrast reached in the noiseless synthetic spectrum.
This simplified noise model was used in \citet{carriongonzalezetal2020} and it allows us to focus on the specifics of radiative transfer rather than on the details of instrument-specific noise sources.

Once the measured spectrum is simulated, we use the Markov Chain Monte Carlo (MCMC) sampler \texttt{emcee} \citep{foremanmackeyetal2013} to explore the 7D parameter space.
This 7D space comprises the six atmospheric parameters in the vector $\boldsymbol{p}$ and the planet radius.
The sampler tests atmospheric configurations and obtains the corresponding synthetic spectrum by interpolating from a pre-computed grid of $\sim$300$\,$000 atmospheric configurations at each specified phase angle \citep[see][]{carriongonzalezetal2020}.
Each test spectrum is compared to the simulated measurement with the figure of merit $\chi^2$ \citep{bevington-robinson2003}, computed from Eqs. (\ref{eq:Fp_Fstar}) and (\ref{eq:sigma_noise}) as:
\begin{equation}
 \label{eq:chisq_individual}
 \chi^2(\alpha) =
\sum_{i=1}^N\left(\frac{{F_p}/{F_{\star}}(\alpha,\lambda_i)_{test}-{F_p}/{F_{\star}}(\alpha,\lambda_i)_{measured}}
 {\sigma_m} 
 \right)^2  
\end{equation}
where $N$ is the number of data points in the spectrum.
Similarly, to perform simultaneous retrieval of $n$ observations at phase angles $\alpha_1,\,...\,\alpha_n$ we also start by simulating the corresponding planet-to-star contrasts of the measurements ($F_p/F_\star(\alpha_1),\,...\,F_p/F_\star(\alpha_n)$).
For that, we add noise to each of the synthetic spectra with the desired $S/N$, obtaining $\sigma_{m_1}\,...\,\sigma_{m_n}$.
At each test atmospheric configuration, the sampler obtains the synthetic spectrum of the planet at each phase angle ($F_p/F_\star(\alpha_1)_{test}$, $F_p/F_\star(\alpha_2)_{test}$) by interpolation.
For this work we recalculated the pre-computed grid of $\sim300\,000$ atmospheric configurations by solving the radiative-transfer equation for each configuration at $\alpha=37^\circ$, $85^\circ$ and $123^\circ$.
This results in a grid of $\sim300\,000 \times 3$ spectra.
For a total of $n$ combined observations, each of them with an individual $\chi^2_j$, the combined $\chi^2$ is computed as:

\begin{equation}
 \label{eq:chisq_combined}
 \chi^2 =
\frac{
\sum_{j=1}^n \left( \chi^2_j \right)
 }{n}
\end{equation}
Unless stated otherwise, we will assume $S/N$=10 at all phase angles.
We note that the integration time needed to achieve a certain $S/N$ will increase with $\alpha$ because the apparent brightness of the planet will decrease.
Whether this will be feasible for future direct-imaging space telescopes needs to be addressed with noise models specific to the corresponding facilities.
In this work, we omit such considerations on integration times to focus on the theoretical opportunities of multi-phase observations.

\section{Results} \label{sec:results}

\begin{table*}[t]
\caption{Retrieval results for different phase angles and their combinations, for all three cloud scenarios. The quoted values correspond to the median of the marginalized posterior probability distribution for each model parameter, with the upper and lower uncertainties given by the 16\% and 84\% quantiles.} 
\label{table:results_retrievals}
\begin{center}
 \begin{tabular}{ c  c  c  c  c  c  c  c  c } 
 \hline \hline
   & $\alpha$ &  log$_{10}$($R_p/R_{N}$)  &  log$_{10}$($\tau_{c}$)  &  $\Delta_{c}$ [$H_g$]  &  log$_{10}$($\tau_{\rm{c\rightarrow TOA}}$)  &  $r_{\rm{eff}}$ [$\mu m$]  &  $\omega_0$  &  log$_{10}$($f_{\rm{CH_4}}$) \\
  \hline
  \multirow{8}{*}{\rotatebox{90}{No-cloud}} & 37$^\circ$ &  $-0.16^{+0.17}_{-0.11}$  &  $-0.16^{+0.95}_{-0.78}$  &  $3.14^{+2.19}_{-1.51}$  &  $-1.98^{+1.06}_{-0.89}$  &  $5.18^{+3.30}_{-3.39}$  &  $0.75^{+0.17}_{-0.17}$  &  $-2.30^{+0.62}_{-0.87}$ \\
  & 85$^\circ$ &  $-0.17^{+0.15}_{-0.11}$  &  $-0.22^{+0.98}_{-0.74}$  &  $3.12^{+2.21}_{-1.50}$  &  $-1.94^{+1.05}_{-0.91}$  &  $5.19^{+3.30}_{-3.38}$  &  $0.75^{+0.17}_{-0.17}$  &  $-2.25^{+0.60}_{-0.85}$ \\
  & 123$^\circ$ &  $-0.19^{+0.09}_{-0.08}$  &  $-0.34^{+1.15}_{-0.66}$  &  $3.08^{+2.15}_{-1.47}$  &  $-1.84^{+1.01}_{-0.93}$  &  $5.16^{+3.32}_{-3.39}$  &  $0.75^{+0.17}_{-0.17}$  &  $-2.20^{+0.57}_{-0.76}$ \\
  & 37$^\circ$ ($S/N$=20) &  $-0.17^{+0.12}_{-0.06}$  &  $-0.34^{+0.75}_{-0.65}$  &  $3.17^{+2.19}_{-1.54}$  &  $-2.01^{+1.11}_{-0.88}$  &  $5.17^{+3.31}_{-3.40}$  &  $0.76^{+0.16}_{-0.17}$  &  $-2.16^{+0.39}_{-0.37}$ \\
  & (37$^\circ$+85$^\circ$) &  $-0.17^{+0.16}_{-0.11}$  &  $-0.20^{+0.96}_{-0.75}$  &  $3.12^{+2.19}_{-1.51}$  &  $-1.96^{+1.06}_{-0.90}$  &  $5.16^{+3.31}_{-3.38}$  &  $0.75^{+0.17}_{-0.17}$  &  $-2.27^{+0.61}_{-0.86}$ \\
  & (37$^\circ$+123$^\circ$) &  $-0.20^{+0.09}_{-0.08}$  &  $-0.47^{+0.85}_{-0.57}$  &  $3.05^{+2.21}_{-1.45}$  &  $-1.85^{+1.08}_{-0.97}$  &  $5.09^{+3.36}_{-3.35}$  &  $0.75^{+0.17}_{-0.17}$  &  $-2.29^{+0.58}_{-0.77}$ \\
  & (37$^\circ$+85$^\circ$+123$^\circ$) &  $-0.19^{+0.09}_{-0.09}$  &  $-0.45^{+0.85}_{-0.58}$  &  $3.06^{+2.21}_{-1.46}$  &  $-1.85^{+1.08}_{-0.97}$  &  $5.07^{+3.37}_{-3.36}$  &  $0.75^{+0.17}_{-0.17}$  &  $-2.29^{+0.58}_{-0.78}$ \\
  & True values  &  $-0.22$  &  $-1.30$  &  $2$  &  $-2.04$  &  $0.50$  &  $0.90$  &  $-2.30$  \\
  \hline
 \multirow{8}{*}{\rotatebox{90}{Thin-cloud}} & 37$^\circ$ &  $-0.23^{+0.20}_{-0.11}$  &  $-0.05^{+0.95}_{-0.85}$  &  $3.17^{+2.18}_{-1.54}$  &  $-1.99^{+1.05}_{-0.88}$  &  $5.18^{+3.28}_{-3.40}$  &  $0.74^{+0.17}_{-0.17}$  &  $-2.49^{+0.71}_{-0.94}$ \\
  & 85$^\circ$ &  $-0.23^{+0.21}_{-0.11}$  &  $-0.02^{+0.97}_{-0.86}$  &  $3.17^{+2.17}_{-1.54}$  &  $-1.99^{+1.04}_{-0.86}$  &  $5.15^{+3.32}_{-3.39}$  &  $0.74^{+0.17}_{-0.17}$  &  $-2.64^{+0.78}_{-0.96}$ \\
  & 123$^\circ$ &  $-0.22^{+0.13}_{-0.08}$  &  $-0.07^{+1.11}_{-0.83}$  &  $3.21^{+2.19}_{-1.57}$  &  $-2.07^{+1.06}_{-0.79}$  &  $5.02^{+3.38}_{-3.34}$  &  $0.75^{+0.17}_{-0.17}$  &  $-2.94^{+0.93}_{-0.90}$ \\
  & 37$^\circ$ ($S/N$=20) &  $-0.24^{+0.16}_{-0.07}$  &  $-0.21^{+0.75}_{-0.73}$  &  $3.17^{+2.22}_{-1.54}$  &  $-2.04^{+1.09}_{-0.85}$  &  $5.13^{+3.32}_{-3.39}$  &  $0.76^{+0.16}_{-0.18}$  &  $-2.36^{+0.45}_{-0.40}$ \\
  & (37$^\circ$+85$^\circ$) &  $-0.23^{+0.20}_{-0.11}$  &  $-0.05^{+0.96}_{-0.84}$  &  $3.15^{+2.16}_{-1.52}$  &  $-1.97^{+1.03}_{-0.88}$  &  $5.17^{+3.31}_{-3.38}$  &  $0.74^{+0.17}_{-0.17}$  &  $-2.57^{+0.74}_{-0.94}$ \\
  & (37$^\circ$+123$^\circ$) &  $-0.23^{+0.12}_{-0.09}$  &  $-0.01^{+0.83}_{-0.80}$  &  $3.18^{+2.16}_{-1.54}$  &  $-1.96^{+1.02}_{-0.89}$  &  $4.88^{+3.46}_{-3.30}$  &  $0.77^{+0.16}_{-0.18}$  &  $-2.69^{+0.73}_{-0.89}$ \\
  & (37$^\circ$+85$^\circ$+123$^\circ$) &  $-0.23^{+0.13}_{-0.09}$  &  $-0.05^{+0.84}_{-0.79}$  &  $3.12^{+2.16}_{-1.50}$  &  $-1.93^{+1.00}_{-0.92}$  &  $4.94^{+3.42}_{-3.31}$  &  $0.76^{+0.17}_{-0.18}$  &  $-2.70^{+0.73}_{-0.90}$ \\
  & True values  &  $-0.22$  &  $0.0$  &  $2$  &  $-2.04$  &  $0.50$  &  $0.90$  &  $-2.30$  \\
  \hline
  \multirow{8}{*}{\rotatebox{90}{Thick-cloud}} & 37$^\circ$ &  $-0.43^{+0.33}_{-0.11}$  &  $0.13^{+1.07}_{-0.98}$  &  $3.36^{+2.18}_{-1.67}$  &  $-2.30^{+1.06}_{-0.71}$  &  $4.89^{+3.45}_{-3.31}$  &  $0.78^{+0.15}_{-0.18}$  &  $-4.09^{+1.29}_{-0.60}$ \\
  & 85$^\circ$ &  $-0.38^{+0.31}_{-0.10}$  &  $0.07^{+1.10}_{-0.94}$  &  $3.37^{+2.17}_{-1.67}$  &  $-2.30^{+1.05}_{-0.71}$  &  $4.86^{+3.47}_{-3.33}$  &  $0.79^{+0.15}_{-0.18}$  &  $-4.02^{+1.31}_{-0.62}$ \\
  & 123$^\circ$ &  $-0.24^{+0.15}_{-0.07}$  &  $0.15^{+1.05}_{-0.99}$  &  $3.33^{+2.18}_{-1.65}$  &  $-2.25^{+1.03}_{-0.72}$  &  $4.95^{+3.42}_{-3.36}$  &  $0.78^{+0.16}_{-0.18}$  &  $-3.49^{+1.25}_{-0.86}$ \\
  & 37$^\circ$ ($S/N$=20) &  $-0.49^{+0.24}_{-0.05}$  &  $-0.28^{+1.25}_{-0.69}$  &  $3.35^{+2.18}_{-1.66}$  &  $-2.28^{+1.05}_{-0.73}$  &  $4.89^{+3.46}_{-3.32}$  &  $0.80^{+0.15}_{-0.19}$  &  $-4.37^{+1.34}_{-0.37}$ \\
  & (37$^\circ$+85$^\circ$) &  $-0.37^{+0.34}_{-0.14}$  &  $0.47^{+0.86}_{-1.26}$  &  $3.37^{+2.19}_{-1.68}$  &  $-2.31^{+1.05}_{-0.69}$  &  $4.71^{+3.52}_{-3.23}$  &  $0.80^{+0.14}_{-0.18}$  &  $-3.90^{+1.51}_{-0.72}$ \\
  & (37$^\circ$+123$^\circ$) &  $-0.16^{+0.12}_{-0.12}$  &  $1.04^{+0.45}_{-0.46}$  &  $3.62^{+2.06}_{-1.80}$  &  $-2.37^{+0.60}_{-0.58}$  &  $4.04^{+3.78}_{-2.78}$  &  $0.80^{+0.12}_{-0.18}$  &  $-2.77^{+0.85}_{-1.17}$ \\
  & (37$^\circ$+85$^\circ$+123$^\circ$) &  $-0.20^{+0.14}_{-0.14}$  &  $0.98^{+0.49}_{-0.52}$  &  $3.55^{+2.08}_{-1.76}$  &  $-2.35^{+0.66}_{-0.61}$  &  $3.97^{+3.84}_{-2.80}$  &  $0.83^{+0.11}_{-0.18}$  &  $-2.94^{+0.96}_{-1.21}$ \\
  & True values  &  $-0.22$  &  $1.30$  &  $2$  &  $-2.04$  &  $0.50$  &  $0.90$  &  $-2.30$  \\
  \hline
\end{tabular}
\end{center}
\end{table*}

Below, we analyse the retrieval results and the potential gain from multi-phase measurements.
To that end, we first study the retrievals of single-phase observations at phase angles $\alpha$=37$^\circ$, 85$^\circ$ and 123$^\circ$ separately (Sect. \ref{subsec:results_individual}).
Then, we explore several combinations of phase angles (Sect. \ref{subsec:results_combined}).

The phase angle $\alpha$=85$^\circ$ is representative of orbital positions near quadrature, where the maximum planet-star angular separation is reached.
This phase is expected to be accessible for several known exoplanets with different possible configurations of the Roman Telescope's coronagraph \citep{carriongonzalezetal2021}.
Correspondingly, $\alpha$=37$^\circ$ and 123$^\circ$ are realistic lower and upper limits of the phase angles that will be observable with future direct-imaging space telescopes.
This range of phases may only be accessible to the Roman Telescope's coronagraph for a small number of exoplanets and in the best-case scenario of IWA, OWA and $C_{min}$ considered in \citet{carriongonzalezetal2021}.

\subsection{Retrievals for single-phase measurements} \label{subsec:results_individual}

  \begin{figure*}
    \includegraphics[width=0.16\linewidth]{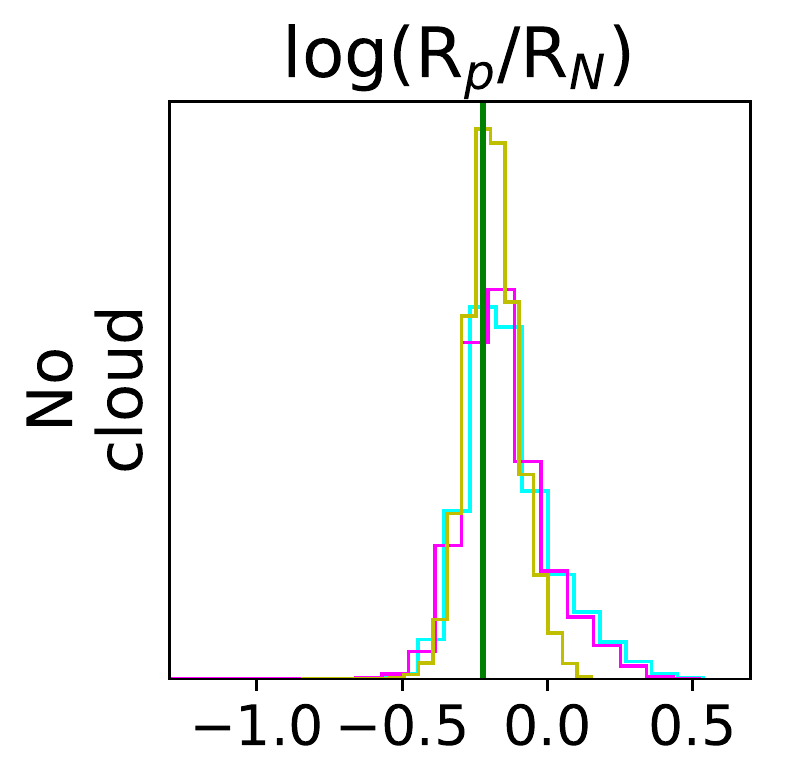}
    \includegraphics[width=0.13\linewidth]{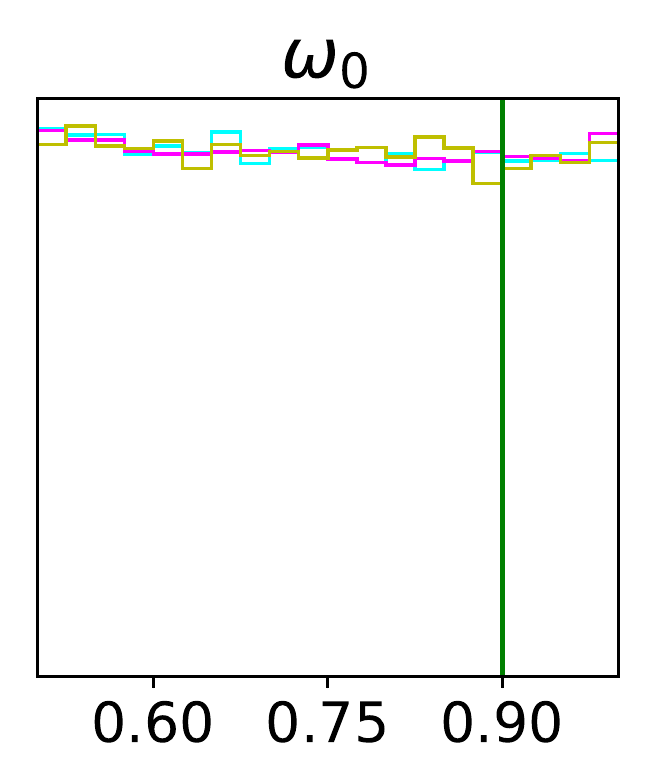}
    \includegraphics[width=0.13\linewidth]{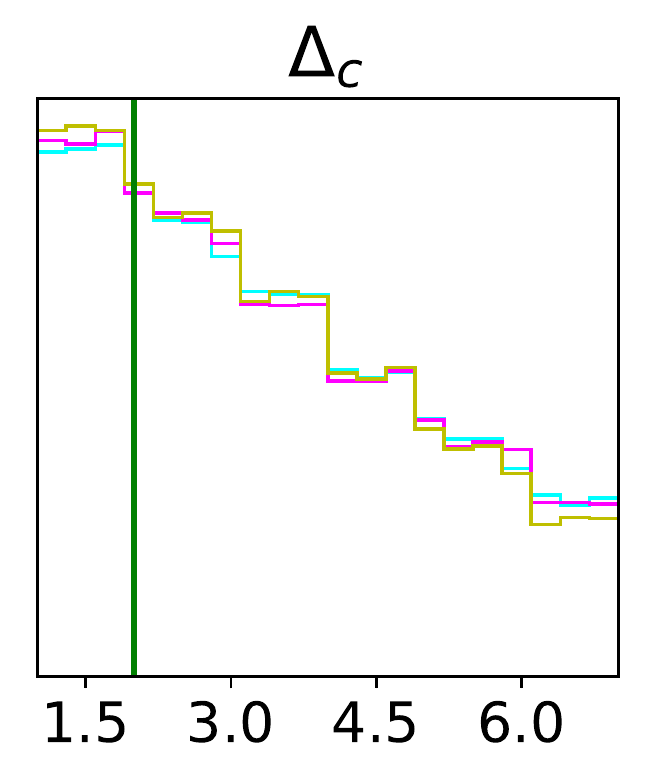}
    \includegraphics[width=0.13\linewidth]{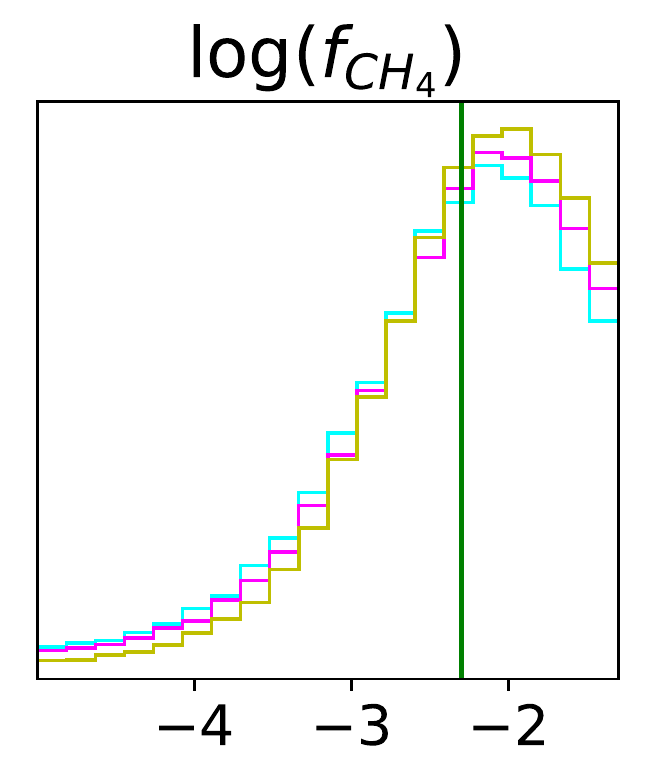}
    \includegraphics[width=0.14\linewidth]{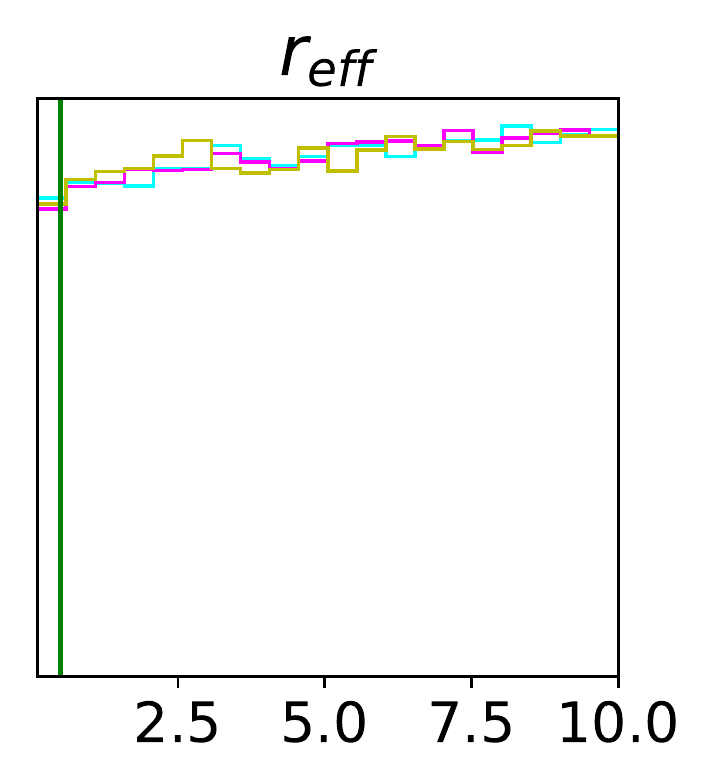}
    \includegraphics[width=0.13\linewidth]{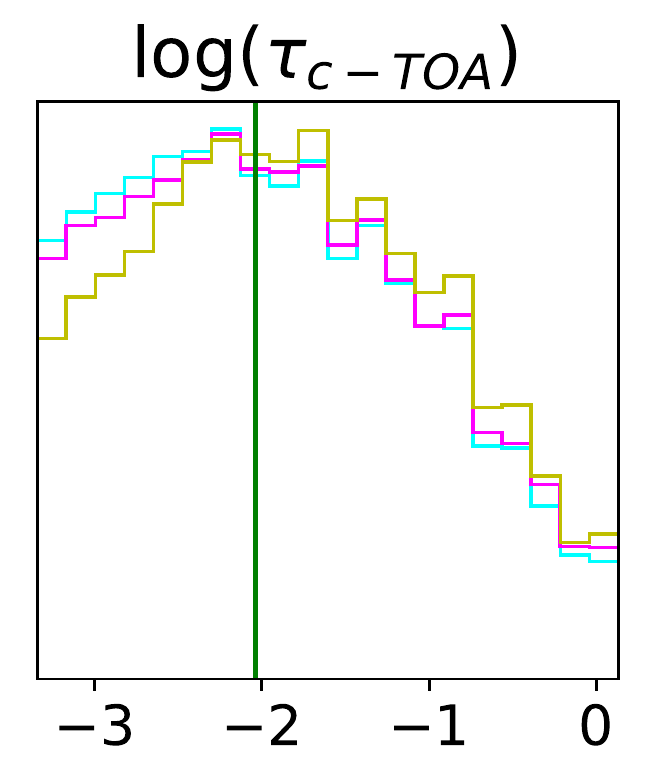}
    \includegraphics[width=0.13\linewidth]{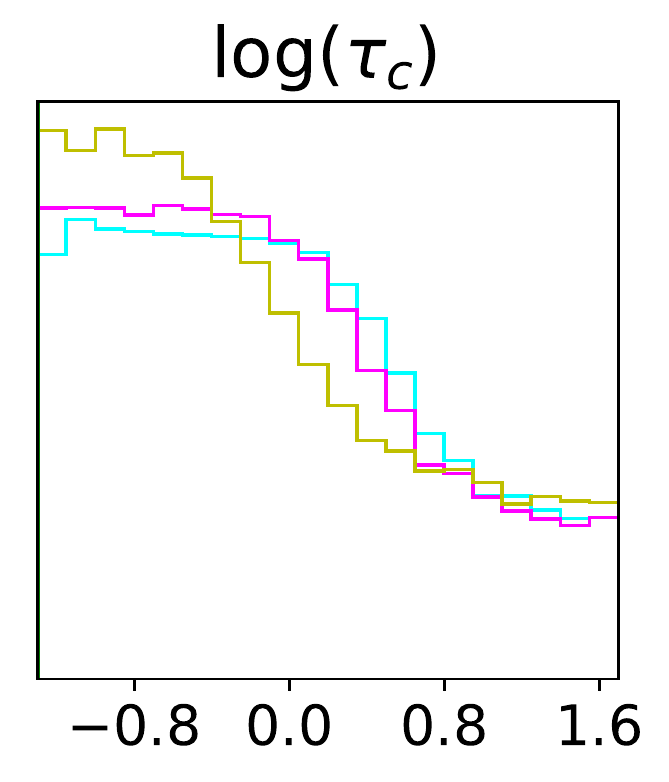}
    \\
    \includegraphics[width=0.16\linewidth]{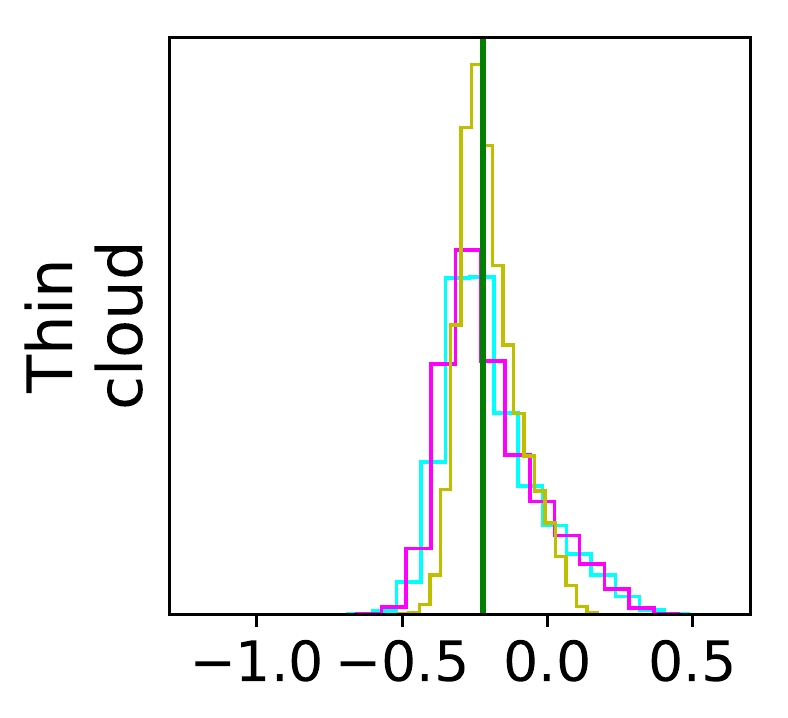}
    \includegraphics[width=0.13\linewidth]{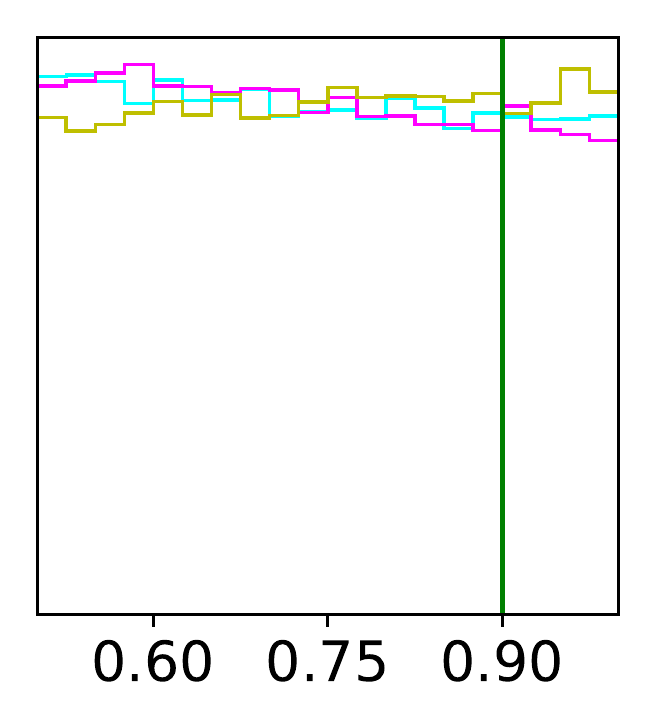}
    \includegraphics[width=0.13\linewidth]{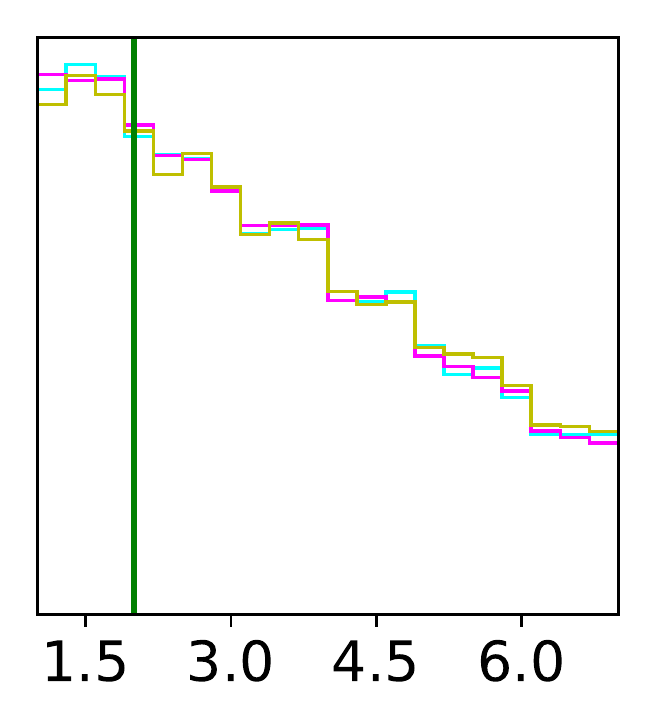}
    \includegraphics[width=0.13\linewidth]{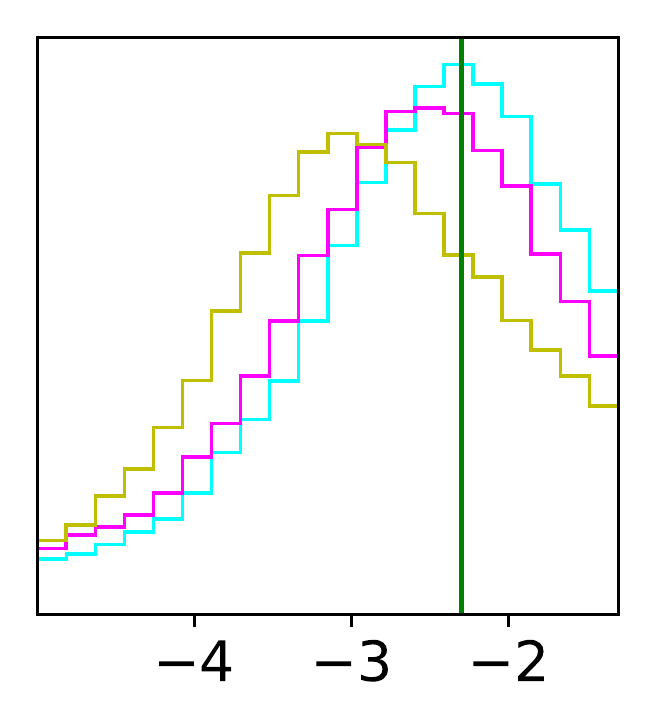}
    \includegraphics[width=0.14\linewidth]{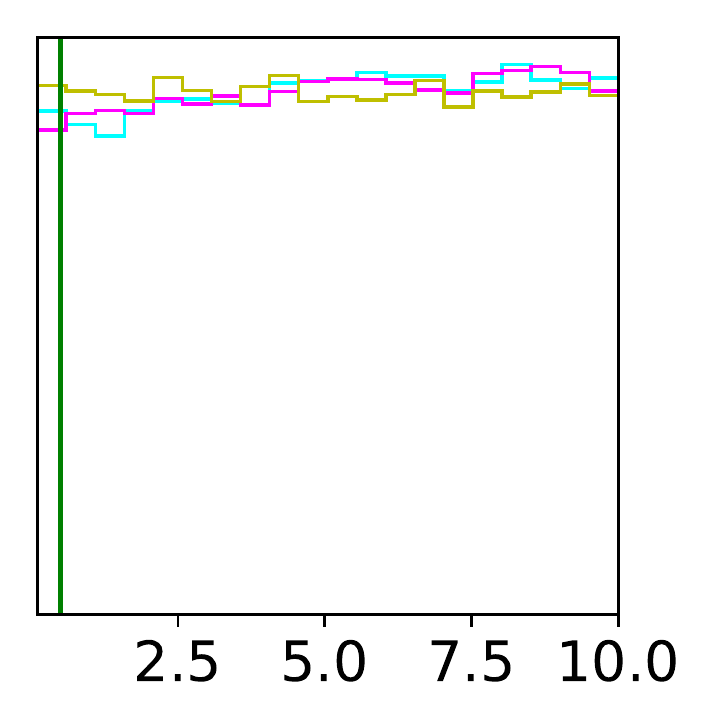}
    \includegraphics[width=0.13\linewidth]{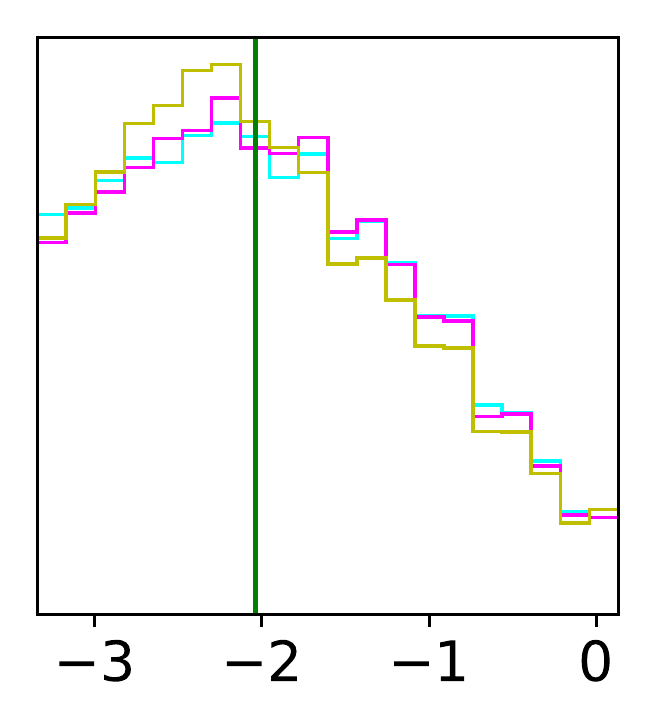}
    \includegraphics[width=0.13\linewidth]{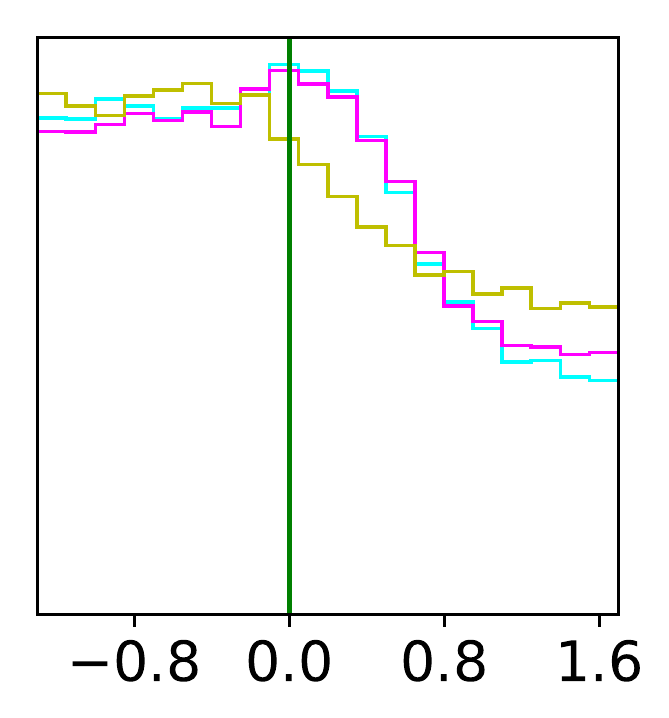}
    \\
    \includegraphics[width=0.16\linewidth]{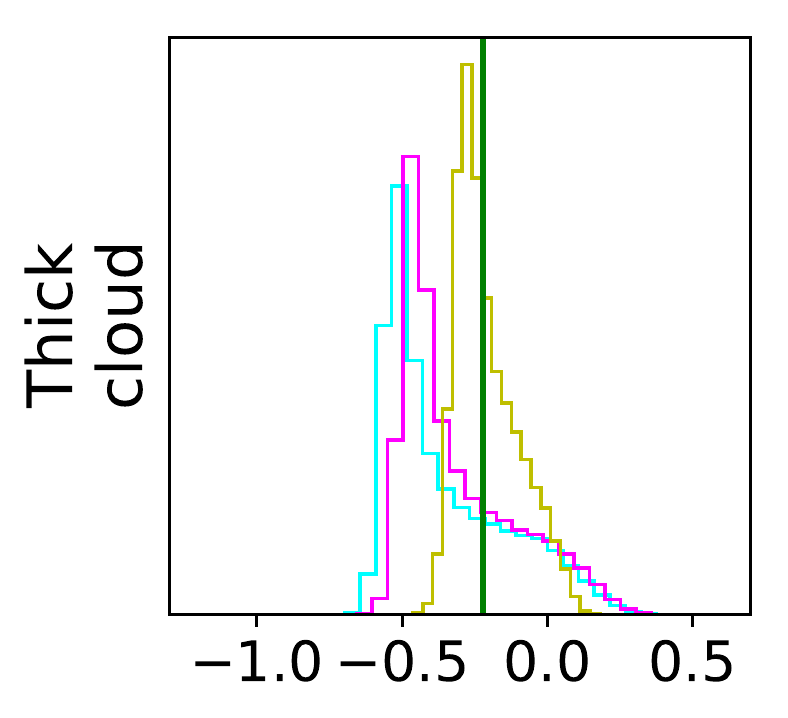}
    \includegraphics[width=0.13\linewidth]{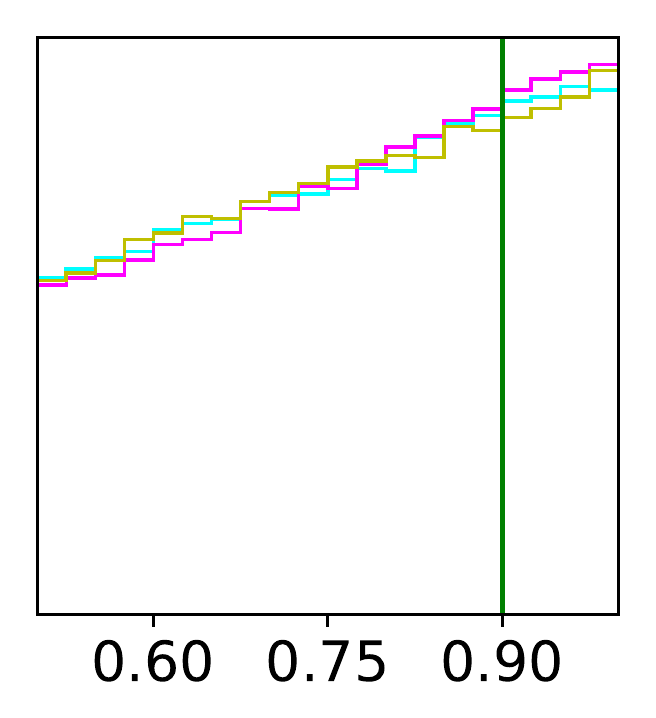}
    \includegraphics[width=0.13\linewidth]{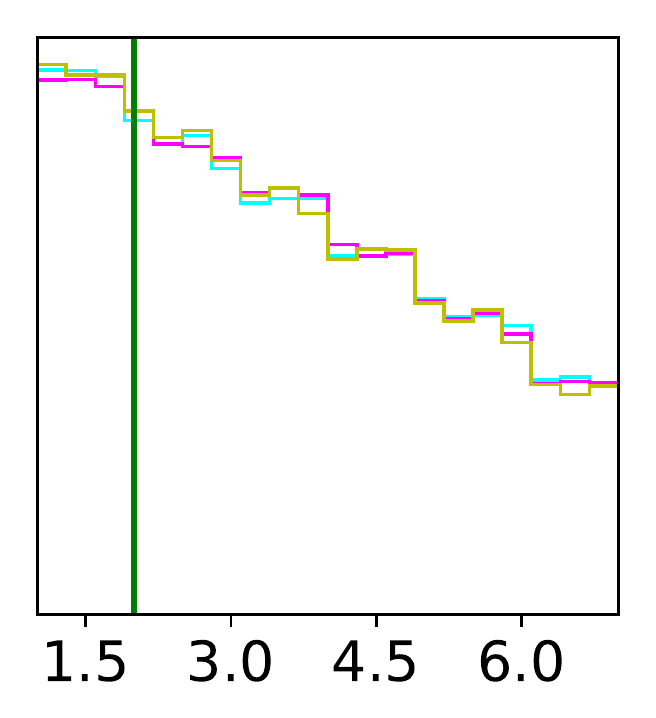}
    \includegraphics[width=0.13\linewidth]{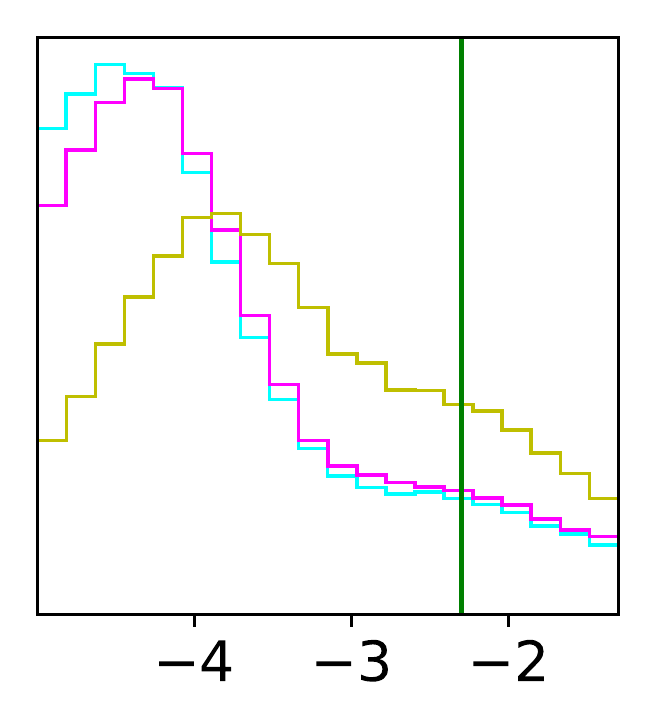}
    \includegraphics[width=0.14\linewidth]{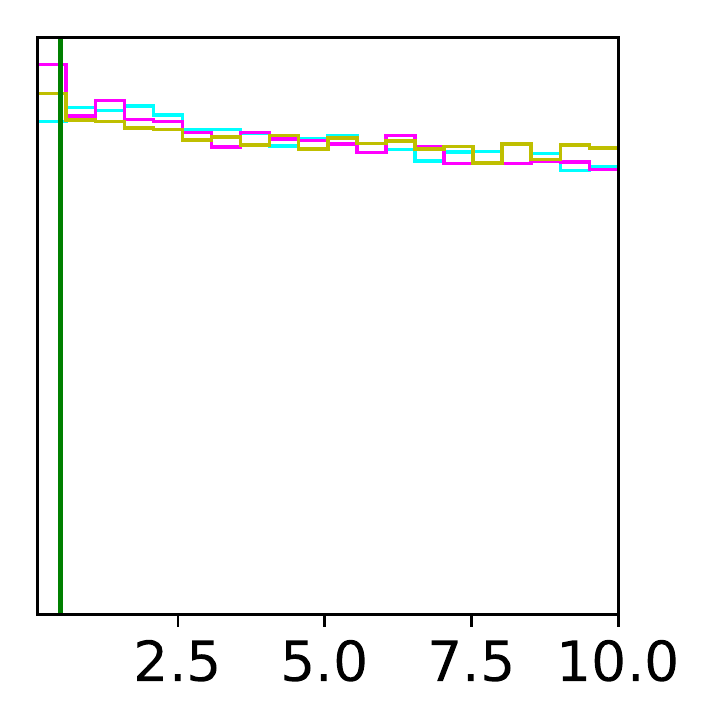}
    \includegraphics[width=0.13\linewidth]{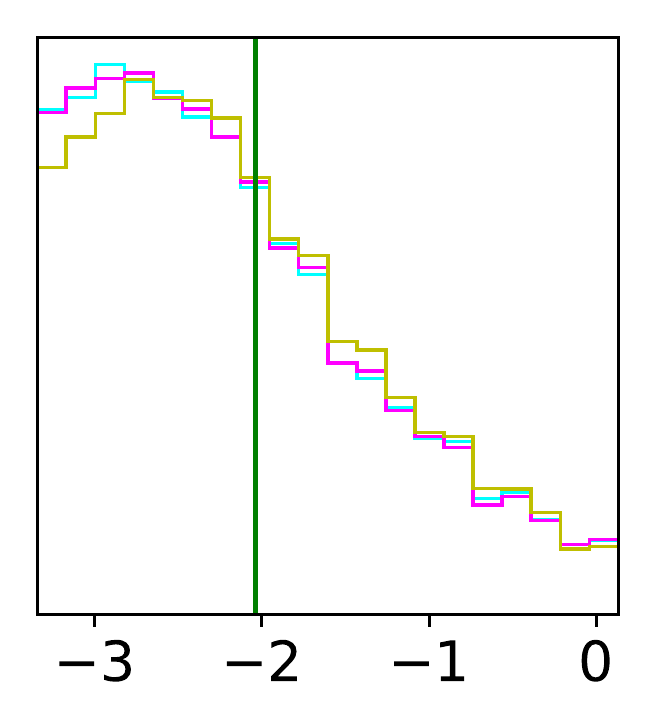}
    \includegraphics[width=0.13\linewidth]{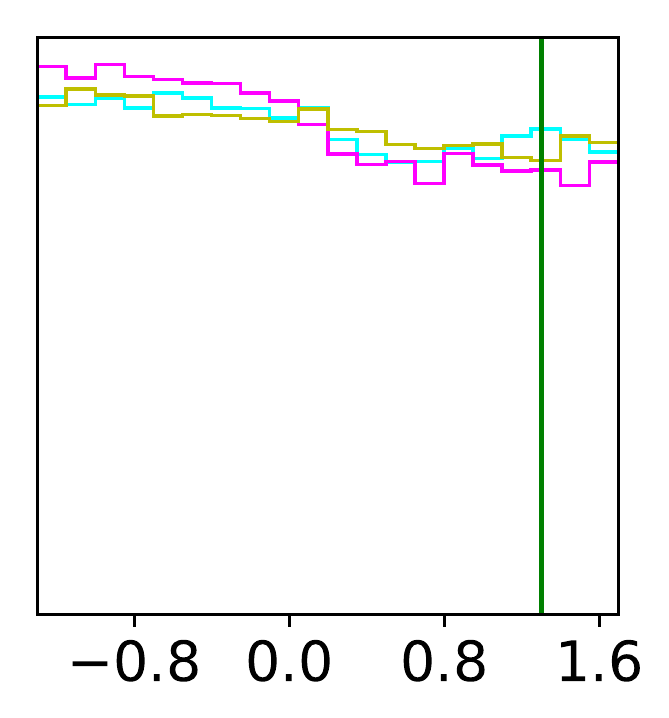}
    \caption{\label{fig:individual_retrievals_1D}
    Marginalized posterior probability distributions of each model parameter for single-phase observations in the no-, thin- and thick-cloud scenarios (top, middle and bottom rows, respectively).
    The single-phase observations are obtained at phase angles 37$^\circ$ (cyan lines), 85$^\circ$ (magenta lines) and 123$^\circ$ (yellow lines).
    In all cases, the signal-to-noise ratio is $S/N$=10.
    Green lines mark the true values of the model parameters (see Sect. \ref{subsec:model_atmosphere}) for this scenario.
    The 2D posterior probability distributions for these retrievals can be found in Figs. \ref{fig:results_individual_nocloud}$-$\ref{fig:results_individual_thickcloud}.}
    \end{figure*}

  \begin{figure*}
    \includegraphics[width=0.16\linewidth]{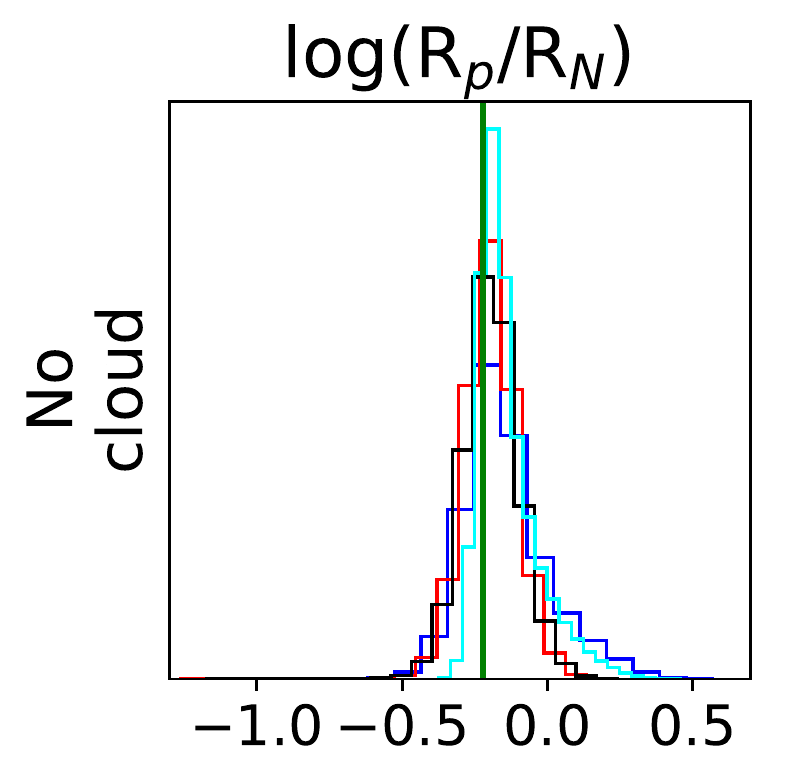}
    \includegraphics[width=0.13\linewidth]{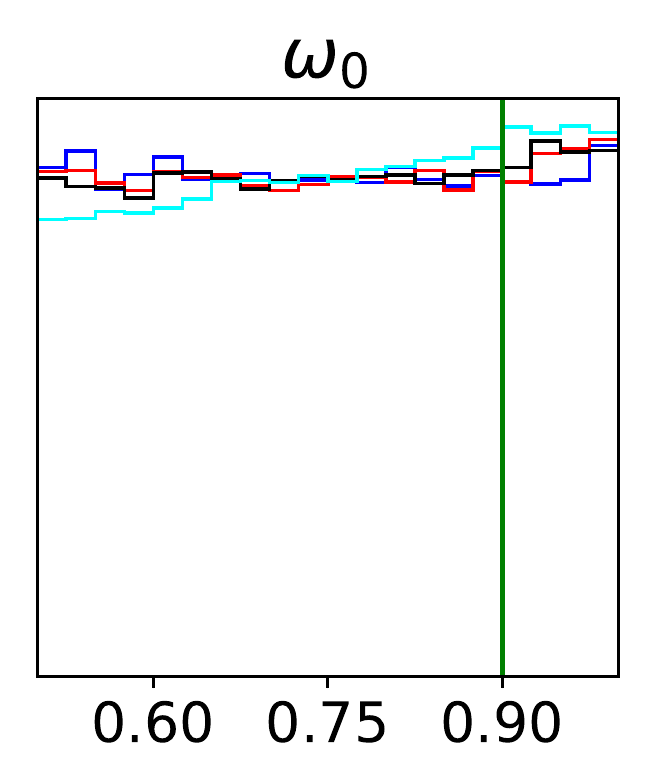}
    \includegraphics[width=0.13\linewidth]{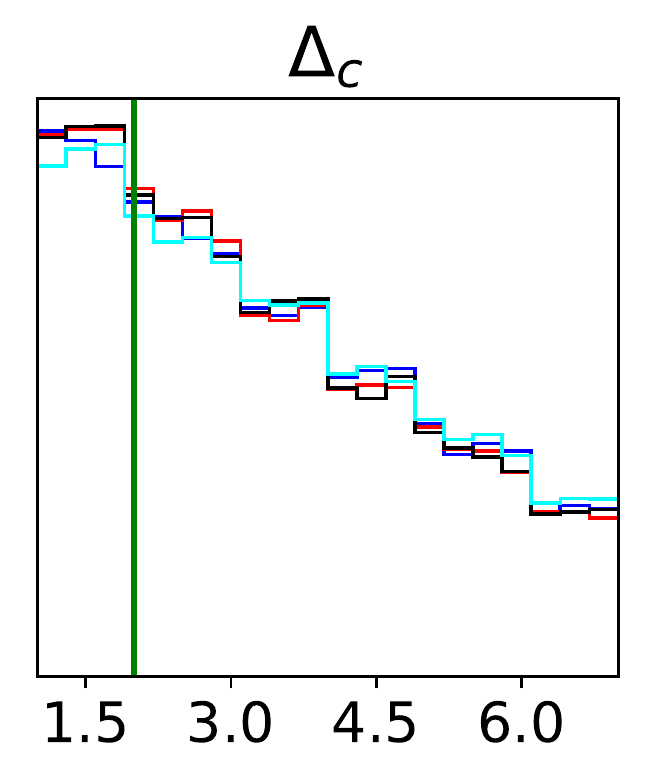}
    \includegraphics[width=0.13\linewidth]{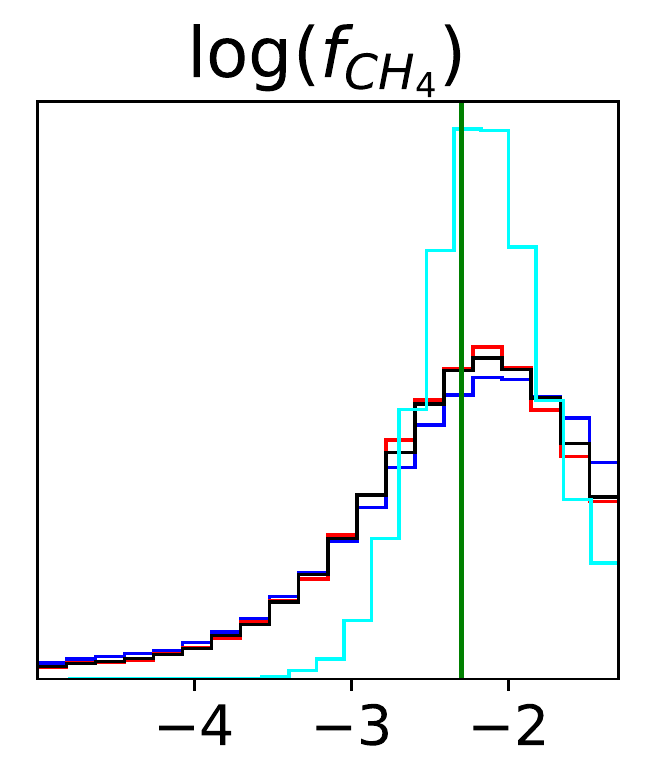}
    \includegraphics[width=0.14\linewidth]{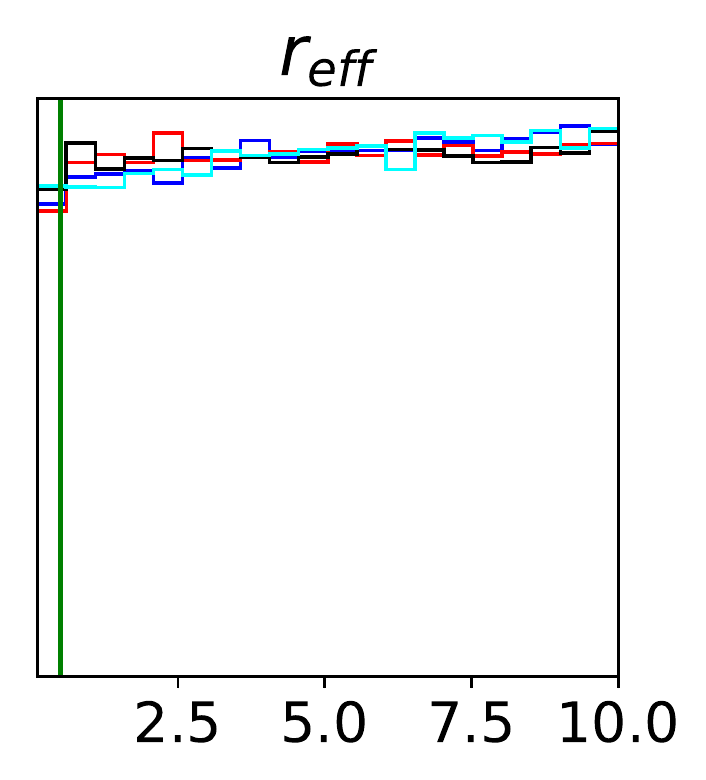}
    \includegraphics[width=0.13\linewidth]{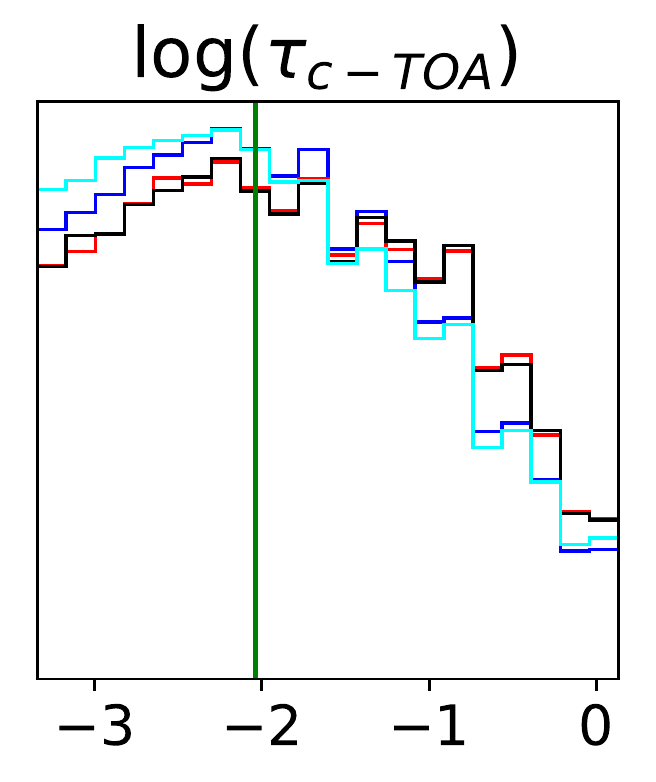}
    \includegraphics[width=0.13\linewidth]{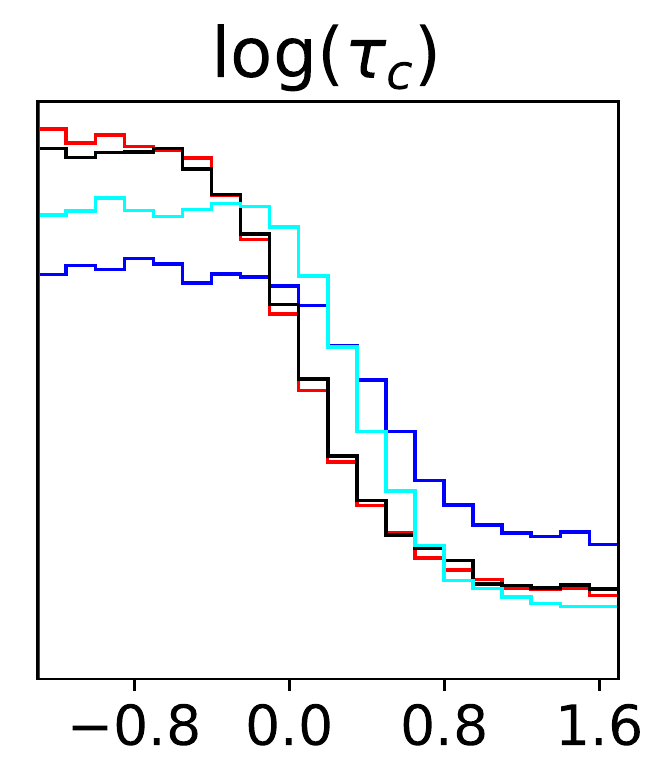}
    \\
    \includegraphics[width=0.16\linewidth]{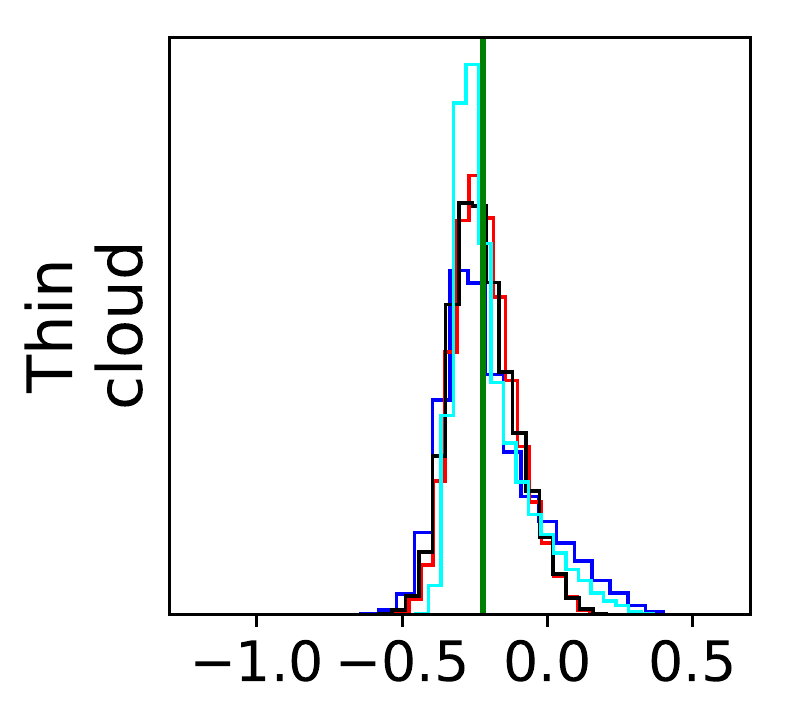}
    \includegraphics[width=0.13\linewidth]{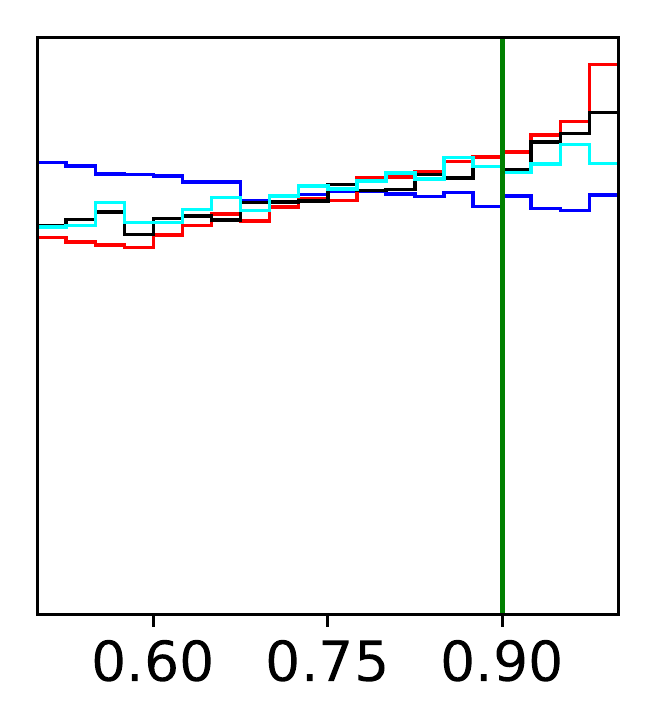}
    \includegraphics[width=0.13\linewidth]{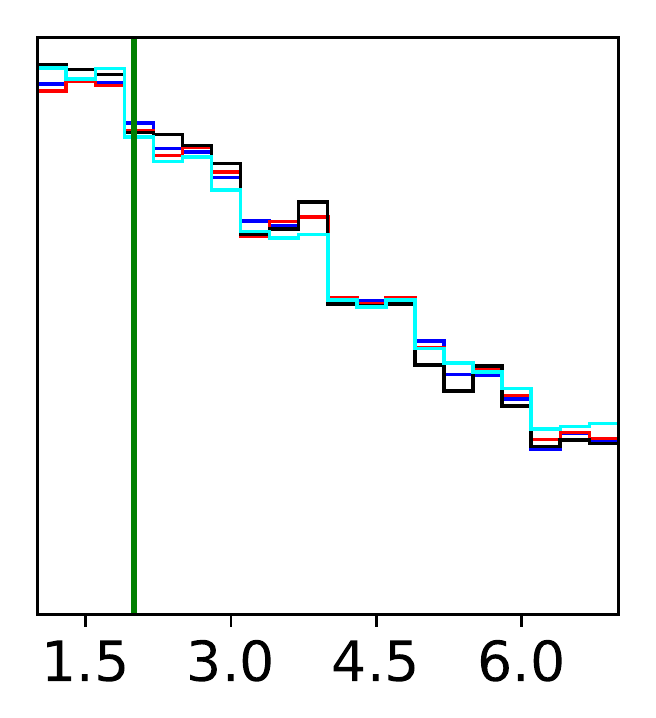}
    \includegraphics[width=0.13\linewidth]{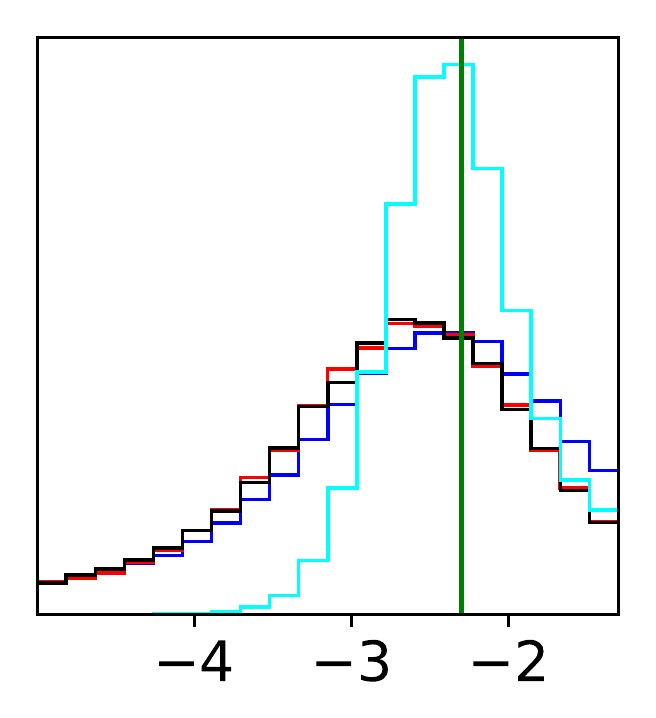}
    \includegraphics[width=0.14\linewidth]{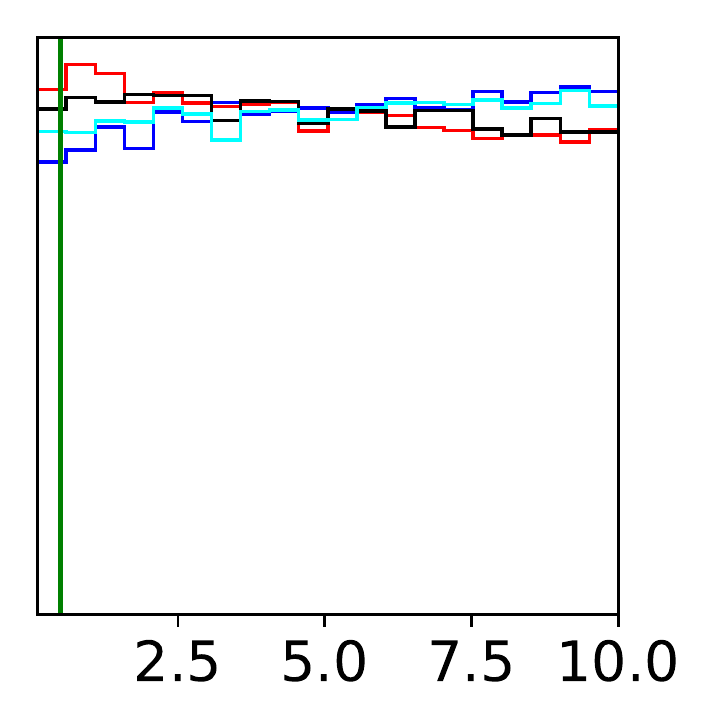}
    \includegraphics[width=0.13\linewidth]{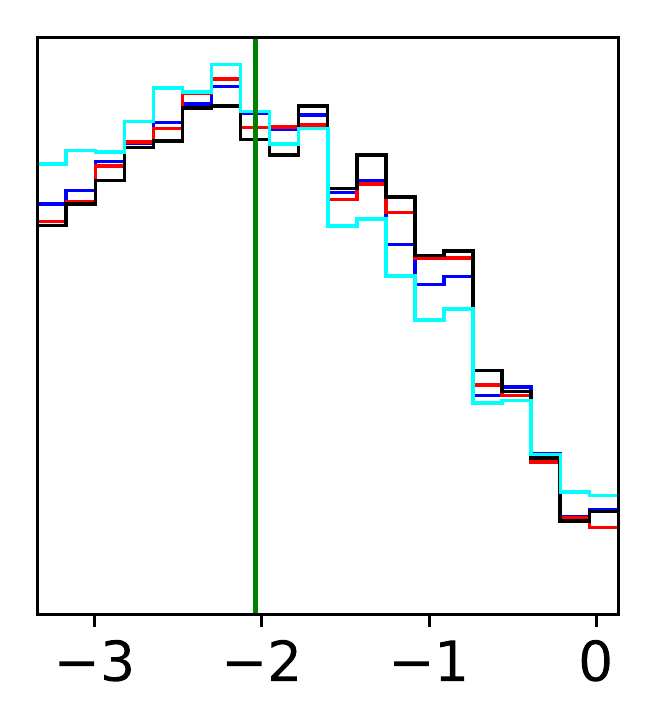}
    \includegraphics[width=0.13\linewidth]{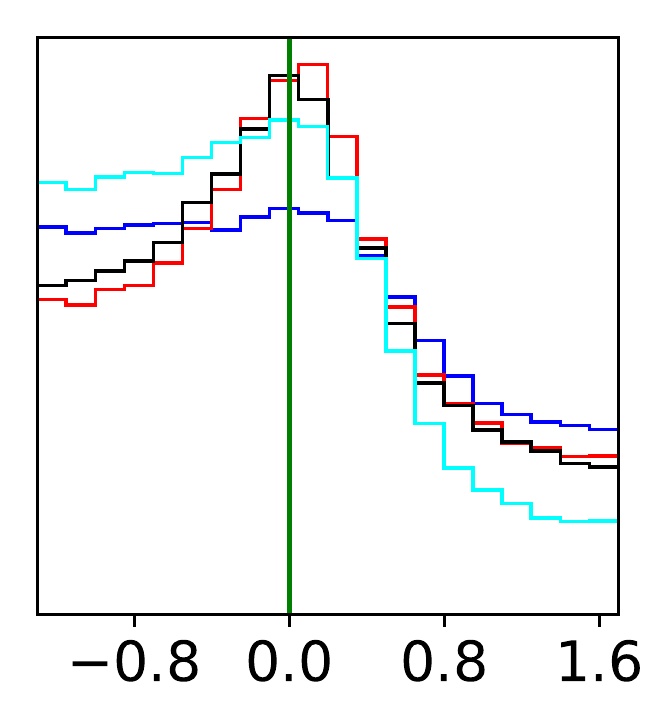}
    \\
    \includegraphics[width=0.16\linewidth]{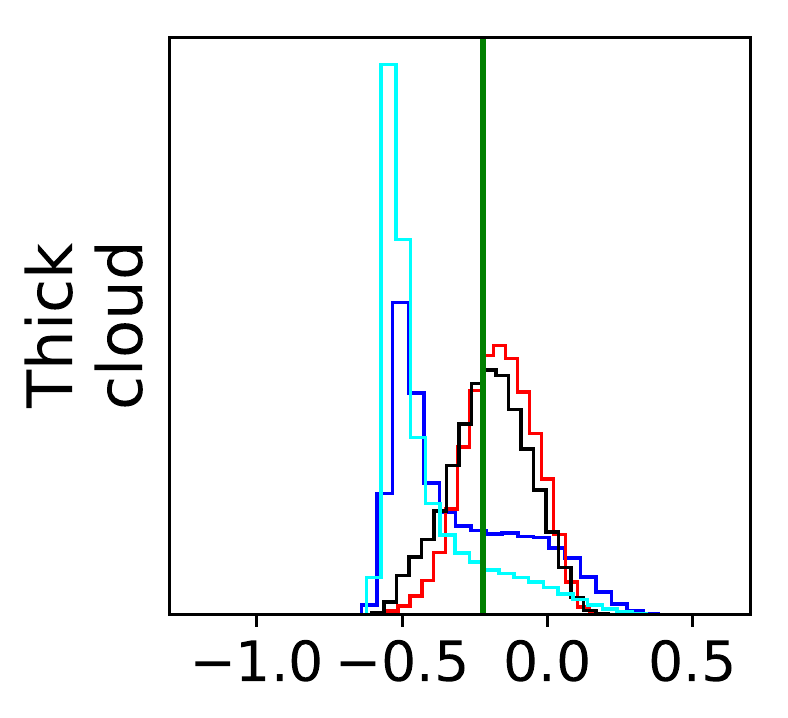}
    \includegraphics[width=0.13\linewidth]{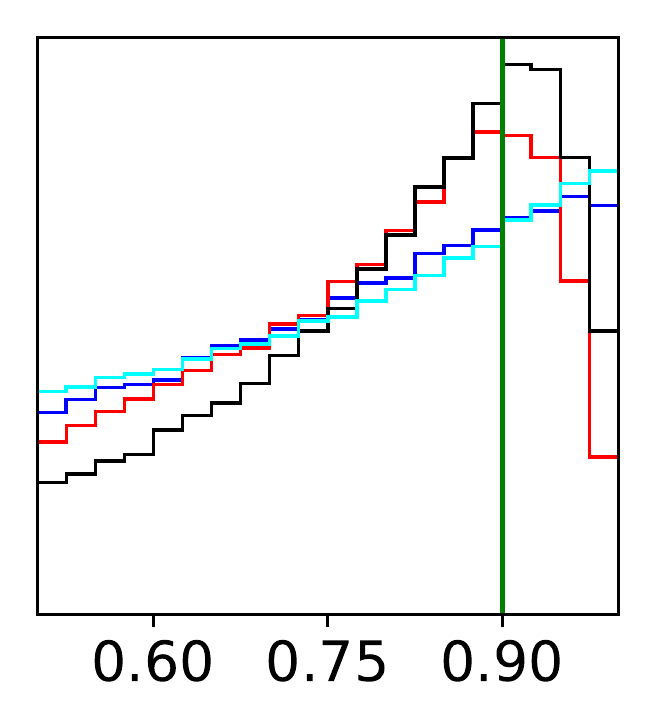}
    \includegraphics[width=0.13\linewidth]{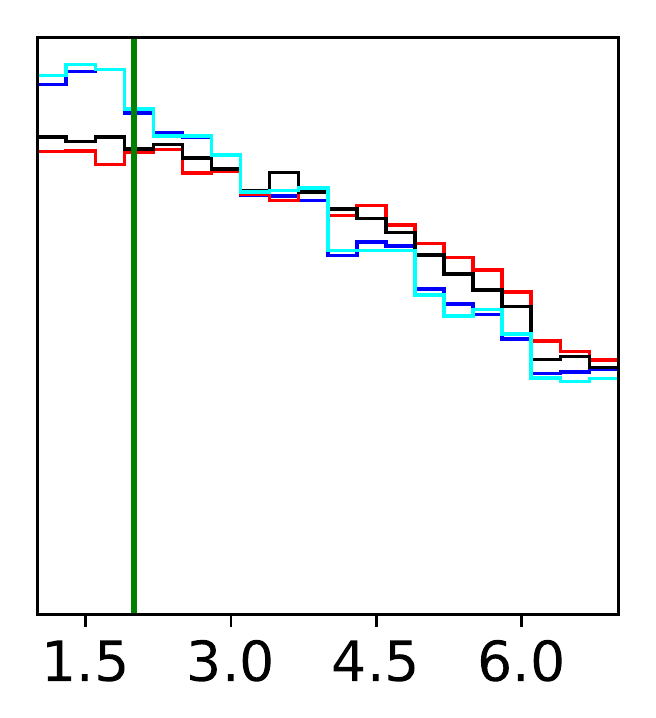}
    \includegraphics[width=0.13\linewidth]{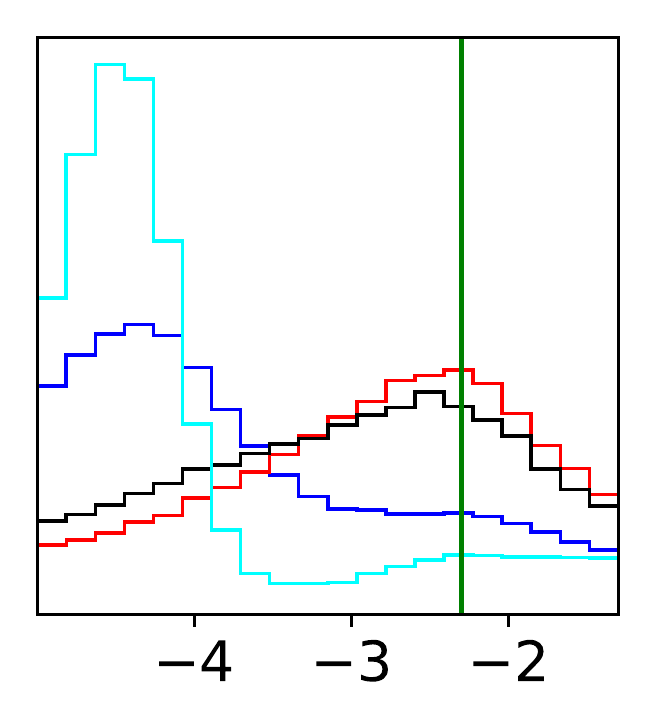}
    \includegraphics[width=0.14\linewidth]{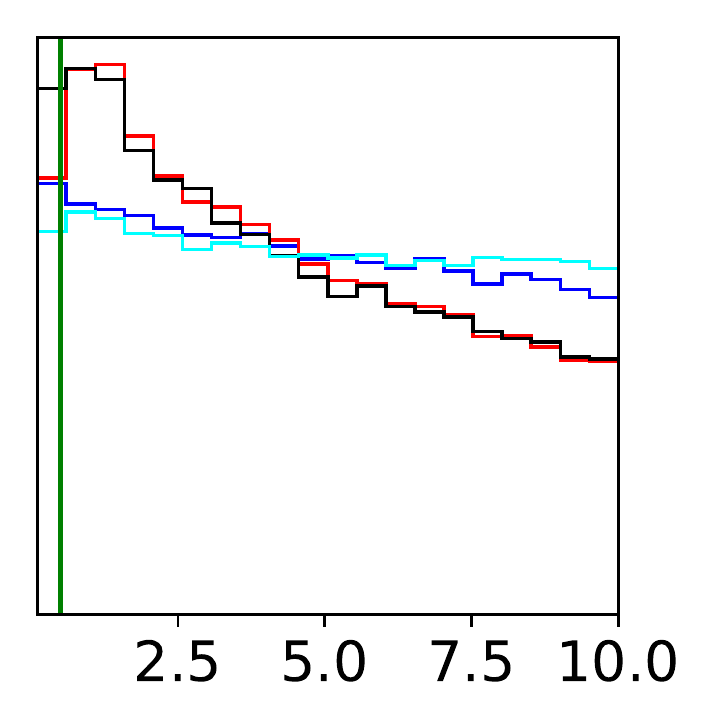}
    \includegraphics[width=0.13\linewidth]{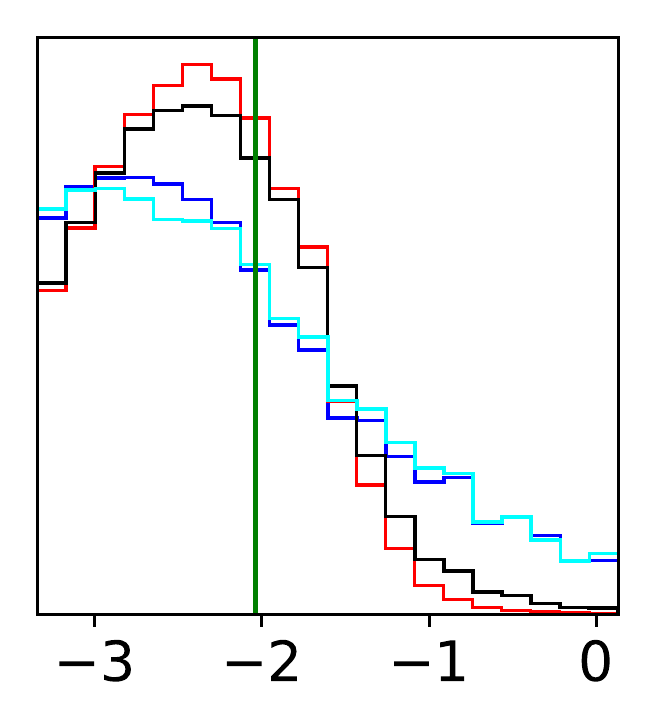}
    \includegraphics[width=0.13\linewidth]{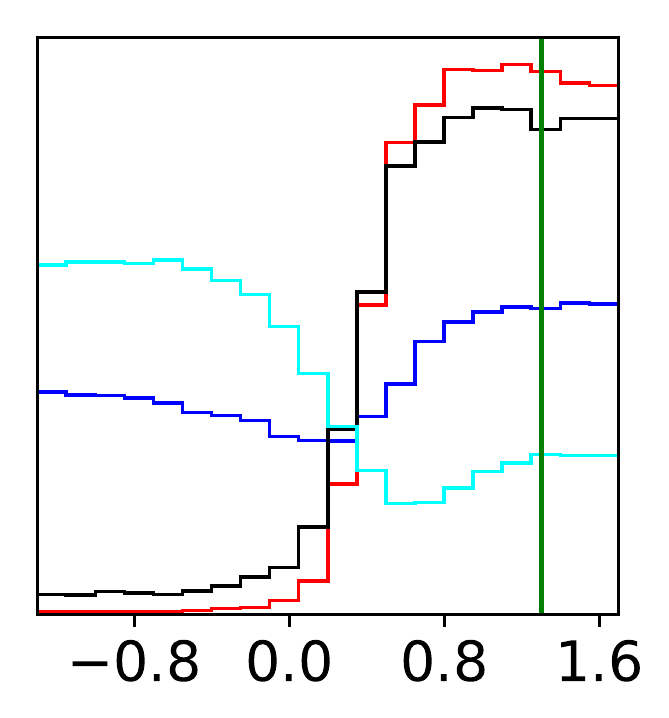}
    \caption{\label{fig:combined_retrievals_1D}
    Marginalized posterior probability distributions of each model parameter for several combinations of observations in the no-, thin- and thick-cloud scenarios (top, middle and bottom row, respectively).
    Blue lines mark the combination (37$^\circ$+85$^\circ$); red lines, (37$^\circ$+123$^\circ$) and black lines, (37$^\circ$+85$^\circ$+123$^\circ$).
    For reference, cyan lines indicate the results of a single-phase observation at $\alpha=37^\circ$ but doubling the signal-to-noise ratio ($S/N$=20).
    Green lines mark the true values of the model parameters (see Sect. \ref{subsec:model_atmosphere}) for this observation.
    The 2D posterior probability distributions for these retrievals can be found in Figs. \ref{fig:results_combined_nocloud}$-$\ref{fig:results_combined_thickcloud}.}
    \end{figure*}

For each of the three cloud scenarios, we carried out single-phase retrievals for observations at $\alpha$=37$^\circ$, 85$^\circ$ and 123$^\circ$ with $S/N$=10 in all cases.
Figures \ref{fig:results_individual_nocloud}$-$\ref{fig:results_individual_thickcloud} show the posterior probability distributions of the model parameters for each cloud scenario.
Contour lines display the 1$\sigma$ confidence level and the 1D histograms show the marginalized probability distribution for each parameter\footnote{In order to compare retrievals with different number of samples, instead of plotting the total count of samples we plot at each histogram bin the corresponding number of counts divided by the total count and by the bin width. This way, the integral under the histogram equals 1 in all retrievals.} (also displayed in Fig. \ref{fig:individual_retrievals_1D}).
For comparison, green lines indicate the true value given as input.
The median values of these marginalized distributions, as well as their upper and lower uncertainties corresponding to the 84\% and 16\% quantiles (or, equivalently, to 1-$\sigma$ for Gaussian errors), are listed in Table \ref{table:results_retrievals}.

As a general result we find that none of the single-phase retrievals can confirm the presence or absence of clouds regardless of the actual cloud coverage of the planet.
Indeed, the retrieval results for the cloud optical thickness are indistinguishable in the no- and thin-cloud scenarios (Fig. \ref{fig:individual_retrievals_1D}) and thus it is challenging to favour one scenario over the other.
Moreover, high values of $\tau_c$ cannot be completely ruled out from the retrieval results in these two scenarios. 
As for the thick-cloud  scenario, the posterior probability distribution of $\tau_c$ (Fig. \ref{fig:individual_retrievals_1D}, bottom) is flat. 
In case of a real observation, such a retrieval result would be compatible both with cloud-free or cloudy atmospheres.
This extends for nonzero values of $\alpha$, the findings for $\alpha$=0$^\circ$ in \citet{carriongonzalezetal2020} about the correlations between model parameters that are triggered if $R_p$ is unknown.

To understand the fundamentals of these correlations and how they change with $\alpha$ depending on the cloud coverage, it is useful to analyse the retrieval results in each of the cloud scenarios.
This provides insight valuable for real observations, in which case prior information on the true cloud coverage will be unavailable.

In the single-phase analyses, the aerosol properties ($r_{\rm{eff}}$, $\omega_0$) and the cloud geometrical extension ($\Delta_c$, normalized to $H_g$) remain largely unconstrained for all cloud scenarios.
The vertical location of the cloud ($\tau_{\rm{c\rightarrow TOA}}$) is reasonably well constrained in the thin-cloud scenario (Fig. \ref{fig:individual_retrievals_1D}).
This estimate is somewhat less accurate in the thick-cloud scenario due to stronger parameter correlations which tend to place the cloud higher up in the atmosphere than the true value.

The correlations between the planet radius and the cloud properties propagate to other parameters affecting in particular the methane abundance.
Indeed, the most accurate estimates of $f_{\rm{CH_4}}$ are retrieved in the no-cloud scenario while $f_{\rm{CH_4}}$ is severely underestimated in the thick-cloud scenario at all phases (Fig. \ref{fig:individual_retrievals_1D}).
This was also found in \citet{carriongonzalezetal2020} for a single-phase observation at $\alpha$=0$^\circ$.

Remarkably, we find that it is possible to constrain the planet radius at least better than a factor of two in all the scenarios and phase angles explored, similar to the findings by \citet{carriongonzalezetal2020} for $\alpha$=0$^\circ$.
However, in this work we also find that the retrievals of $R_p$ improve as the phase angle increases.
For our analysis at $\alpha$=123$^\circ$, we estimate values of $R_p/R_N$ of $0.65^{+0.14}_{-0.11}$ for the no-cloud scenario, $0.60^{+0.21}_{-0.10}$ for the thin-cloud one and $0.58^{+0.23}_{-0.09}$ for the thick-cloud one.
This means that at this large phase angle the maximum deviation from the true value ($R_p/R_N$=0.6) is of only 35\%, regardless of the cloud coverage of the planet.
In Sect. \ref{sec:discussion} we discuss the possible physical reasons.
Constraining $R_p$ to within a certain level of uncertainty also provides more restricted priors of this parameter for subsequent retrievals. 
As shown in \citet{carriongonzalezetal2020}, this can partially break some of the parameter correlations, which results in more accurate retrievals.

Overall, we find that the retrieval results change with $\alpha$ more dramatically in the thick-cloud scenario.
This supports the idea that the information content in a spectrum depends on the phase angle $\alpha$ at which it was obtained and that this connection occurs through the scattering properties of the atmosphere. 
These ideas are further explored in Sect. \ref{sec:discussion}.
We also conclude that if $R_p$ and the cloud properties are a priori unknown, a single-phase observation at $S/N$=10 cannot break the degeneracies between model parameters and accurately constrain the atmospheric properties of an exoplanet.
This holds true for a broad range of phase angles representative of the capabilities of future direct-imaging space telescopes.

\subsection{Retrievals for multi-phase combinations of measurements} \label{subsec:results_combined}

Building upon the analysis of single-phase retrievals, we next proceed to the analysis of simultaneous multi-phase retrievals.
Following a similar approach to that in \citet{damianoetal2020}, we assumed an observation at a small phase angle (in our case, $\alpha$=37$^\circ$ with $S/N$=10) and tested the improvement in the retrievals produced by different observing strategies.
As a comparison, we also doubled the $S/N$ to 20 for a single-phase measurement at $\alpha$=37$^\circ$ to test if such an observing strategy yields similar results to the combination of multiple phases.

We first explored the combination of small and moderate phase angles (37$^\circ$+85$^\circ$) and subsequently, the effect of adding phases larger than quadrature with the combinations (37$^\circ$+123$^\circ$) and (37$^\circ$+85$^\circ$+123$^\circ$).
Figures \ref{fig:results_combined_nocloud}$-$\ref{fig:results_combined_thickcloud} show the corresponding posterior probability distributions for the three cloud scenarios.
The results of the marginalized probability distribution for each parameter are displayed in Fig. \ref{fig:combined_retrievals_1D} and summarized in Table \ref{table:results_retrievals}.

We find that the best strategy to discern the presence or absence of clouds is to combine observations at small phase angles with others at phases larger than quadrature.
In all cloud scenarios, the combinations (37$^\circ$+123$^\circ$) (red lines in Fig. \ref{fig:combined_retrievals_1D}) and (37$^\circ$+85$^\circ$+123$^\circ$) (black lines) yield the most accurate estimates of $\tau_c$.
Both combinations allow us to more robustly distinguish between cloudy and cloud-free atmospheres and, furthermore, more accurately estimate the true value of the cloud's optical thickness (Table \ref{table:results_retrievals}).
Regarding the other strategies explored, in the no- and thin-cloud scenarios we find only minor differences between a combination (37$^\circ$+85$^\circ$) and a single-phase observation at $\alpha$=37$^\circ$ with $S/N$=20 (see Table \ref{table:results_retrievals} and Fig. \ref{fig:combined_retrievals_1D}).
In the thick-cloud scenario, where the correlations between $R_p$ and the cloud are stronger, combining (37$^\circ$+85$^\circ$) produces a better retrieval for $\tau_c$ than a single observation at $\alpha$=37$^\circ$ with $S/N$=20.
This suggests that the complementary information content from multiple phases helps improve the retrievals more effectively than increasing the $S/N$ of a single-phase observation.

In the scenarios with a cloud layer (either thin or thick), combining small and large phases helps constrain the aerosol optical properties $r_{\rm{eff}}$ and $\omega_0$ (Fig. \ref{fig:combined_retrievals_1D}).
The results are similar for the (37$^\circ$+123$^\circ$) and (37$^\circ$+85$^\circ$+123$^\circ$) combinations, which are both slightly more accurate than either a combination of (37$^\circ$+85$^\circ$) or a single-phase observation at 37$^\circ$ with $S/N$=20.
As expected, the improvements in the retrievals of $r_{\rm{eff}}$ and $\omega_0$ are more noticeable in the thick-cloud scenario, as the aerosol properties have a larger impact on the spectra.
The retrievals of the cloud's vertical location also improve when combining (37$^\circ$+123$^\circ$) or (37$^\circ$+85$^\circ$+123$^\circ$), especially in the thick-cloud scenario.
The geometrical extension of the cloud $\Delta_c$ remains unconstrained in all cases.

The retrieved methane abundances depend significantly on the cloud scenario, as expected from the impact of the $R_p-\tau_c-f_{\rm{CH_4}}$ degeneracies in each case.
For the no- and thin-cloud scenarios, a single-phase observation at $\alpha$=37$^\circ$ with $S/N$=20 appears as the best strategy to reduce the uncertainties in the retrieved $f_{\rm{CH_4}}$.
However, this strategy cannot be generalized.
For instance,  in the case of $\alpha$=37$^\circ$ and $S/N$=20 the above parameter degeneracies in the thick-cloud scenario result in a $f_{\rm{CH_4}}$ estimate that deviates significantly from the true value.
On the other hand, in all cloud scenarios the combinations (37$^\circ$+123$^\circ$) and (37$^\circ$+85$^\circ$+123$^\circ$) yield estimates of $f_{\rm{CH_4}}$ which are on the same order of magnitude as the true value.

For the retrieved value of $R_p$, combining small and large phase angles is again the only strategy that works for all cloud scenarios.
For the no- and thin-cloud scenarios, the resulting estimates are similarly accurate (with maximum deviations of about 60\%) for all the cases explored here: the single-phase observation at $\alpha$=37$^\circ$ with $S/N$=20 and the combinations (37$^\circ$+85$^\circ$), (37$^\circ$+123$^\circ$) and (37$^\circ$+85$^\circ$+123$^\circ$).
However, in the thick-cloud scenario there is a significant underestimation of $R_p$ both for the (37$^\circ$+85$^\circ$) combination and for $\alpha$=37$^\circ$ with $S/N$=20 (Fig. \ref{fig:combined_retrievals_1D}).
In this case, only adding the observation at $\alpha$=123$^\circ$ breaks the degeneracies and recovers an estimate with a maximum deviation from the true value of about 50\% (Table \ref{table:results_retrievals}).

By comparing our single-phase retrievals (Fig. \ref{fig:individual_retrievals_1D}) with our multi-phase retrievals (Fig. \ref{fig:combined_retrievals_1D}) we find a general improvement in the latter that cannot be anticipated from the single-phase results.
This indicates that combining the marginalized posterior probability distributions of single-phase retrievals is not equivalent to a simultaneous multi-phase retrieval.
Hence our multi-phase results cannot be reproduced by applying the intersection criterion used by \citet{nayaketal2017}.
Because this intersection approach works on the individual posterior probability distributions for each parameter, it likely misses the subtleties of fitting several spectra simultaneously in the multi-dimensional parameter space.

In summary, combining small and large phase angles proves to be the most general strategy to break some of the parameter correlations in all cloud scenarios.
Both (37$^\circ$+123$^\circ$) and (37$^\circ$+85$^\circ$+123$^\circ$) combinations can constrain reasonably well the planet radius, the optical properties of the cloud and the abundance of the absorbing gas in all cloud scenarios.
This indicates that the improvements are indeed caused by adding the information at $\alpha$=123º (see Sect. \ref{sec:discussion}).
Other strategies such as increasing $S/N$ to 20 can narrow significantly the uncertainties in the retrieved $f_{\rm{CH_4}}$ but only if the cloud thickness is small or null.

\begin{figure*}
    \centering
    \includegraphics[width=18cm]{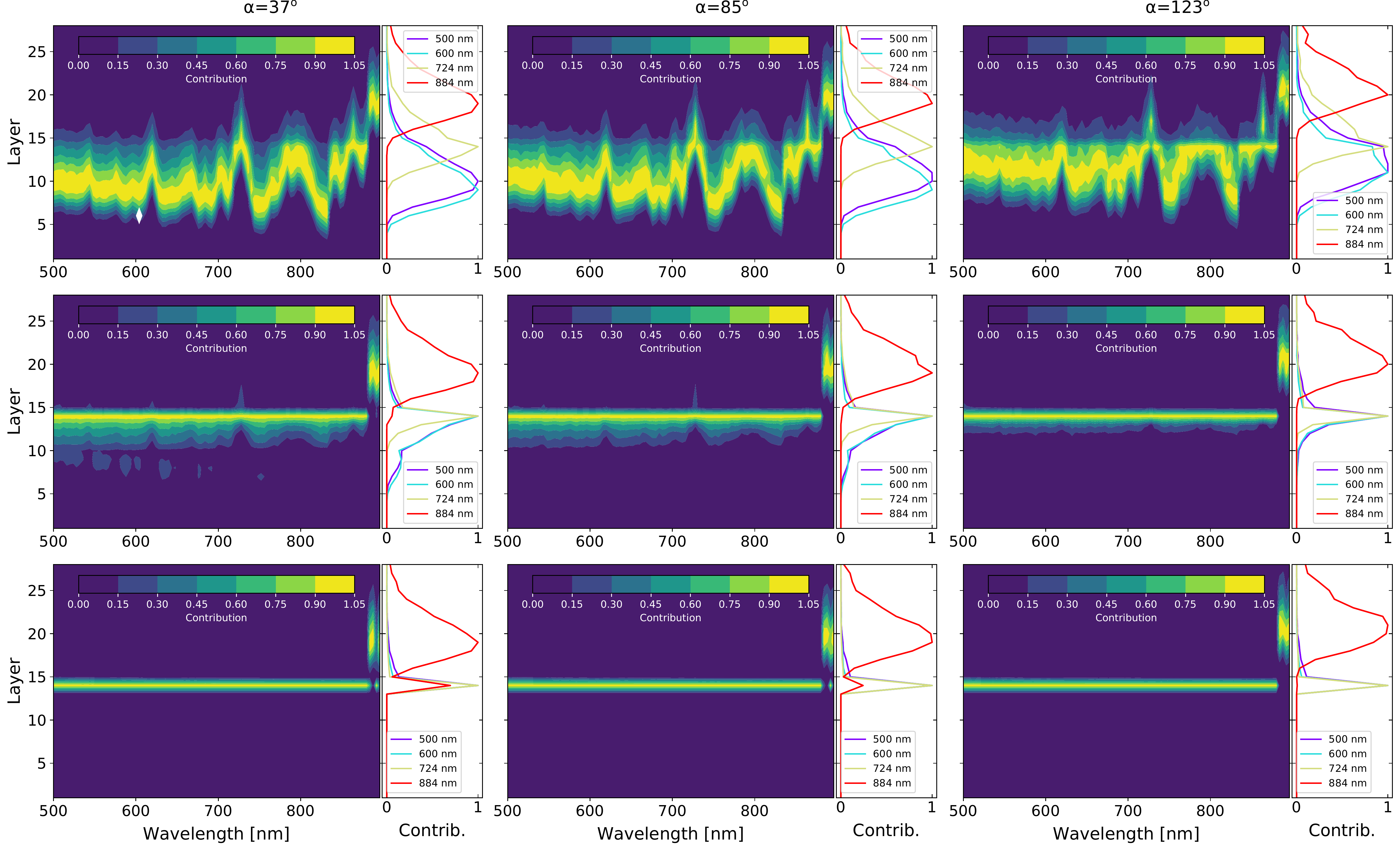}
    \caption{\label{fig:altitude_photons}
    Atmospheric layers probed by the photons in our radiative-transfer computations.
    Layer 28 is the top of the atmosphere and layer 0, the bottom (see \citealt{carriongonzalezetal2020} for details).
    Each row corresponds to a cloud scenario. 
    Upper row: no-cloud; middle: thin-cloud; bottom: thick-cloud.
    Left column: phase angle of 37$^\circ$; middle: $\alpha$=85$^\circ$, right: $\alpha$=123$^\circ$.
    The maximum photon contribution at each wavelength is normalized to 1. 
    However, not all wavelengths have comparable absolute contributions to the spectra, as shown in Fig. \ref{fig:specs}.
    The right-hand-side panels of each subplot show the vertical profiles of these normalized photon contributions for a selection of wavelengths: 500 nm (purple lines), 600 nm (blue lines), 724 nm (yellow lines) and 884 nm (red lines).}
\end{figure*}

\section{Discussion on the physical reasons for the improvement in multi-phase retrievals} \label{sec:discussion}

\begin{figure}
	\centering
	\includegraphics[width=8.5cm]{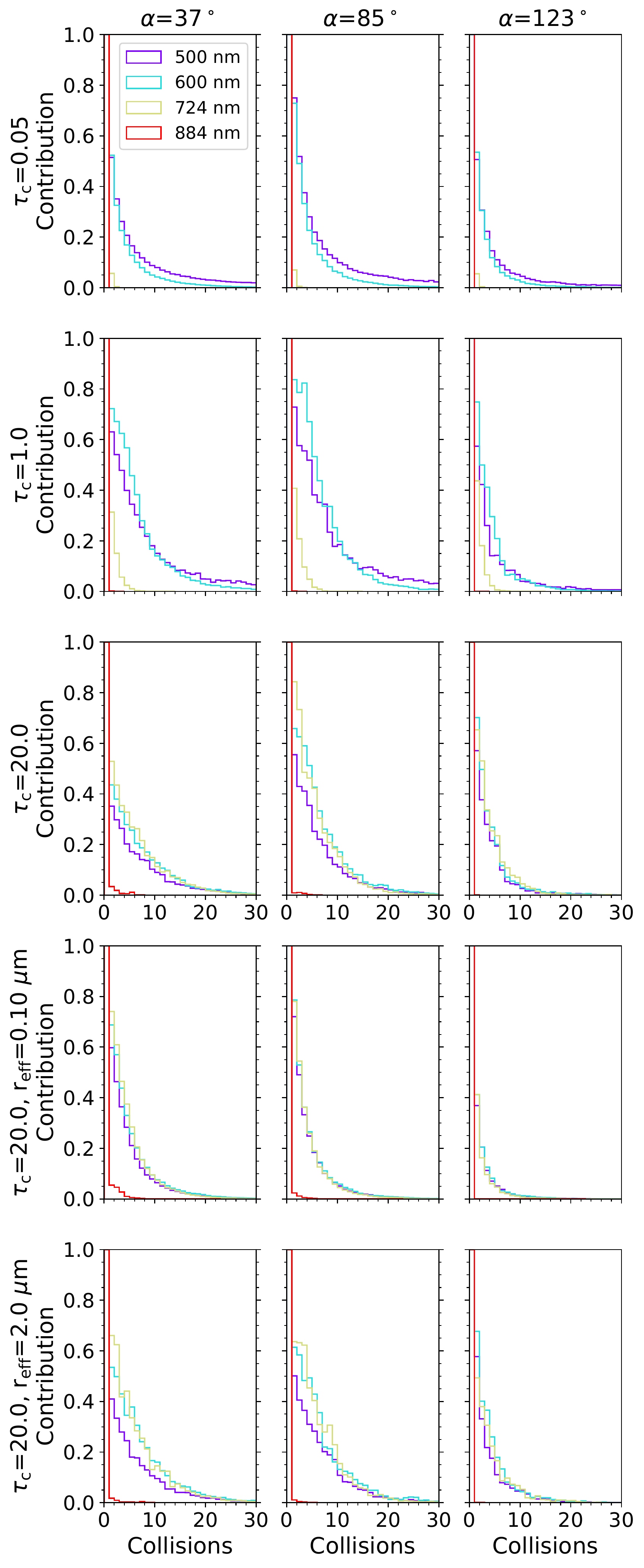} 
    \caption{\label{fig:collisionsVScontribution}
    Contribution to the overall spectral brightness of the successive scattering events that the simulated photons undergo in the atmosphere at $\lambda$=500 nm (purple lines), 600 nm (blue lines), 724 nm (yellow lines) and 884 nm (red lines).
    At each wavelength, the maximum photon contribution to the overall brightness of the spectrum is normalized to 1.
    Each column corresponds to a phase angle: 37$^\circ$ (left), $\alpha$=85$^\circ$ (middle), and $\alpha$=123$^\circ$ (right).
    Each row corresponds to a cloud scenario, as indicated in the labels.
    Also included are the thick-cloud-scenario retrievals for values of $r_{\rm{eff}}$=0.10 and 2.0 $\mu$m (see Sect. \ref{subsec:discussion_reff}).
    }
\end{figure}

From the above we conclude that the information contained in reflected-starlight spectra depends on the phase angle of the observation, which results in improved values of the retrieved atmospheric properties and $R_p$ by combining observations at multiple phases.
In the following we test several hypotheses to determine the physical reasons for this improvement in multi-phase retrievals.

We first analyse whether observations at different phases probe different altitudes of the atmosphere. 
Then, we analyse how the shape of the aerosol scattering phase function affects the multi-phase retrievals.
Finally, we analyse how the retrieval results change if the cloud optical properties are assumed known a priori.

\subsection{Atmospheric altitudes probed at each phase} \label{subsec:discussion_altitude}

We study which atmospheric layers are predominantly probed at each phase angle and whether that depends significantly on $\alpha$.
The exercise aims to reveal whether the different information from each single-phase observation comes from the different range of altitudes probed in each case.
In our multiple-scattering radiative transfer code \citep{garciamunoz-mills2015}, it is possible to trace the trajectory of each simulated photon and the atmospheric layer where it undergoes scattering collisions with the atmospheric medium.
It is also possible to compute how much each collision contributes to the overall brightness \citep[e.g.][]{garciamunozetal2017}.

Figure \ref{fig:altitude_photons} shows, for each cloud scenario and assumed $\alpha$, how the different atmospheric layers contribute to the spectrum.
The photon interactions in the no-cloud scenario come either from Rayleigh scattering with the background gas or from methane absorption.
In contrast, in both the thin and thick cloud scenarios, the majority of scattering events contributing to the spectrum take place at the cloud layer.
This holds true at the three phase angles considered and at all wavelengths except for the methane absorption band at $\sim$884 nm.
At this wavelength, CH$_4$ absorption is very efficient and the photons are extinguished at higher atmospheric levels, before reaching the cloud.
In summary, we find that the simulated photons in the thin- and thick-cloud scenarios mostly probe the atmospheric layers at which the cloud is located, while in the no-cloud scenario a much broader range of altitudes can be probed.

For the wavelengths of 500, 600, 724 and 884 nm, we computed how many collisions the photons undergo on average at each phase angle and their relative contributions to the brightness (Fig. \ref{fig:collisionsVScontribution}).
These wavelengths are representative of spectral regions significantly affected by Rayleigh scattering (500 and 600 nm) and moderate and strong CH$_4$ absorption (724 and 884 nm, respectively).
We find that the number of effective scattering events is lower at $\alpha$=123$^\circ$ than at 37º or 85º.
This trend is more noticeable in the thick-cloud scenario.
For all scenarios and phases, photons at $\lambda$=884 nm (the strongest absorption band) generally undergo only one scattering event before being completely absorbed.
At $\lambda$=724 nm (a weaker absorption band) the number of collisions increases as the cloud optical thickness increases.
This happens because the aerosols are quite reflective ($\omega_0$=0.90) and increase the number of scattering events, preventing the photons from reaching deeper atmospheric layers where they would be absorbed by methane.

In summary, the information contained in the reflected-starlight spectra of a cloudy atmosphere comes mainly from the atmospheric layers where the cloud is located.
Hence, the improvement from multi-phase retrievals (Sect. \ref{subsec:results_combined}) cannot be due to probing different atmospheric depths at each $\alpha$.
We have also found that the number of photon collisions drops at $\alpha$=123$^\circ$, especially in the thick-cloud scenario, making large-phase observations closer to the single-scattering limit.
Plausibly, this tendency towards the single-scattering limit reduces the possible trajectories of the photons within the atmosphere and, in turn, the degeneracies in the inverse exercise of retrieval.
This is consistent with the more accurate $R_p$ retrievals at $\alpha$=123$^\circ$ (Sect. \ref{subsec:results_individual}) and the enhancement in multi-phase retrievals that combine small and large phases (Sect. \ref{subsec:results_combined}).

\subsection{The impact of the aerosol's scattering phase function} \label{subsec:discussion_reff}

  \begin{figure*}
       \centering
    \includegraphics[width=0.13\linewidth]{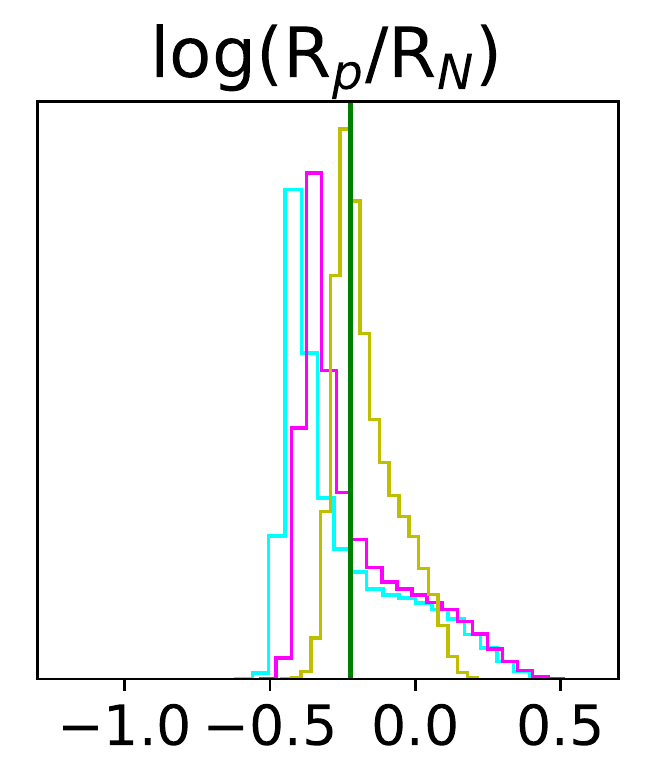}
    \includegraphics[width=0.13\linewidth]{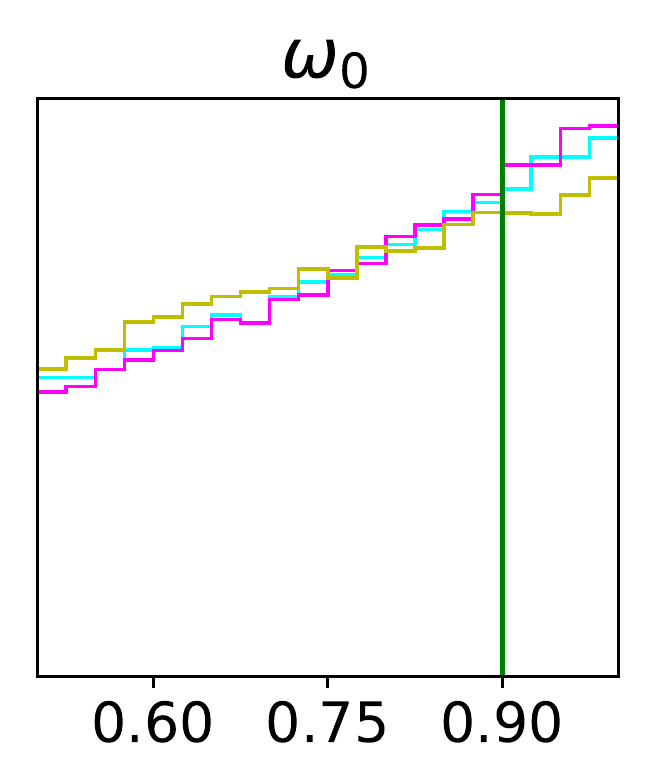}
    \includegraphics[width=0.13\linewidth]{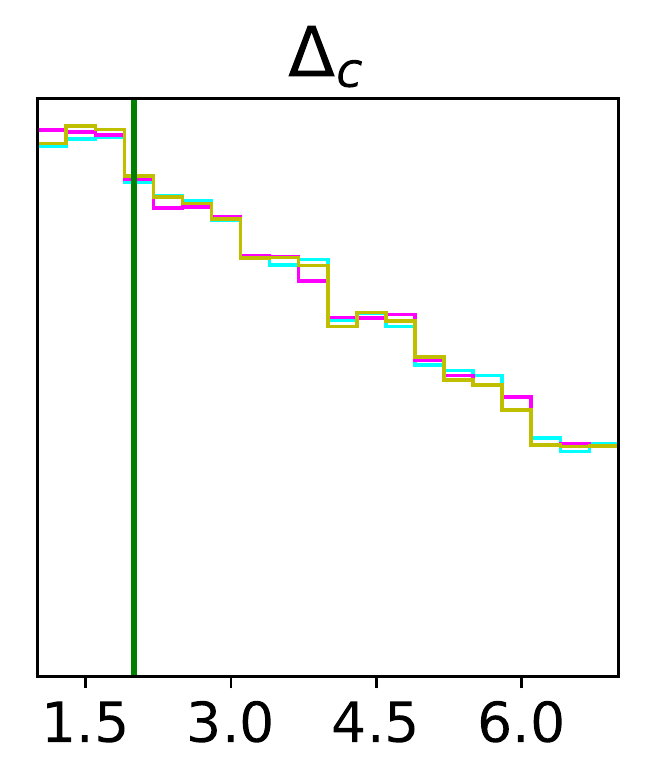}
    \includegraphics[width=0.13\linewidth]{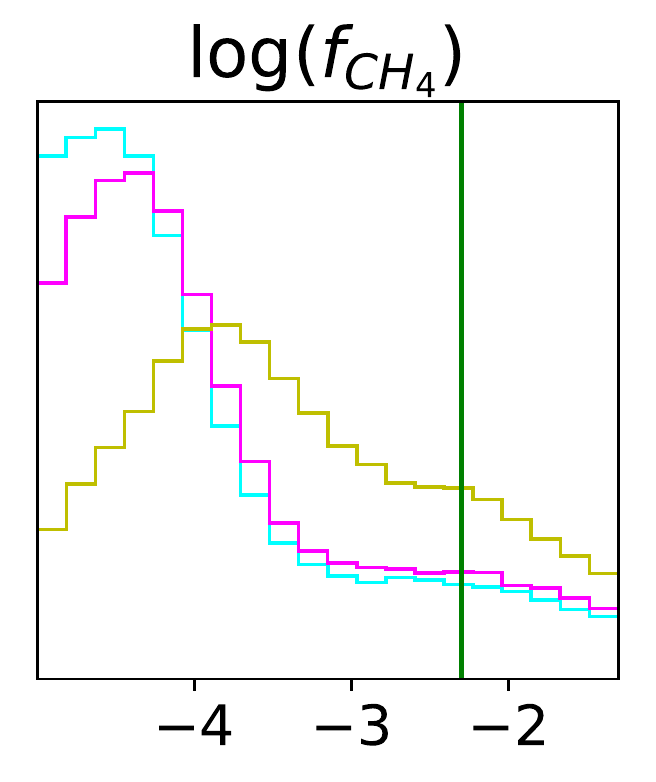}
    \includegraphics[width=0.14\linewidth]{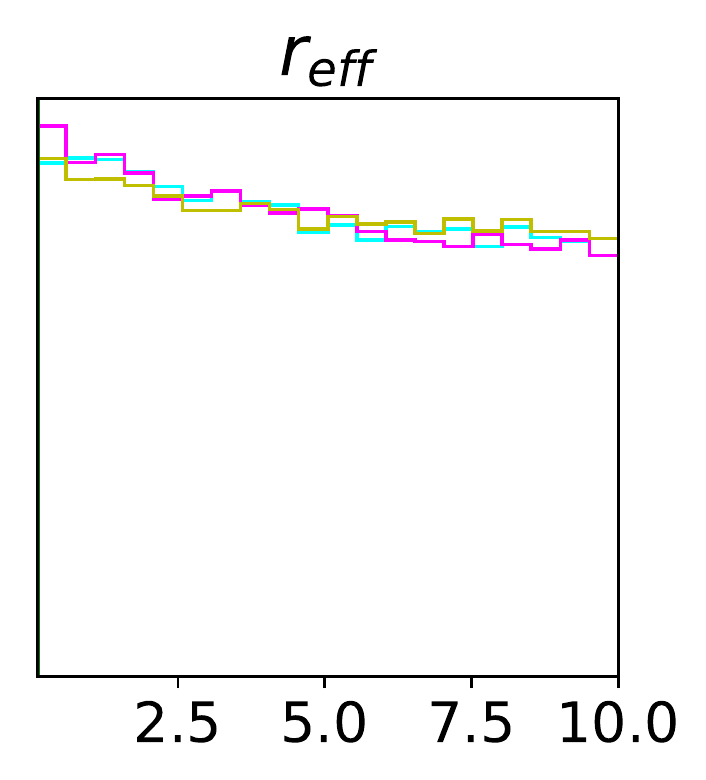}
    \includegraphics[width=0.13\linewidth]{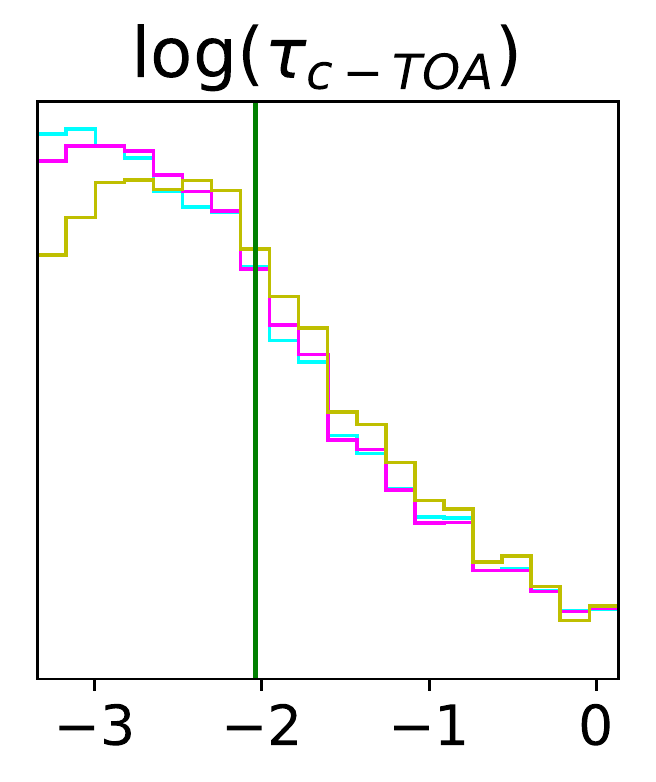}
    \includegraphics[width=0.13\linewidth]{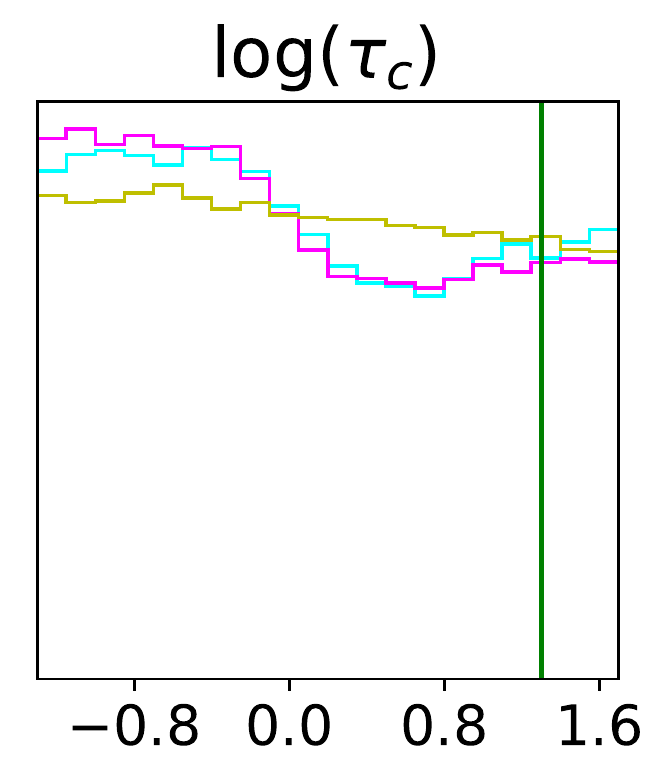}
    \\
    \includegraphics[width=0.13\linewidth]{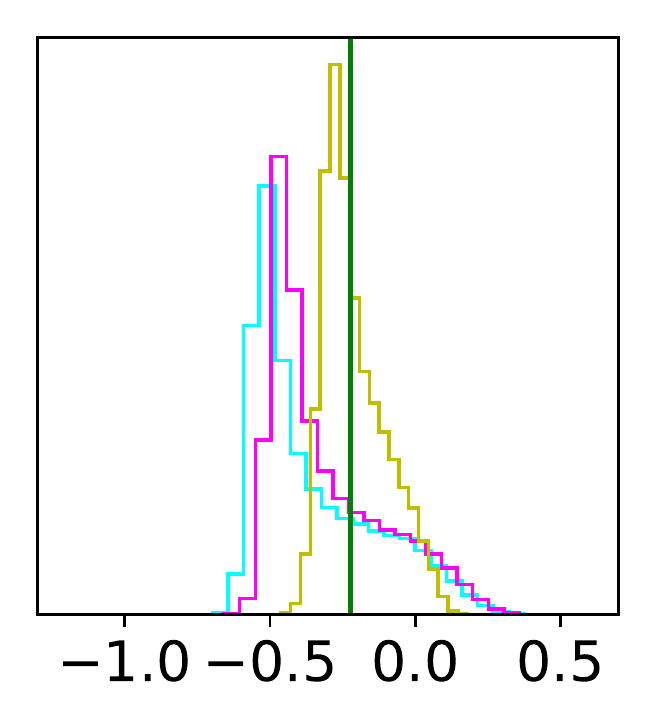}
    \includegraphics[width=0.13\linewidth]{./FIGURES/Thick-cloud/1Dplot_wholegrid_ssa.pdf}
    \includegraphics[width=0.13\linewidth]{./FIGURES/Thick-cloud/1Dplot_wholegrid_deltac.pdf}
    \includegraphics[width=0.13\linewidth]{./FIGURES/Thick-cloud/1Dplot_wholegrid_fch4.pdf}
    \includegraphics[width=0.14\linewidth]{./FIGURES/Thick-cloud/1Dplot_wholegrid_reff.pdf}
    \includegraphics[width=0.13\linewidth]{./FIGURES/Thick-cloud/1Dplot_wholegrid_taucTOA.pdf}
    \includegraphics[width=0.13\linewidth]{./FIGURES/Thick-cloud/1Dplot_wholegrid_tauc.pdf}
    \\
    \includegraphics[width=0.13\linewidth]{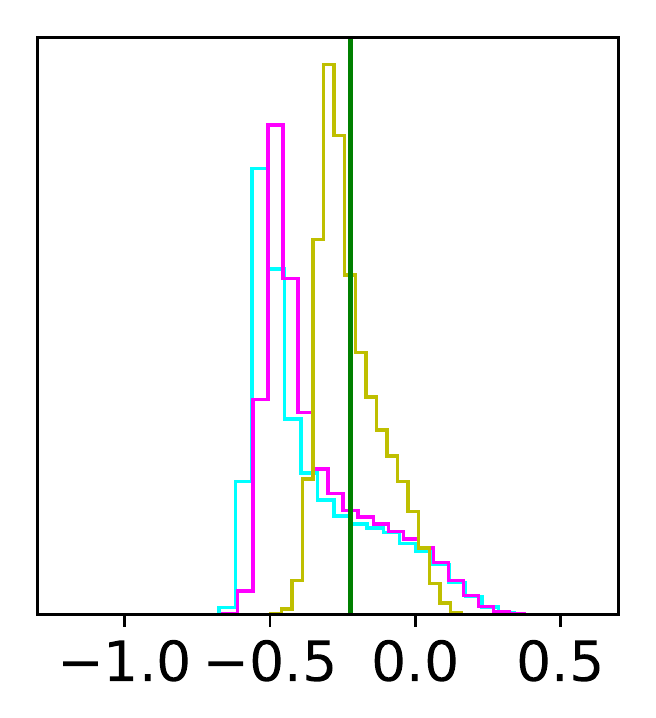}
    \includegraphics[width=0.13\linewidth]{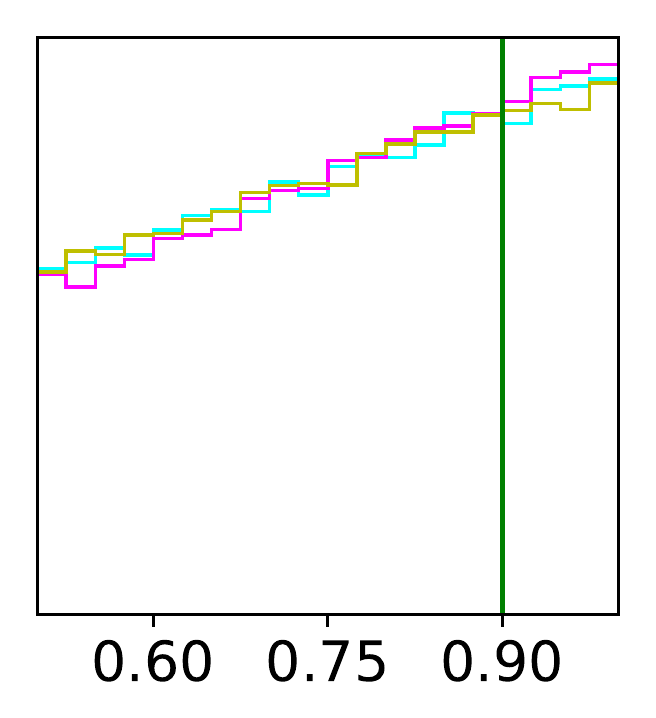}
    \includegraphics[width=0.13\linewidth]{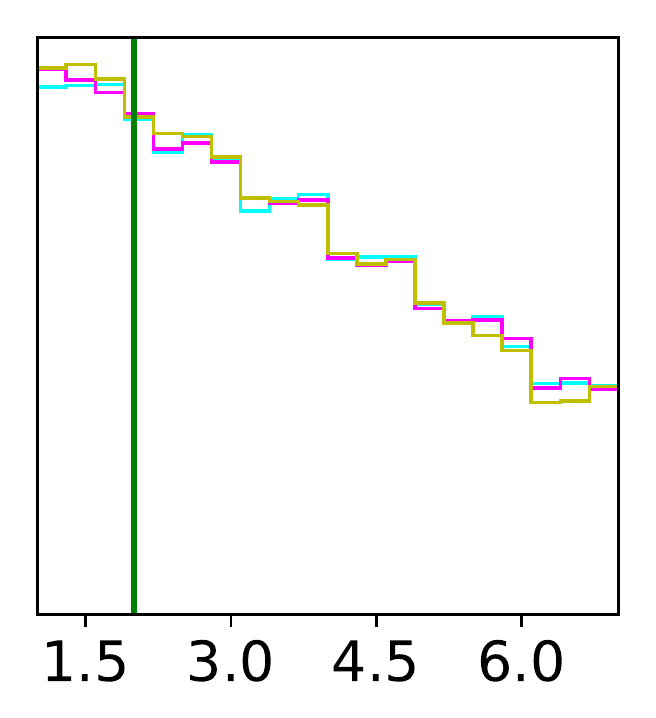}
    \includegraphics[width=0.13\linewidth]{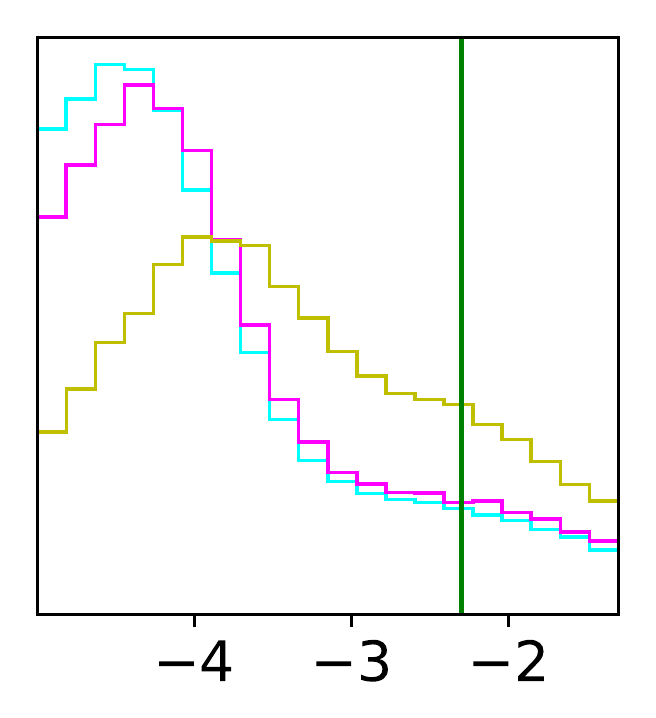}
    \includegraphics[width=0.14\linewidth]{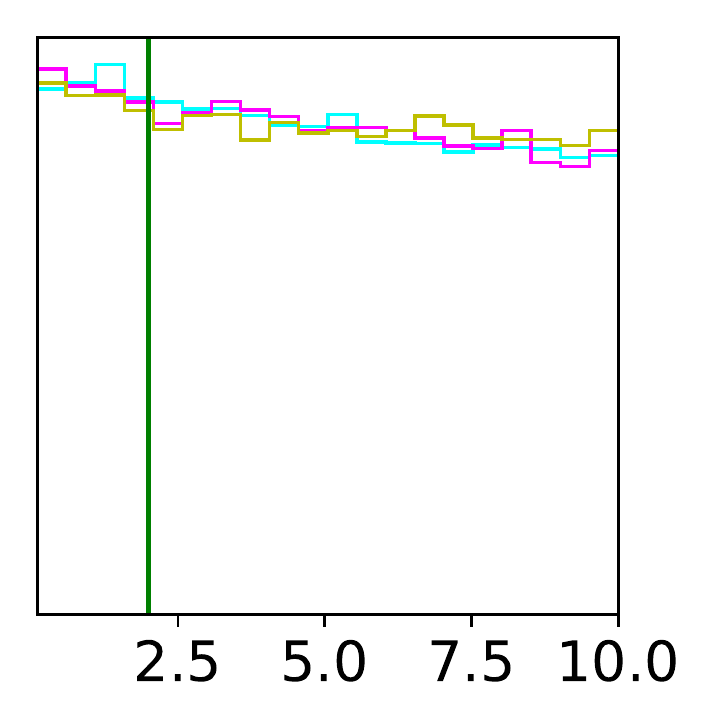}
    \includegraphics[width=0.13\linewidth]{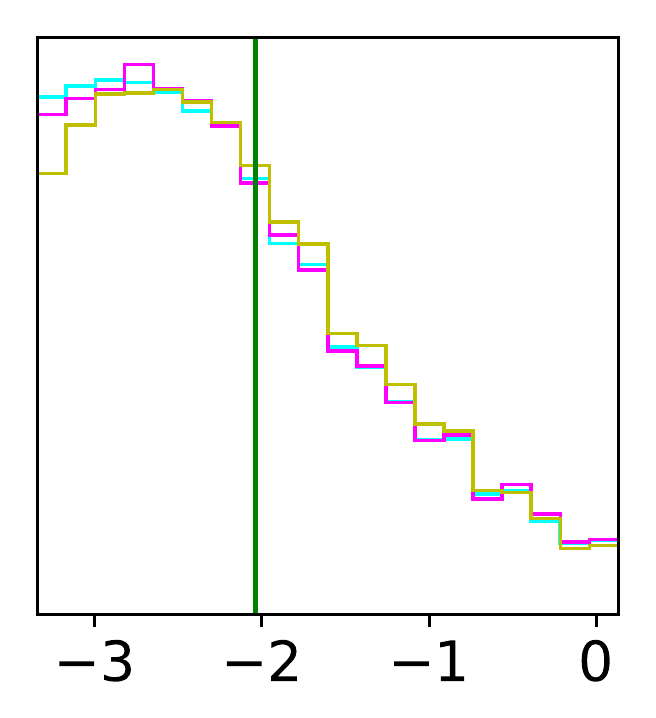}
    \includegraphics[width=0.13\linewidth]{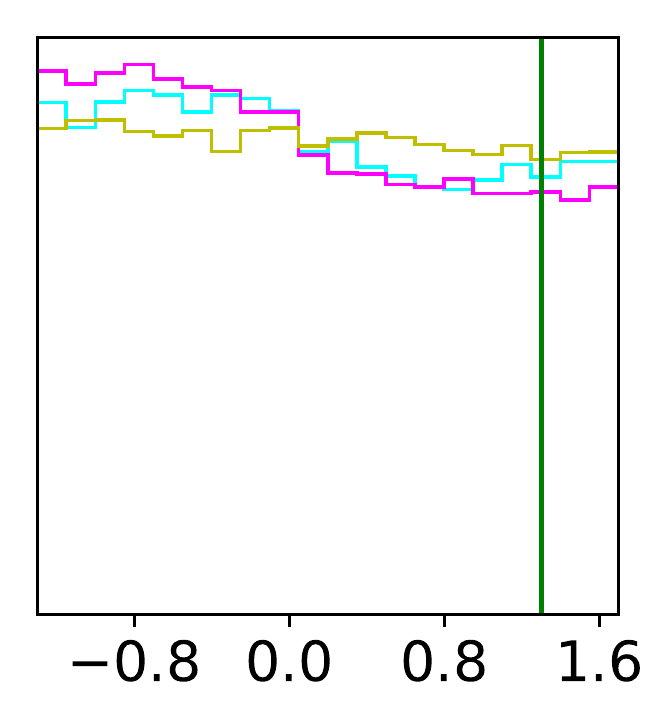}
    \caption{Marginalized posterior probability distributions of each model parameter for single-phase observations in the thick-cloud scenario at phase angles 37$^\circ$ (cyan), 85$^\circ$ (magenta) and 123$^\circ$ (yellow).
    Top row: assuming a true value of $r_{\rm{eff}}=0.10\,\mu m$ for the cloud aerosols.
    Middle row: assuming a true value of $r_{\rm{eff}}=0.50\,\mu m$ for the cloud aerosols, as in Sect. \ref{sec:results}.
    Bottom row: assuming a true value of $r_{\rm{eff}}=2.0\,\mu m$ for the cloud aerosols.
    Vertical green lines mark the true values of the model parameters (see Sect. \ref{subsec:model_atmosphere}).
    }%
    \label{fig:individual_retrievals_1D_several-reff}%
    \end{figure*}

  \begin{figure*}
       \centering
    \includegraphics[width=0.13\linewidth]{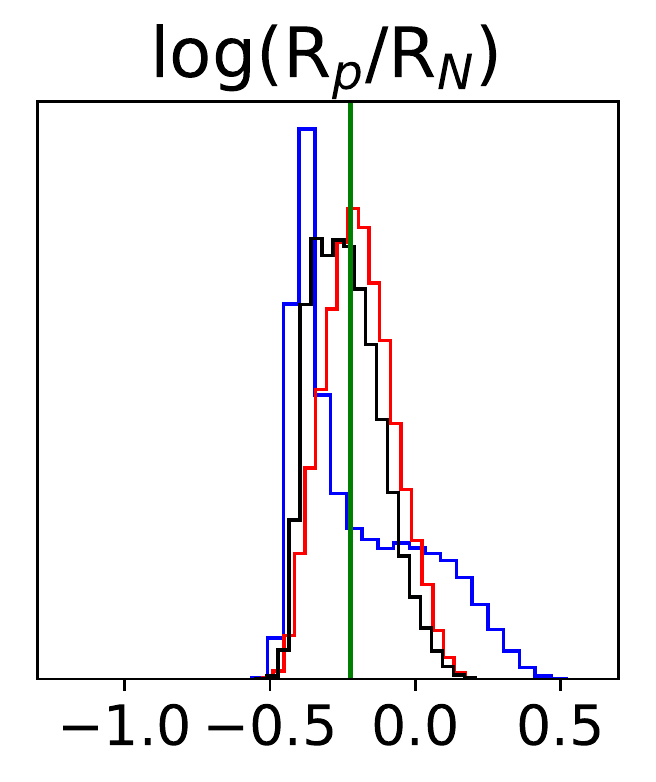}
    \includegraphics[width=0.13\linewidth]{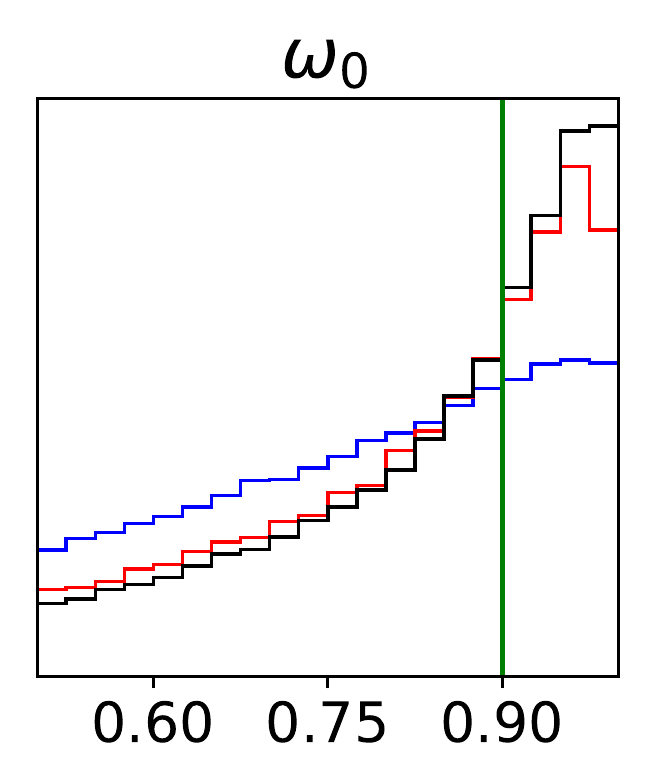}
    \includegraphics[width=0.13\linewidth]{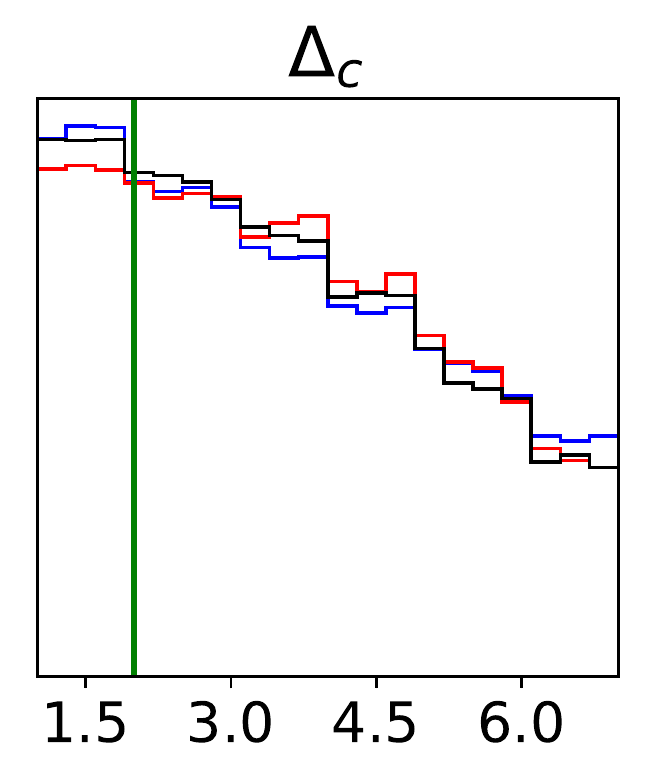}
    \includegraphics[width=0.13\linewidth]{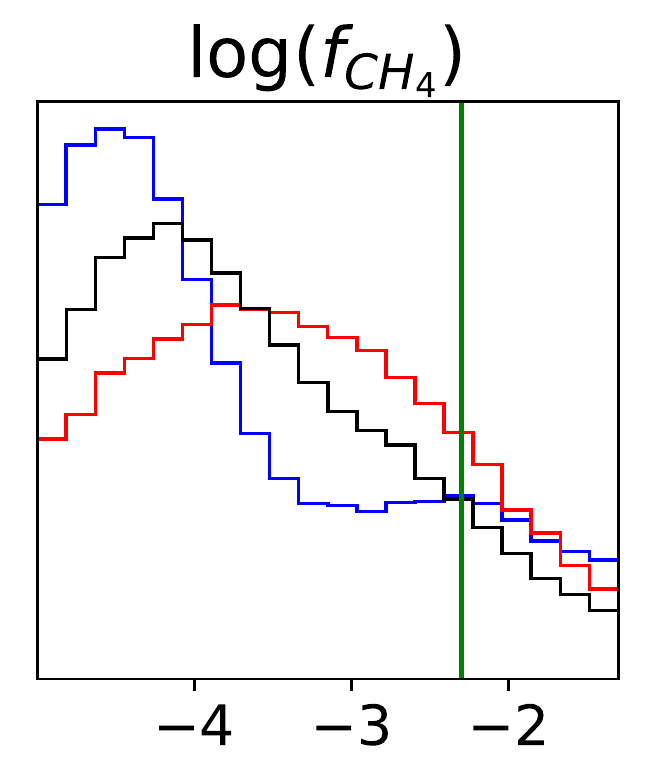}
    \includegraphics[width=0.14\linewidth]{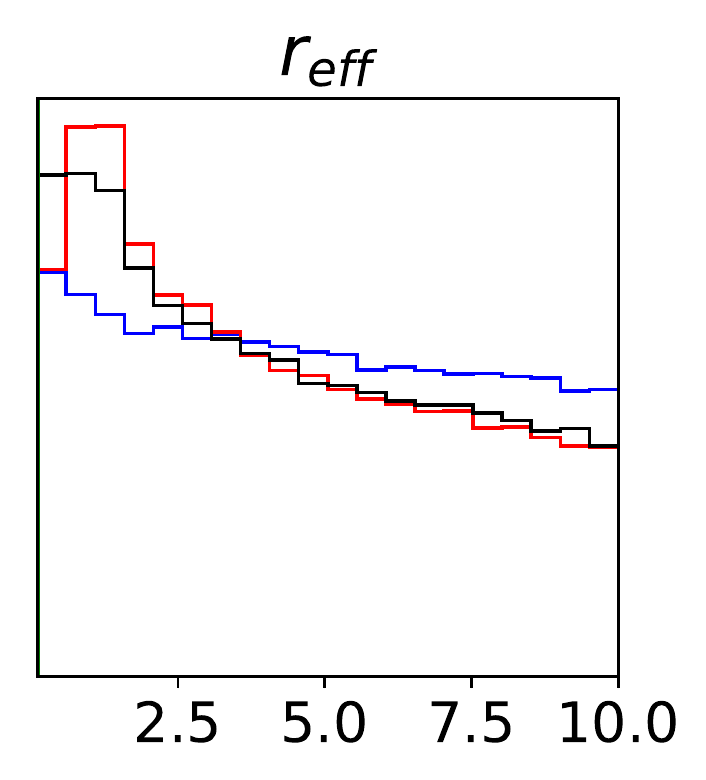}
    \includegraphics[width=0.13\linewidth]{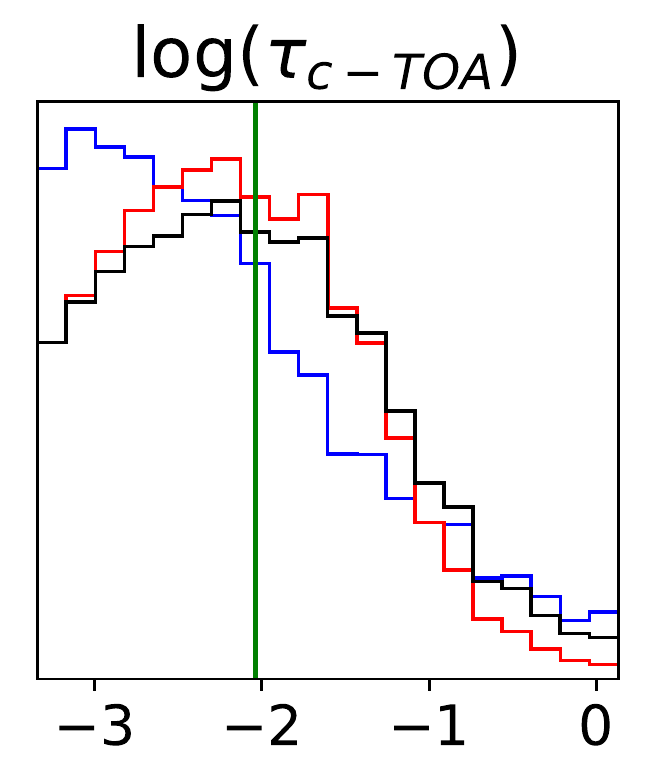}
    \includegraphics[width=0.13\linewidth]{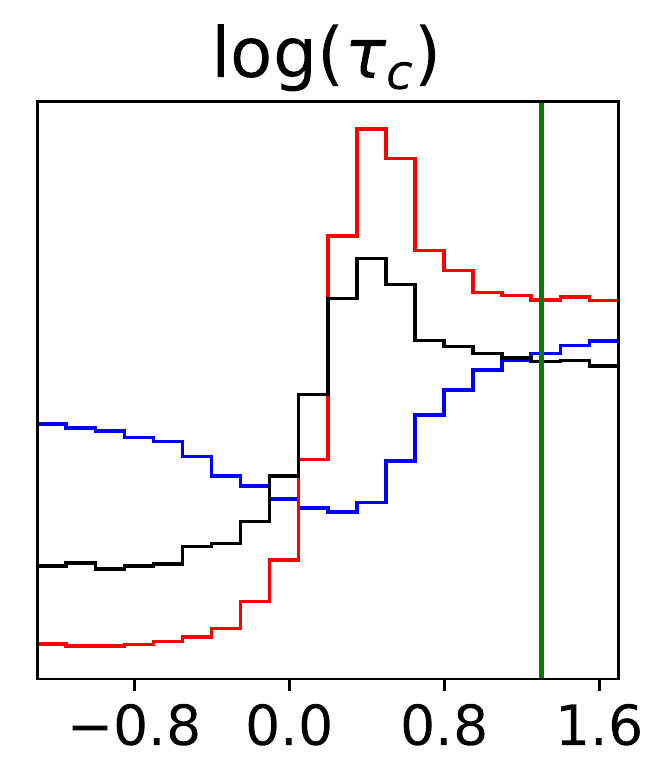}
    \\
    \includegraphics[width=0.13\linewidth]{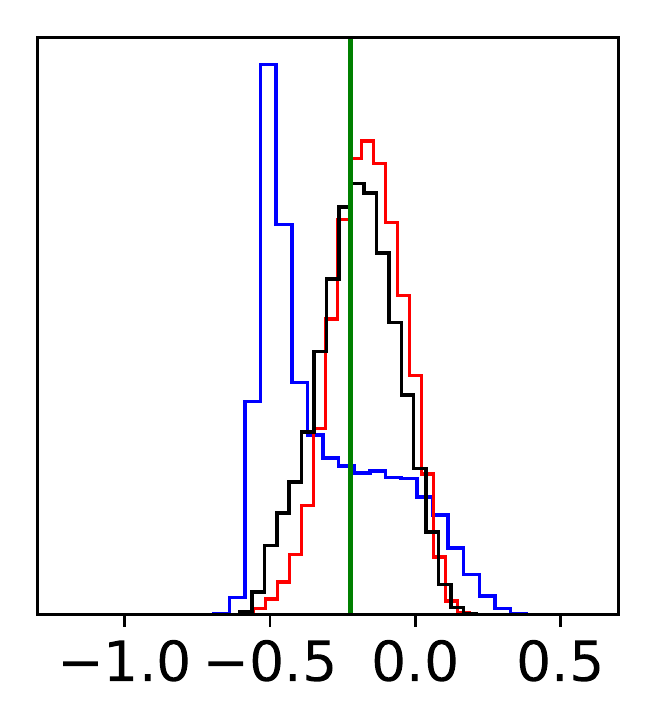}
    \includegraphics[width=0.13\linewidth]{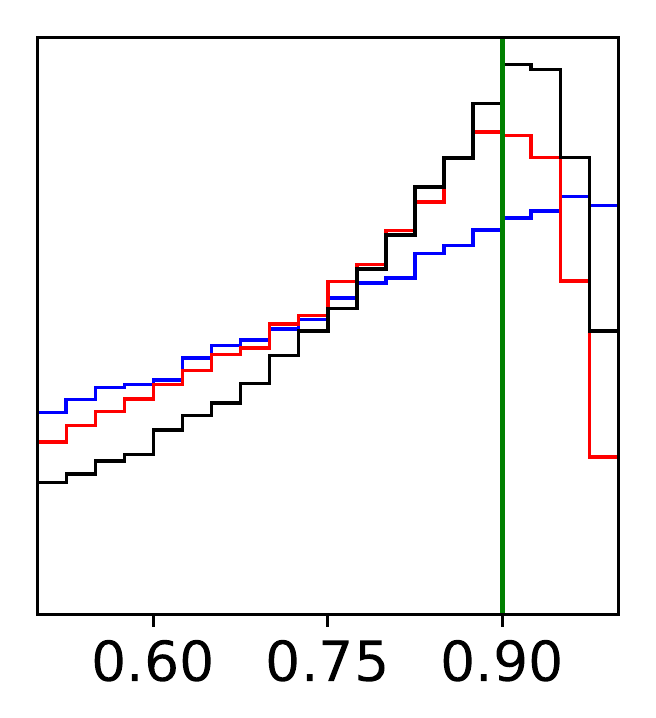}
    \includegraphics[width=0.13\linewidth]{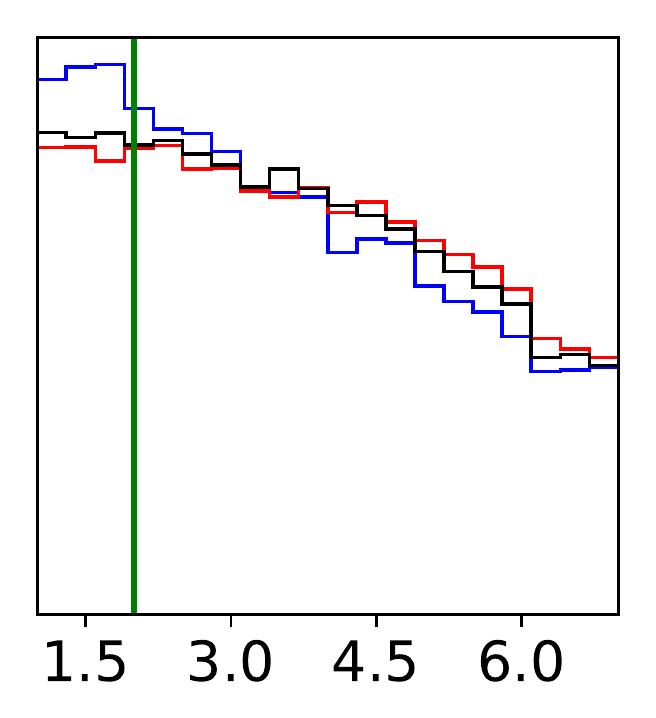}
    \includegraphics[width=0.13\linewidth]{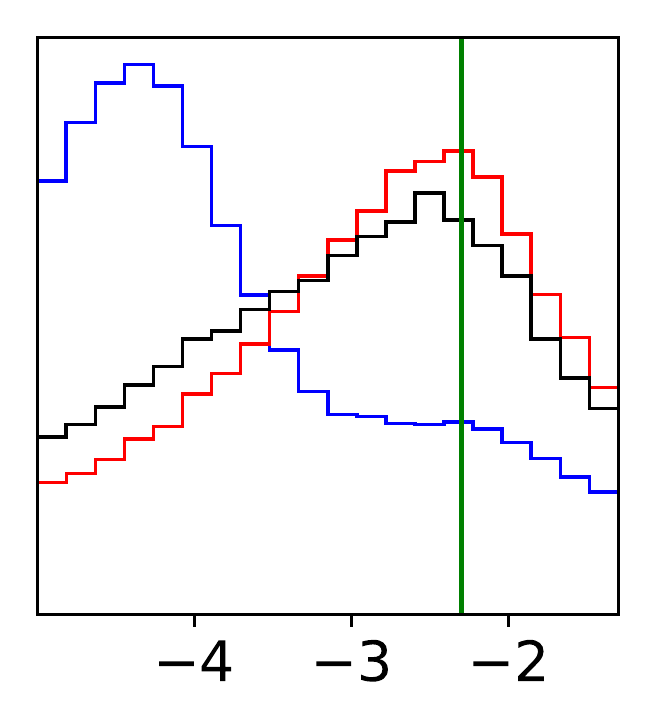}
    \includegraphics[width=0.14\linewidth]{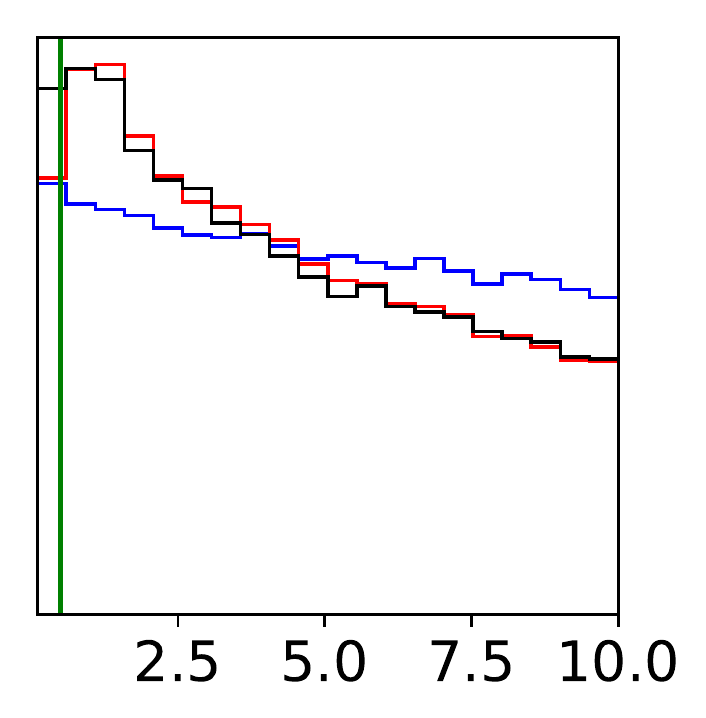}
    \includegraphics[width=0.13\linewidth]{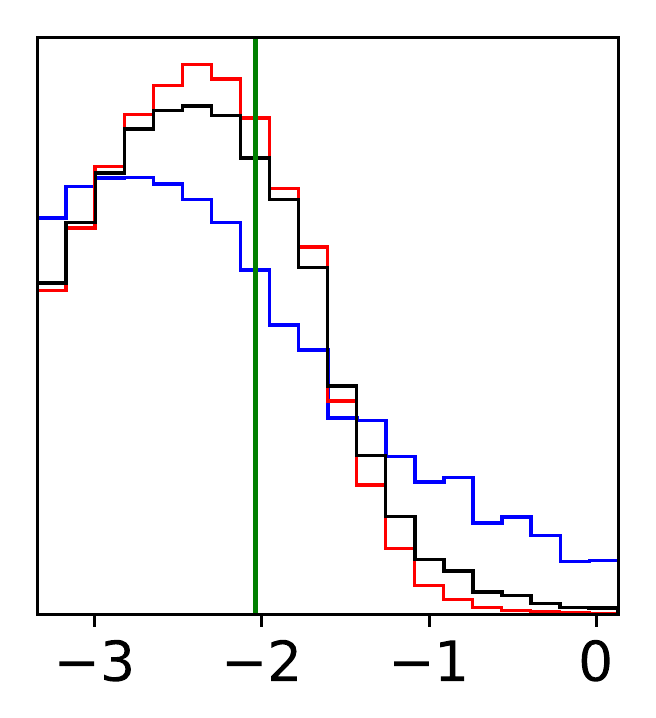}
    \includegraphics[width=0.13\linewidth]{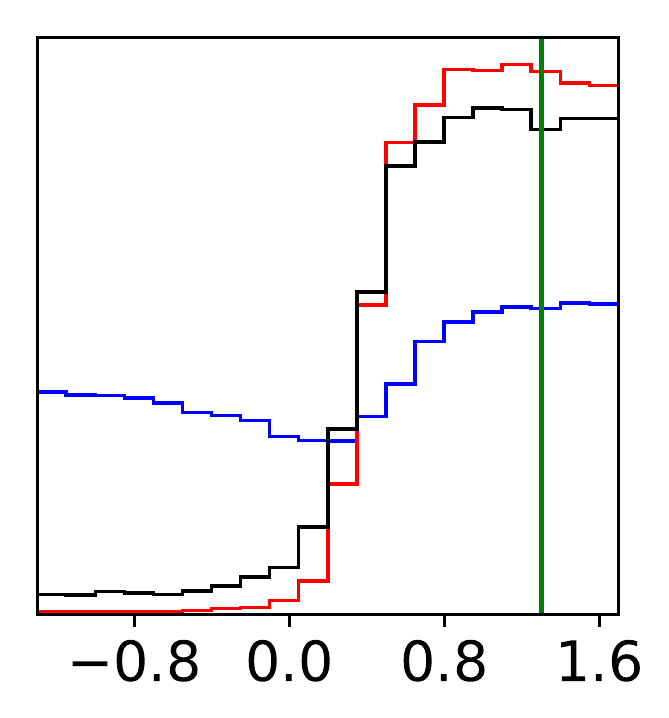}
    \\
    \includegraphics[width=0.13\linewidth]{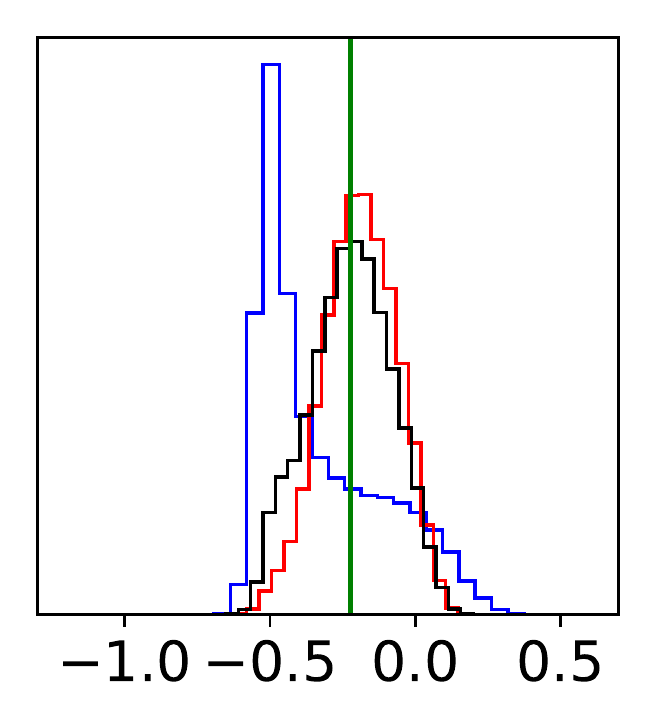}
    \includegraphics[width=0.13\linewidth]{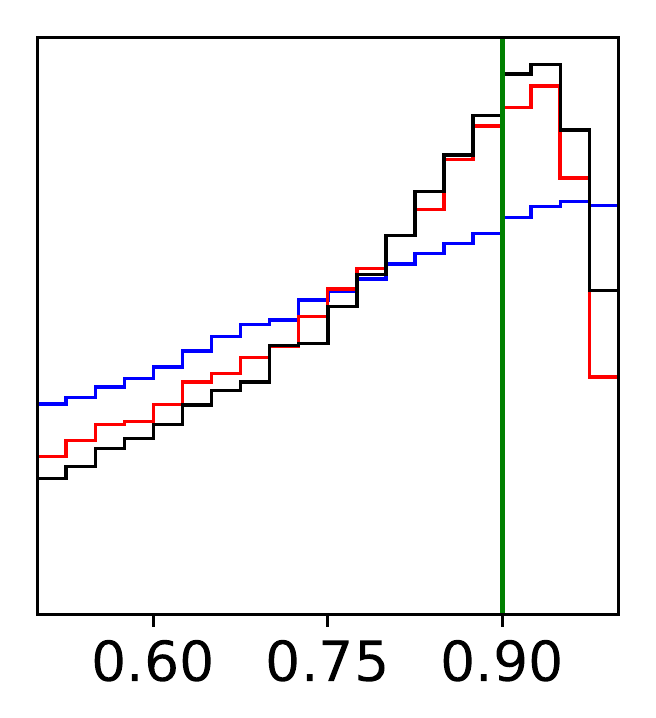}
    \includegraphics[width=0.13\linewidth]{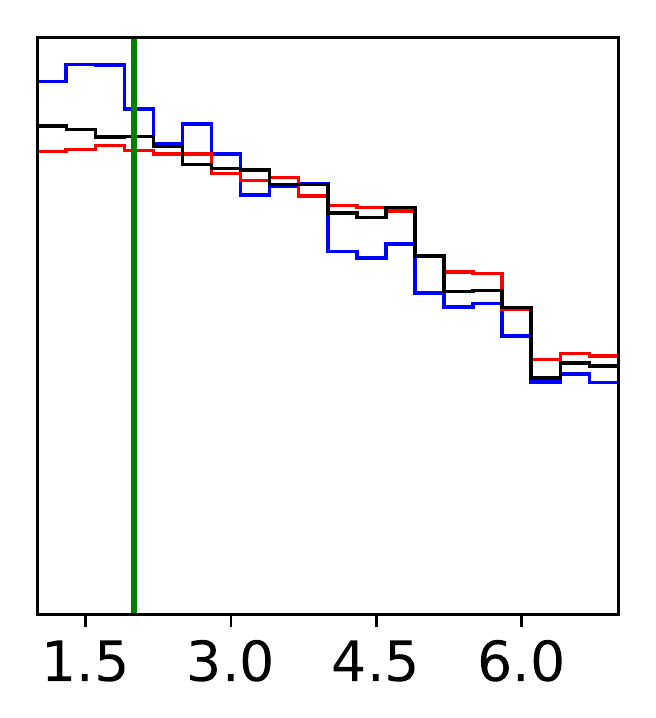}
    \includegraphics[width=0.13\linewidth]{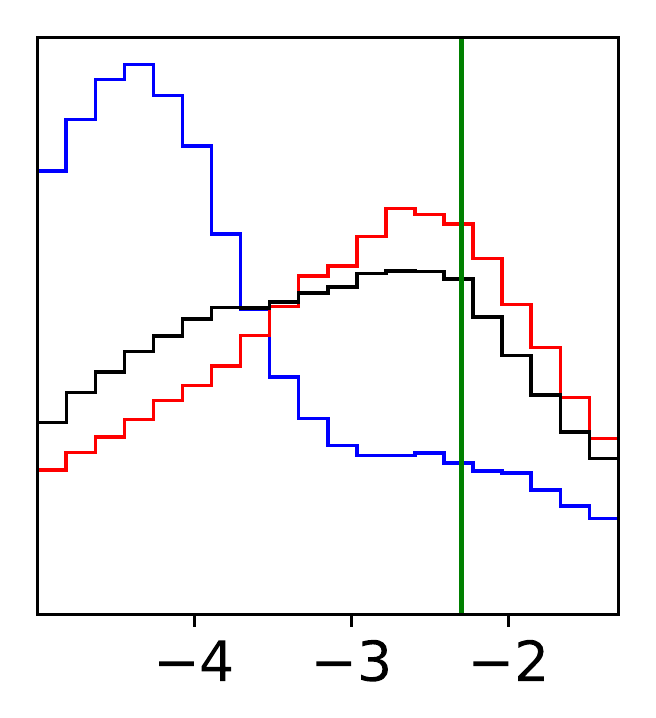}
    \includegraphics[width=0.14\linewidth]{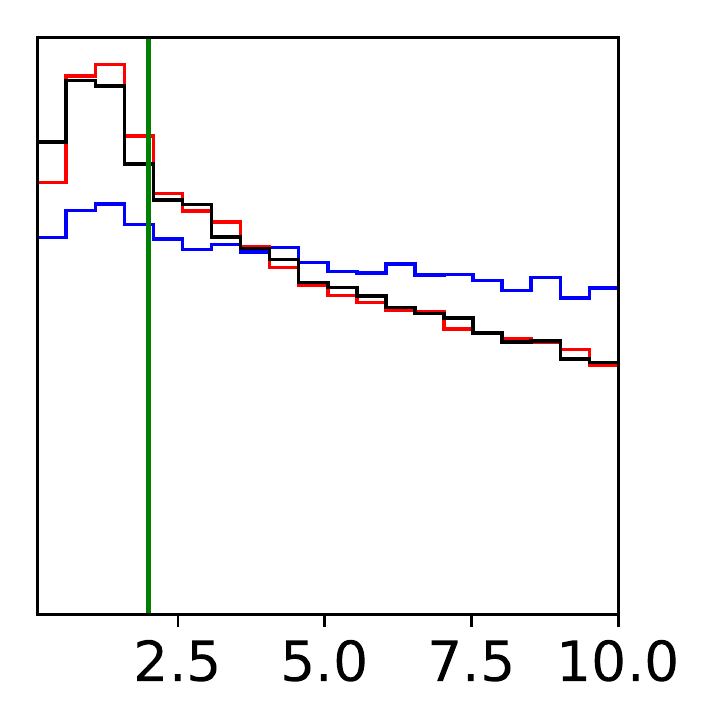}
    \includegraphics[width=0.13\linewidth]{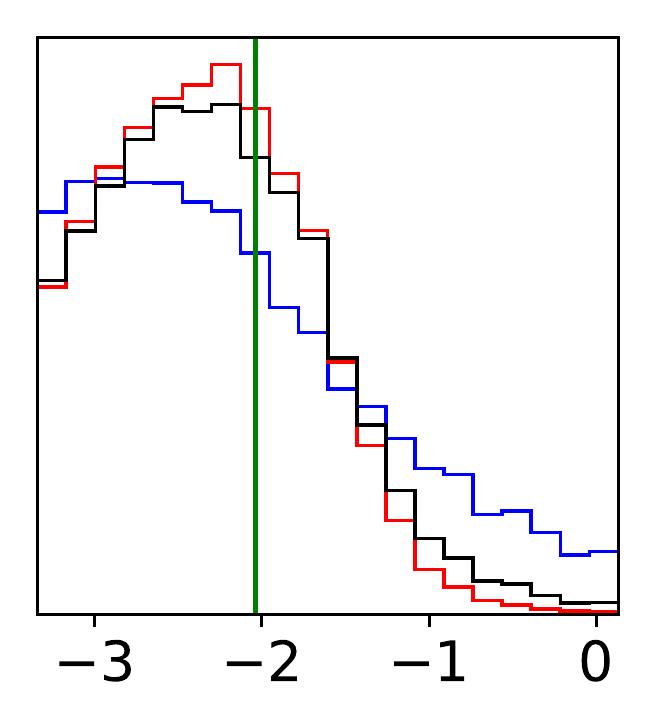}
    \includegraphics[width=0.13\linewidth]{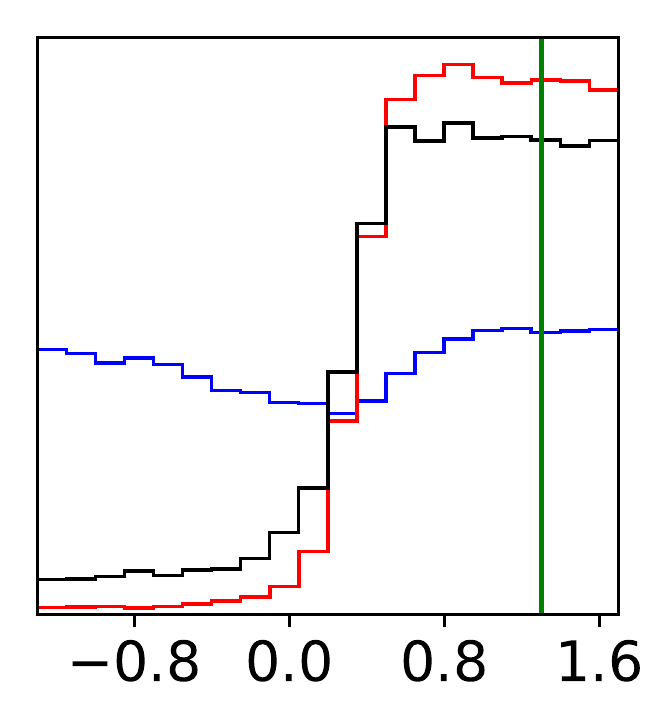}
    \caption{Marginalized posterior probability distributions of each model parameter for several combinations of observations in the thick-cloud atmospheric configuration, as described in Sect. \ref{subsec:model_sampling}.
    Blue lines mark the combination (37$^\circ$+85$^\circ$); red lines, (37$^\circ$+123$^\circ$) and black lines, (37$^\circ$+85$^\circ$+123$^\circ$).
    Top row: assuming a true value of $r_{\rm{eff}}=0.10\,\mu m$ for the cloud aerosols.
    Middle row: assuming a true value of $r_{\rm{eff}}=0.50\,\mu m$ for the cloud aerosols, as in Sect. \ref{sec:results}.
    Bottom row: assuming a true value of $r_{\rm{eff}}=2.0\,\mu m$ for the cloud aerosols.
    Vertical green lines mark the true values of the model parameters (see Sect. \ref{subsec:model_atmosphere}).
    }%
    \label{fig:combined_retrievals_1D_several-reff}%
    \end{figure*}

The findings above suggest that the improvements in multi-phase retrievals are due to the optical properties of the cloud because it is the primary layer being sampled.
Next we aim to determine whether such an improvement is caused by the scattering phase function $p(\theta)$ of the aerosols.
As shown in Fig. \ref{fig:Miescattering}, $p(\theta)$ varies strongly with the scattering angle.
Foreseeably, at small phase angles backscattering ($\theta\to180^\circ$) will be particularly effective, whereas at large phase angles forward scattering ($\theta\to0^\circ$) will be particularly significant.

To test if the dependence of $p(\theta)$ on $\theta$ drives the improvement in multi-phase retrievals, we repeated the thick-cloud-scenario retrieval but assuming two values of the true $r_{\rm{eff}}$ different to the $r_{\rm{eff}}$=0.50 $\mu$m used in the rest of this work.
First, we adopt $r_{\rm{eff}}$=0.10 $\mu$m, which is the smallest value explored in this work and corresponds to a nearly-isotropic scattering phase function (Fig. \ref{fig:Miescattering}).
Then, we adopt $r_{\rm{eff}}$=2.0 $\mu$m as representative of a large particle size with a strong dependence of $p(\theta)$ on $\theta$ and a high forward-scattering component.
If the dependence of $p(\theta)$ with $\theta$ is indeed a cause of the improvement in multi-phase retrievals, this improvement should become more modest as $p(\theta)$ tends to an isotropic function.
For context, the aerosol sizes reported in the literature for the upper cloud layers in the atmospheres of Solar System gas giants are between $r_{\rm{eff}}$=0.40 and 0.75 $\mu$m \citep[e.g.][]{morozhenko-yanovitskij1973, mishchenko1989, perezhoyosetal2012}.

The retrievals for single-phase observations ($\alpha$=37$^\circ$, 85$^\circ$ or 123$^\circ$) are very similar for a thick-cloud scenario with either $r_{\rm{eff}}$=0.10, 0.50 or 2.0 $\mu$m (Fig. \ref{fig:individual_retrievals_1D_several-reff}).
Interestingly, multi-phase retrievals for $r_{\rm{eff}}$=0.10 $\mu$m yield somewhat less accurate estimates than for $r_{\rm{eff}}$=0.50 or 2.0 $\mu$m (Fig. \ref{fig:combined_retrievals_1D_several-reff}). 
This is particularly noticeable for $\tau_c$, $f_{\rm{CH_4}}$ and $\omega_0$ when small and large phases are combined, either (37$^\circ$+123$^\circ$) or (37$^\circ$+85$^\circ$+123$^\circ$).
Indeed, Fig. \ref{fig:combined_retrievals_1D_several-reff} shows that these combinations of small and large phases result in more accurate constraints of $f_{\rm{CH_4}}$ and $\omega_0$ and a sharp lower limit of $\tau_c$ for aerosols with non-isotropic scattering phase functions ($r_{\rm{eff}}$=0.50 or 2.0 $\mu$m).
In contrast, the estimates of these model parameters are less accurate in the case of a nearly-isotropic $p(\theta)$ ($r_{\rm{eff}}$=0.10 $\mu$m).
Nevertheless, even in this least favourable scenario with $r_{\rm{eff}}$=0.10 $\mu$m, combining observations at $\alpha$=37$^\circ$ and 123$^\circ$ still strongly suggests the presence of a cloud layer but fails to constrain its properties.
Both for the $r_{\rm{eff}}$=0.10 $\mu$m and 2.0 $\mu$m scenarios we observe in Fig. \ref{fig:collisionsVScontribution} a decreasing number of scattering events at $\alpha$=123$^\circ$ with respect to smaller phases, similar to the trend found for $r_{\rm{eff}}$=0.50 $\mu$m.

We conclude that the shape of the aerosol scattering phase function affects the 
constraining capability of multi-phase retrievals.
This is consistent with our findings in Sect. \ref{subsec:results_combined} that these improvements are more evident for scenarios with optically thicker clouds and only marginal in the no-cloud scenario.

\subsection{The impact of a priori assumptions in the retrievals} \label{subsec:discussion_knowncloud}

\begin{figure*}
	\centering
 	\includegraphics[width=9cm]{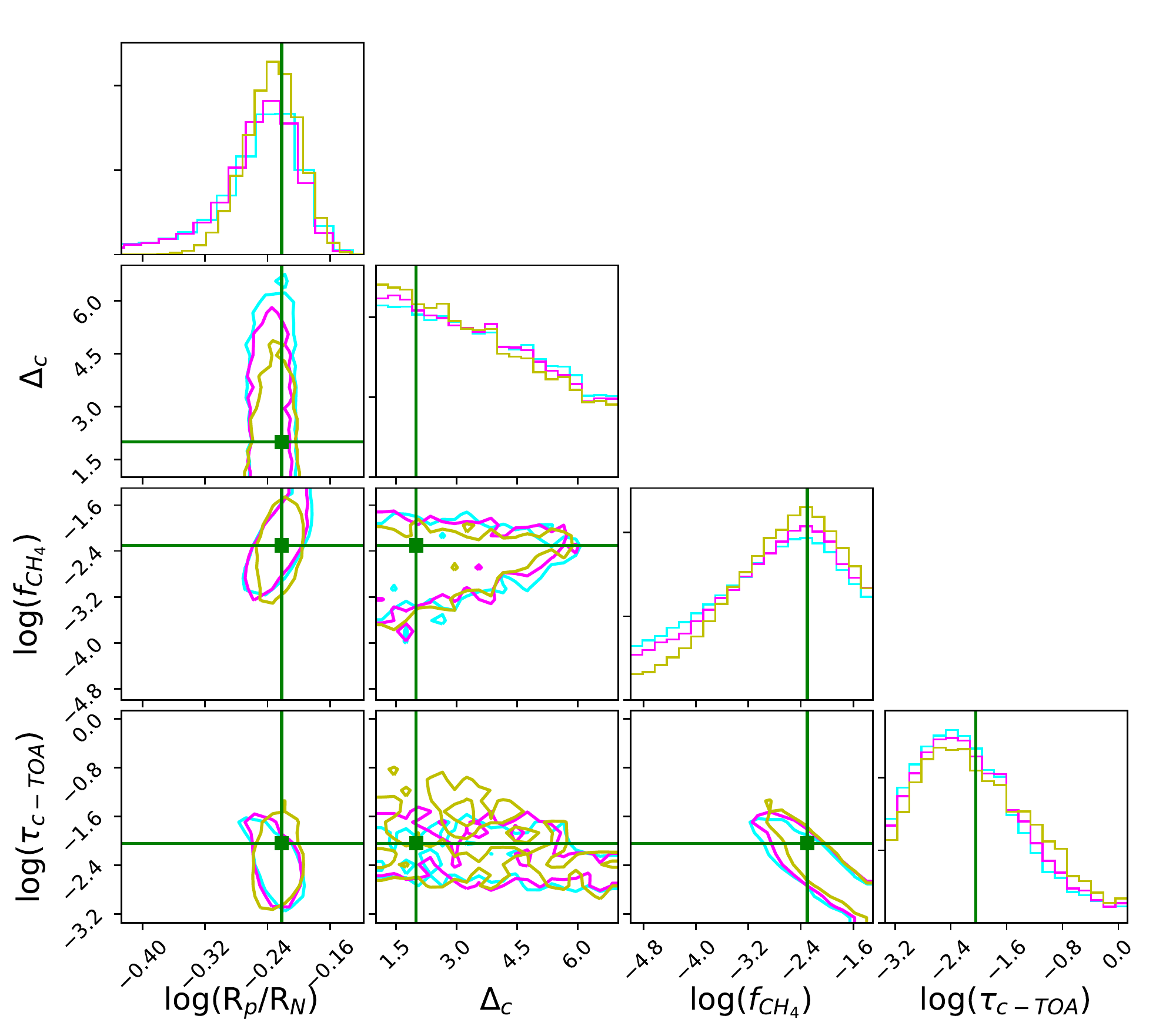} 
	\hfill
	\includegraphics[width=9cm]{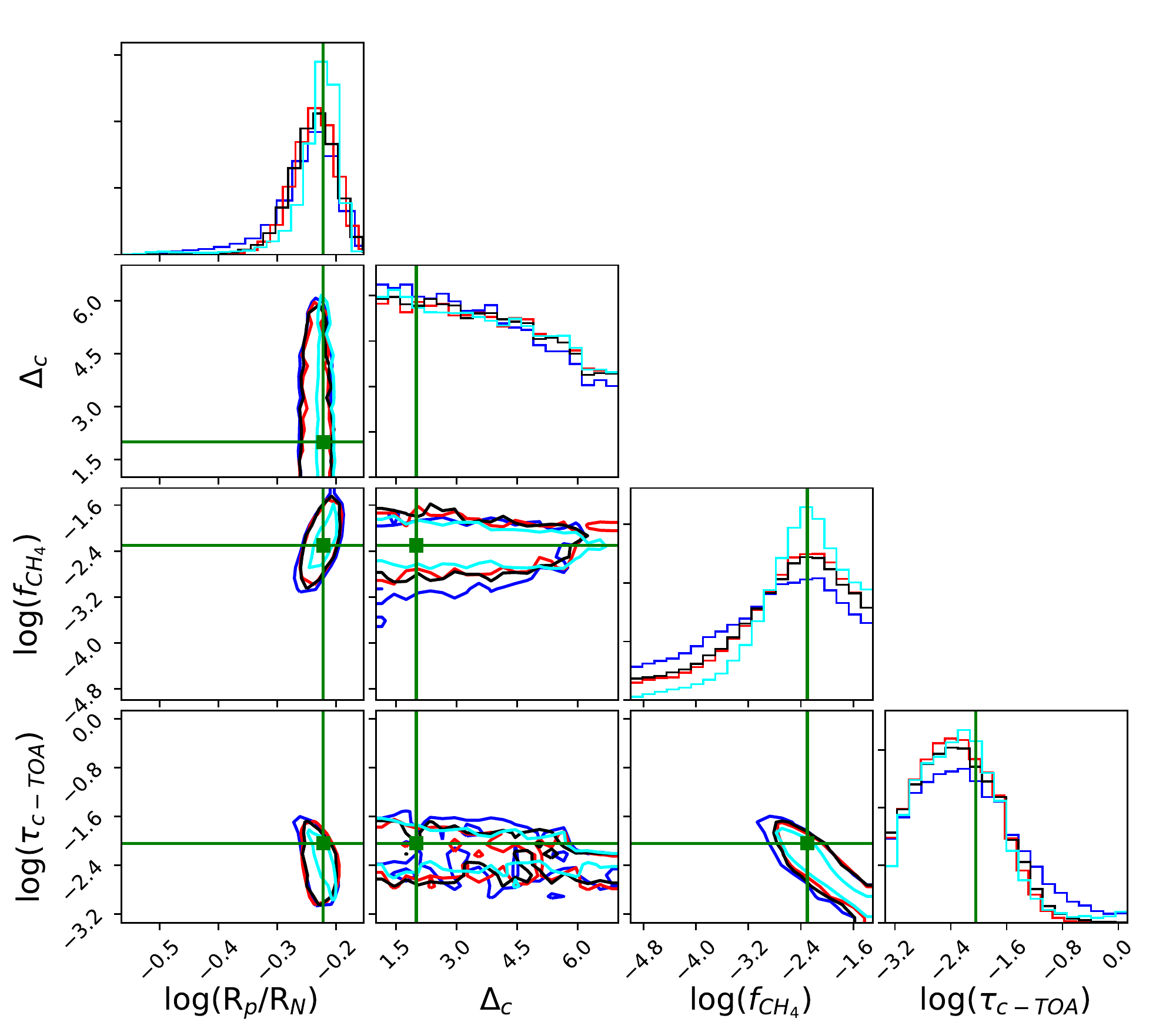} 
  \\
	\caption{Retrieval results for an exercise in which the aerosol properties of the thick-cloud scenario ($\tau_{c}$, $r_{\rm{eff}}$ and $\omega_0$) are assumed known and therefore are not explored by the MCMC retrieval sampler.
	Left panel shows the single-phase retrievals, as in Fig. \ref{fig:results_individual_thickcloud}. 
	Right panel shows the multi-phase retrievals, as in Fig. \ref{fig:results_combined_thickcloud}.}
	\label{fig:retrievals_KnownCloud_Rpfree}
\end{figure*}

\begin{table}
\caption{Values of Rayleigh cross-sections at $\lambda$ $\sim$ 5$\mu$m (in units of 10$^{-27}$cm$^2$) for several possible background gasses. Also included is the underestimation factor in the $f_{CH_4}$ retrievals if a H$_2$-dominated atmosphere is erroneously assumed instead.}
\label{table:cross_sections}
\begin{center}
\begin{tabular}{ c  c  c  c  c  c } 
\hline \hline
  &  H$_2$\tablefoottext{a} &    N$_2$\tablefoottext{a}  &  O$_2$\tablefoottext{a}  & CO$_2$\tablefoottext{a} & H$_2$O\tablefoottext{b}  \\ 
 \\ \hline
 $\sigma_{\rm{Rayleigh}}$    & 1.17 & 5.61 & 4.88 & 17.25 & 3.92 \\
 Error factor       & $-$  & 4.79 & 4.17 & 14.74 & 3.35 \\ 
 \hline
\end{tabular}
\tablefoottext{a}{\citet{shardanand-rao1977}}
\tablefoottext{b}{\citet{sutton-driscoll2004}}
\end{center}
\end{table}

Here we repeat our retrievals for a thick-cloud scenario but without including the cloud optical properties ($\tau_c$, $r_{\rm{eff}}$, $\omega_0$) as free parameters in the retrievals.
By comparing these new calculations to our previous results in Sect. \ref{sec:results}, we aim to understand how the model parameterization and the corresponding assumptions may affect the conclusions of the retrieval exercises.
We thus aim to draw general conclusions of the impact that having or assuming some prior knowledge of the cloud properties has on the main retrieval findings.
Such conclusions are relevant to place our work in the context of recent publications, some of which incorporate such prior knowledge of the cloud properties \citep[e.g.][]{fengetal2018, damianoetal2020}.

In this case we find that several of the degeneracies between parameters described in Sect. \ref{sec:results} no longer occur.
We find that any single-phase observation can constrain the planet radius and the methane abundance reasonably well (Fig. \ref{fig:retrievals_KnownCloud_Rpfree}).
The retrieved cloud-top location $\tau_{\rm{c\rightarrow TOA}}$ places the cloud somewhat higher than the true position.
We find no significant differences between the retrievals of a single-phase observation at $\alpha$=37$^\circ$ with $S/N$=10, with $S/N$=20, or combined with measurements at larger phase angles.
Despite the different modelling approaches, these results agree with the findings in \citet{damianoetal2020} for HabEx-like observations (which are comparable to the spectra used in our work in terms of wavelength coverage and spectral resolution).

If the cloud optical properties are assumed known a priori, we conclude that the prospects for atmospheric characterization (Fig. \ref{fig:retrievals_KnownCloud_Rpfree}) are much more optimistic than if they are considered unknown (Figs. \ref{fig:results_individual_thickcloud}, \ref{fig:results_combined_thickcloud}).
In the latter case, we showed (Sects. \ref{subsec:discussion_altitude} and \ref{subsec:discussion_reff}) that the cloud is a main source of the information contained in the spectrum and that its optical properties affect the uncertainties in the retrieval results.
The prior knowledge of the cloud optical properties also changes the conclusions on which observing strategy would be best to improve the atmospheric retrievals.

We note that additional assumptions in the model might affect the conclusions of the retrievals.
For instance, we have assumed an atmosphere dominated by H$_2$ and He.
This is consistent with the established knowledge of giant planets both in the Solar System and in extrasolar planetary systems.
This is also consistent with the population of exoplanets that the Roman Telescope will be able to directly image.
However, this assumption might not be correct for certain planets in the super-Earth regime that will be observed with the next generation of space telescopes such as HabEx or LUVOIR.

For some of such smaller planets, atmospheres dominated by N$_2$, O$_2$, CO$_2$, H$_2$O or other heavy molecules are in principle possible \citep{grenfelletal2020}. 
This indeterminate nature of the background gas may introduce a systematic uncertainty in the volume mixing ratios of the absorbing gas (CH$_4$ here) that is independent of the S/N in the observations.
If the retrievals are carried out assuming (erroneously) H$_2$ as background gas, the retrieval will underestimate the volume mixing ratio of CH$_4$ by a certain factor depending on the actual background gas (Table \ref{table:cross_sections}).
This will happen in particular when the effect of clouds/hazes on the total optical thickness is negligible (because their opacity is overall small, or because of the wavelengths of the observations) and therefore the atmosphere is effectively clear.
This factor arises from the ratio of Rayleigh cross sections $\sigma_{\rm{(True\,gas)}}$/$\sigma_{H_2}$ that dictates the actual optical thickness of the background gas. 
The approximate value of this factor is given in Table \ref{table:cross_sections} for several possible background gases.
Because the photons only react to optical thickness, misidentifying the background gas will translate in an erroneous estimate of the ratio between the absorbing (e.g. CH$_4$) and the scattering (e.g. H$_2$/He or N$_2$) gases.
This degeneracy is not present in transit observations because the absorption depths scale with the scale height, which will be very different for H$_2$ and heavier-element atmospheres.

\section{Conclusions} \label{sec:conclusions}

Atmospheric retrievals from simulated observations are needed to understand the scientific outcome of future direct-imaging telescopes.
These exercises help understand which model parameters have a larger impact on the spectrum of a planet and reveal correlations among parameters that affect the interpretation of a measurement.
\citet{carriongonzalezetal2020} showed that including the planet radius as a free parameter in the retrievals triggers degeneracies between model parameters that prevent an accurate atmospheric characterization.
In this work, we performed retrievals at multiple phase angles (37$^\circ$, 85$^\circ$ and 123$^\circ$) assuming no prior information on $R_p$ or on the cloud properties.
We extended the study to three cloud scenarios (no-, thin- and thick-cloud) for generality.
With this, we aimed to determine which observing strategy would be more effective to break the parameter degeneracies and constrain the atmospheric properties and $R_p$.

If $R_p$ and the cloud properties of the exoplanet are a priori unknown, we found that no single-phase observation with $S/N$=10 can distinguish between cloudy and cloud-free atmospheres.
This finding, reported in \citet{carriongonzalezetal2020} for $\alpha$=0$^\circ$, is here generalized for a broad range of phase angles.
The retrieval results and thus the information content of the spectra vary with the phase angle (Sect. \ref{subsec:results_individual}). 
In all cloud scenarios, a single-phase observation at $\alpha$=123$^\circ$ was found to constrain $R_p$ remarkably better than at smaller phase angles, with a maximum deviation of 35\% with respect to the true value.

We performed simultaneous multi-phase retrievals and found that the combination of small (37$^\circ$) and large (123$^\circ$) phases is the only strategy that can break some of the correlations between $R_p$, $f_{\rm{CH_4}}$ and the cloud properties in all cloud scenarios.
We find that this is the most general observing strategy to identify the presence or absence of clouds and constrain atmospheric properties such as the methane abundance and the cloud optical properties.
We tested other strategies, such as increasing $S/N$ to 20 for a single-phase observation at $\alpha$=37$^\circ$ or combining small (37$^\circ$) and moderate (85$^\circ$) phases.
However, these failed in breaking the parameter correlations in the scenarios with thicker clouds.

We found that the combination of small and large phases produces an improvement in the retrievals that is more noticeable for the thick-cloud scenario (Table \ref{table:results_retrievals}).
Furthermore, we verified that this improvement is more modest if the cloud aerosols have a nearly-isotropic scattering phase function (Sect. \ref{subsec:discussion_reff}).
We therefore concluded that the optical properties of the cloud, and in particular the scattering phase function of the aerosols, have a significant impact on the information content of the spectra at each phase and on multi-phase retrievals.
The idea that the shape of the phase curve is sensitive to the optical properties of the cloud aerosols is not new, as discussed above for Solar System observations. 
This idea has previously been proposed for the characterization of transiting exoplanets \citep[e.g][]{garciamunoz-cabrera2018}.
Here we investigated its application for future direct-imaging efforts.
We also ruled out that the changes in the information content of the spectra were caused by probing different altitudes of the atmosphere at each phase (Sect. \ref{subsec:discussion_altitude}).
Indeed, for cloudy atmospheres we found that, at all phase angles, the photons contributing to the spectrum mainly probe the atmospheric layers where the cloud is located.

We identify a degeneracy in the retrieval of the absorbing gas abundance (CH$_4$ here, but could be other molecule) that will affect the interpretation of reflected-starlight spectra of super-Earths.
This degeneracy arises from the different Rayleigh cross sections of plausible background gases (e.g. H$_2$ or heavier elements) and may introduce a systematic uncertainty in the retrieved absorbing gas abundance.
This will likely not affect the range of planets observed by the Roman Telescope, mainly giant ones.
However, it will play a role in future observations of planets in the super-Earth to mini-Neptune regime, whose atmospheres might not be H$_2$-dominated but might instead resemble those of the Earth, Venus or Titan.

Our single-phase retrievals at different phase angles are consistent with the results in \citet{nayaketal2017}.
However, our multi-phase retrievals differ from the combinations of measurements in \citet{nayaketal2017} based on their intersection criterion.
We conclude that simultaneous retrievals are needed to accurately model the effects of combining multiple measurements.
Furthermore, we tested in Sect. \ref{subsec:discussion_knowncloud} how our conclusions change if we do not include the cloud optical properties 
as free parameters in the retrievals.
This is consistent to some extent with the assumptions in \citet{fengetal2018} or \citet{damianoetal2020}, although the modelling approaches are somewhat different.
We confirmed that in this case the retrieval results are significantly more optimistic than if no prior knowledge is assumed on the cloud properties.
As prior knowledge of the cloud properties will generally not be available, we conclude that additional correlations between model parameters will occur, reducing the accuracy of the retrievals as shown in Sect. \ref{sec:results}.

Our findings are useful to develop the retrieval methods to be applied on future reflected-starlight observations and to establish the model parameters that need to be included.
They also help in the target prioritization for direct-imaging missions.
For instance, the possibility of constraining the radius of a non-transiting exoplanet with an error smaller than 35\% underlines the importance of observing exoplanets at phase angles larger than quadrature.
Furthermore, we conclude that exoplanets with broad intervals of observable phase angles should be considered prime targets for atmospheric characterization through optical phase-curve measurements.

\begin{acknowledgements}
      The authors acknowledge the support of the DFG priority program SPP 1992 “Exploring the Diversity of Extrasolar Planets (GA 2557/1-1)”. 
      NCS acknowledges the support by FCT - Fundação para a Ciência e a Tecnologia through national funds and by FEDER through COMPETE2020 - Programa Operacional Competitividade e Internacionalização by these grants: UID/FIS/04434/2019; UIDB/04434/2020; UIDP/04434/2020; PTDC/FIS-AST/32113/2017 \& POCI-01-0145-FEDER-032113; PTDC/FIS-AST/28953/2017 \& POCI-01-0145-FEDER-028953.
\end{acknowledgements}


\begin{appendix} %First appendix

\section{Posterior probability distributions of single-phase observations} \label{sec:appendix_retrieval_cornerplots_individual}
\subsection{No-cloud scenario, single-phase observations}
\begin{figure*}[t]
   \centering
   \includegraphics[width=18.cm]{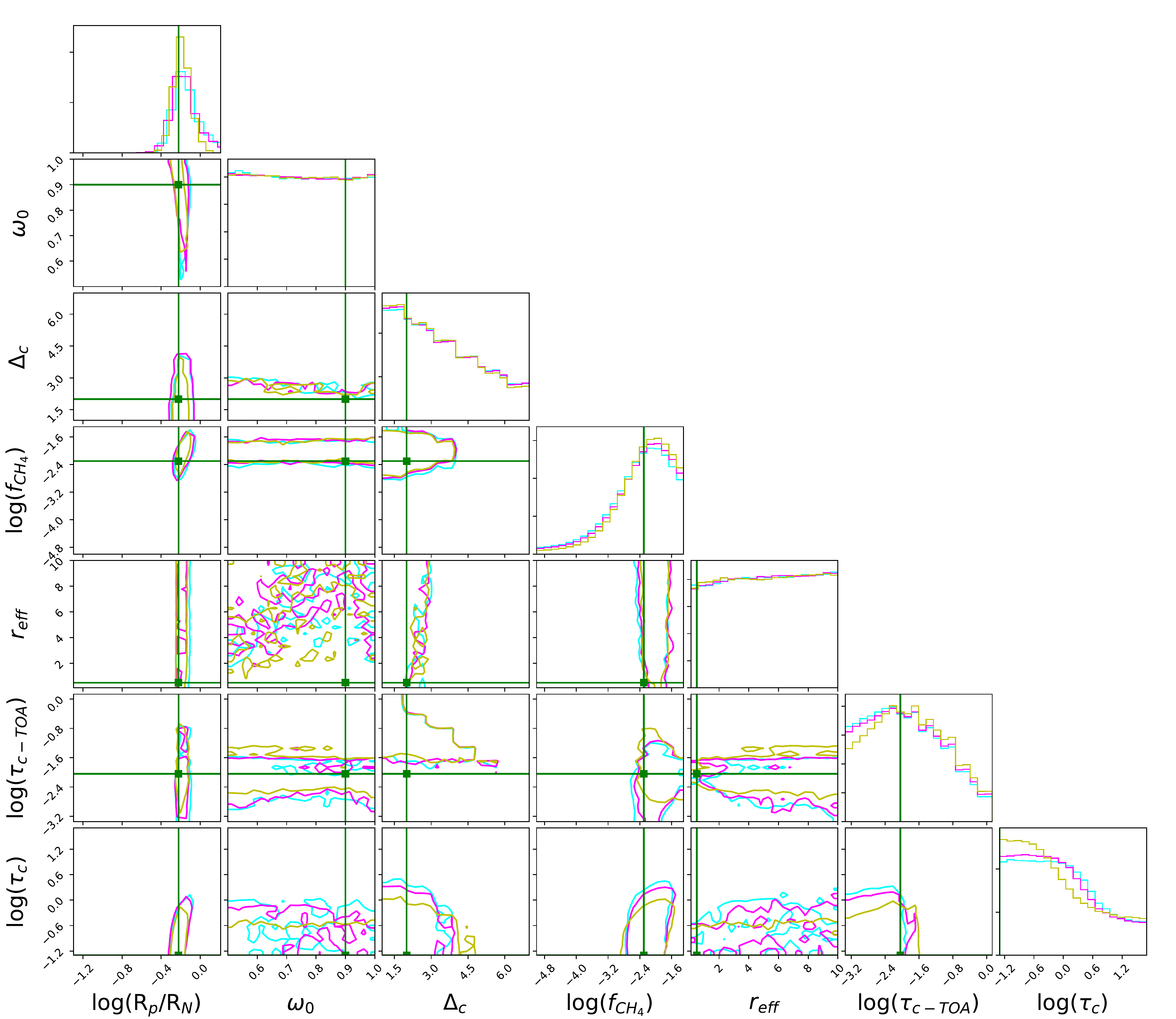}
      \caption{\label{fig:results_individual_nocloud}
      Posterior probability distributions of the model parameters for single-phase observations in the no-cloud atmospheric configuration at phase angles 37$^\circ$ (cyan), 85$^\circ$ (magenta) and 123$^\circ$ (yellow).
      In all cases, the signal-to-noise ratio is $S/N$=10.
      Green lines mark the true values of the model parameters (see Sect. \ref{subsec:model_atmosphere}) for this scenario.
      Two-dimensional subplots show the correlations between pairs of parameters.
      Contour lines correspond to the 1 $\sigma$ confidence levels.
      A detailed description of the parameter correlations with additional $\sigma$ levels in the 2D subplots can be found in \citet{carriongonzalezetal2020} for measurements at $\alpha$=0$^\circ$.
      }
   \end{figure*}

\subsection{Thin-cloud scenario, single-phase observations}
\begin{figure*}[t]
   \centering
   \includegraphics[width=18.cm]{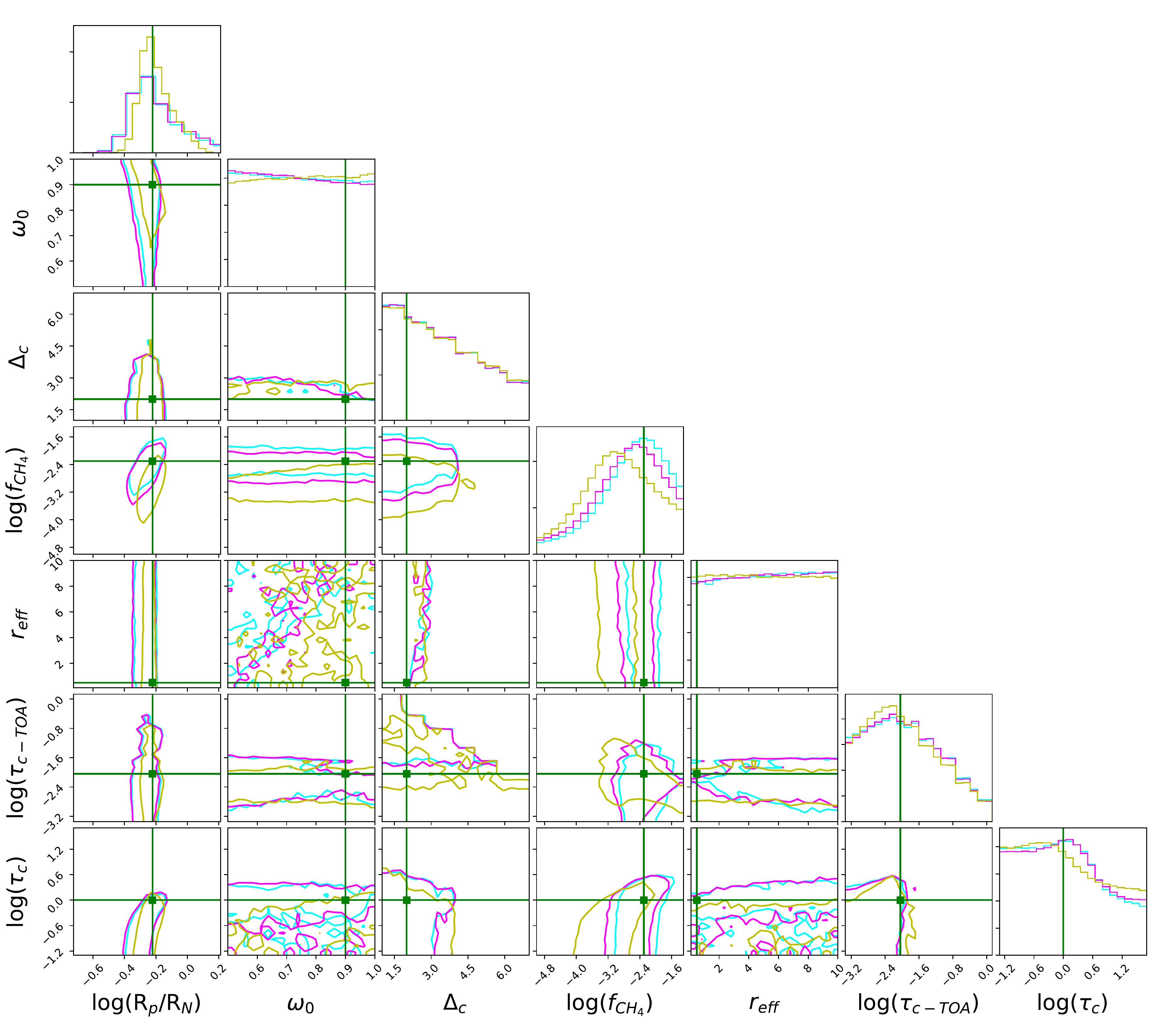}
      \caption{\label{fig:results_individual_thincloud}
      As Fig. \ref{fig:results_individual_nocloud}, but for the thin-cloud scenario.}
   \end{figure*}

\subsection{Thick-cloud scenario, single-phase observations}
\begin{figure*}
   \centering
   \includegraphics[width=18.cm]{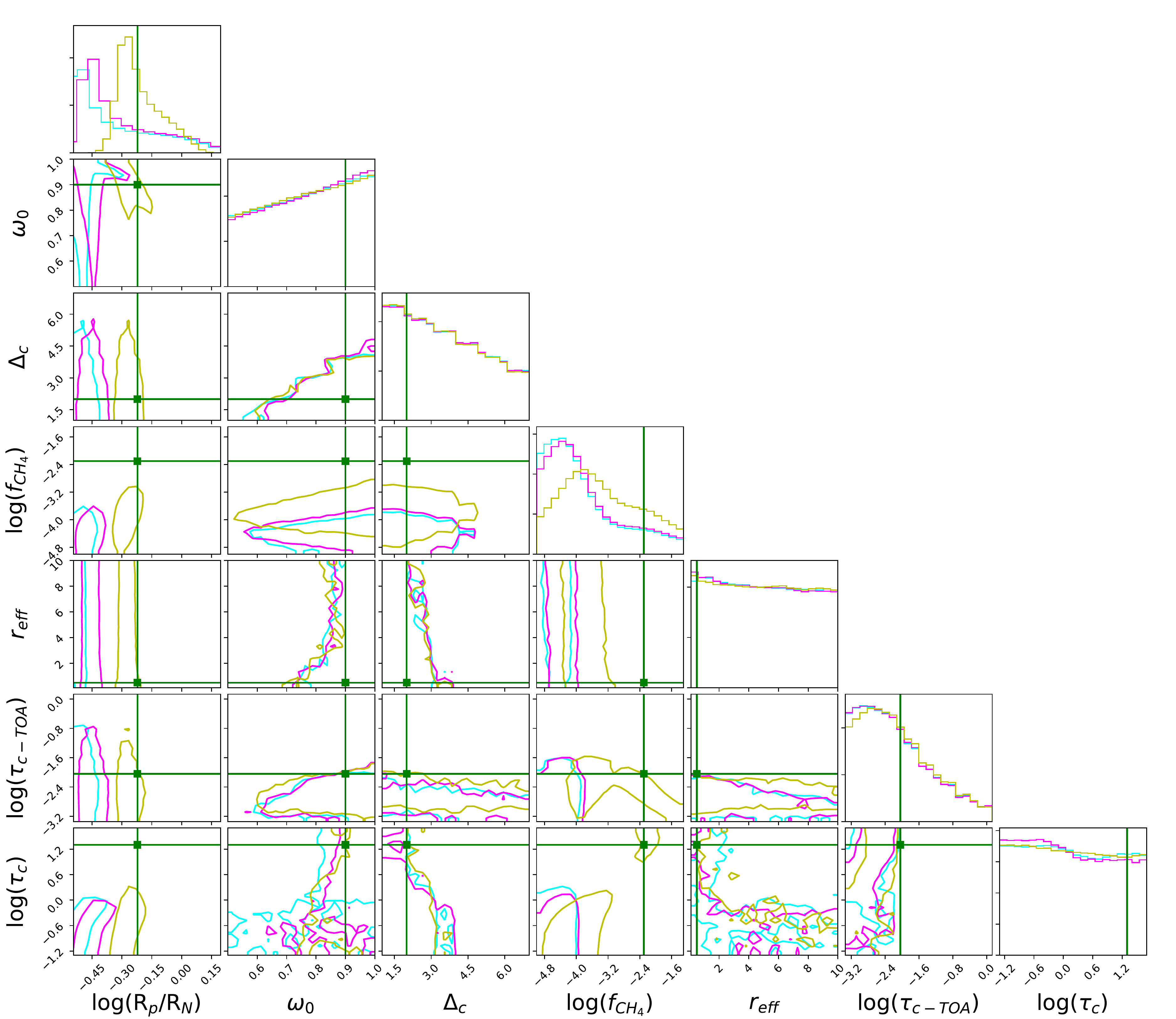}
      \caption{\label{fig:results_individual_thickcloud} 
      As Fig. \ref{fig:results_individual_nocloud}, but for the thick-cloud scenario.}
\end{figure*}

\section{Posterior probability distributions of combined multi-phase observations} \label{sec:appendix_retrieval_cornerplots_combined}
\subsection{No-cloud scenario, combination of multi-phase observations}
\begin{figure*}[t]
   \centering
   \includegraphics[width=18.cm]{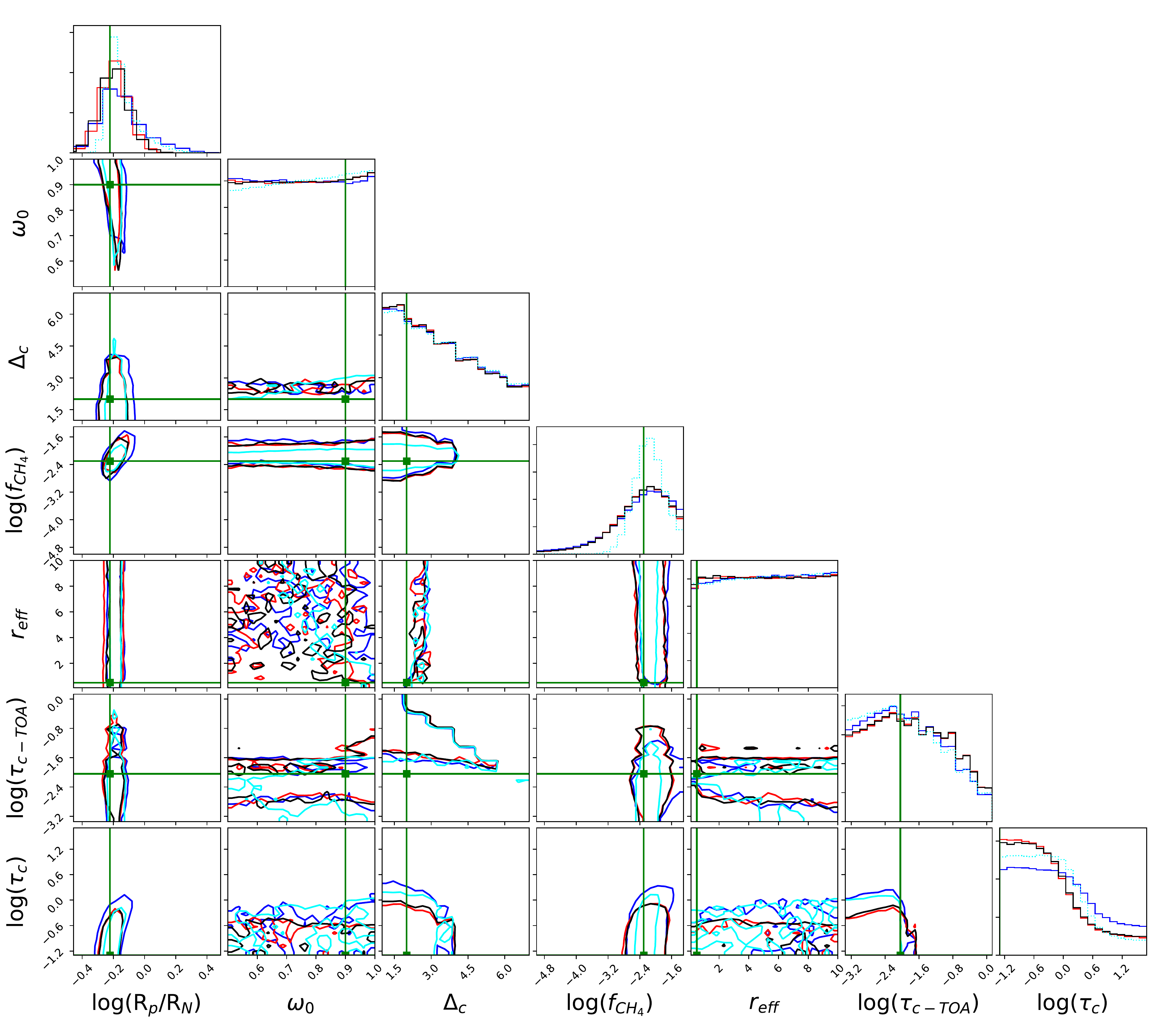}
      \caption{\label{fig:results_combined_nocloud}
      Posterior probability distributions of the model parameters for several combinations of observations of the no-cloud atmospheric configuration, as described in Sect. \ref{subsec:model_sampling}.
      Blue lines indicate the combination (37$^\circ$+85$^\circ$); red lines, (37$^\circ$+123$^\circ$) and black lines, (37$^\circ$+85$^\circ$+123$^\circ$).
      For reference, cyan lines indicate the results of a single-phase observation at $\alpha=37^\circ$ but doubling the signal-to-noise ratio ($S/N$=20).
      Green lines mark the true values of the model parameters (see Sect. \ref{subsec:model_atmosphere}) for this observation.
      Two-dimensional subplots show the correlations between pairs of parameters.
      Contour lines correspond to the 1 $\sigma$ confidence levels.
      }
   \end{figure*}

\subsection{Thin-cloud scenario, combination of multi-phase observations}
\begin{figure*}[t]
   \centering
   \includegraphics[width=18.cm]{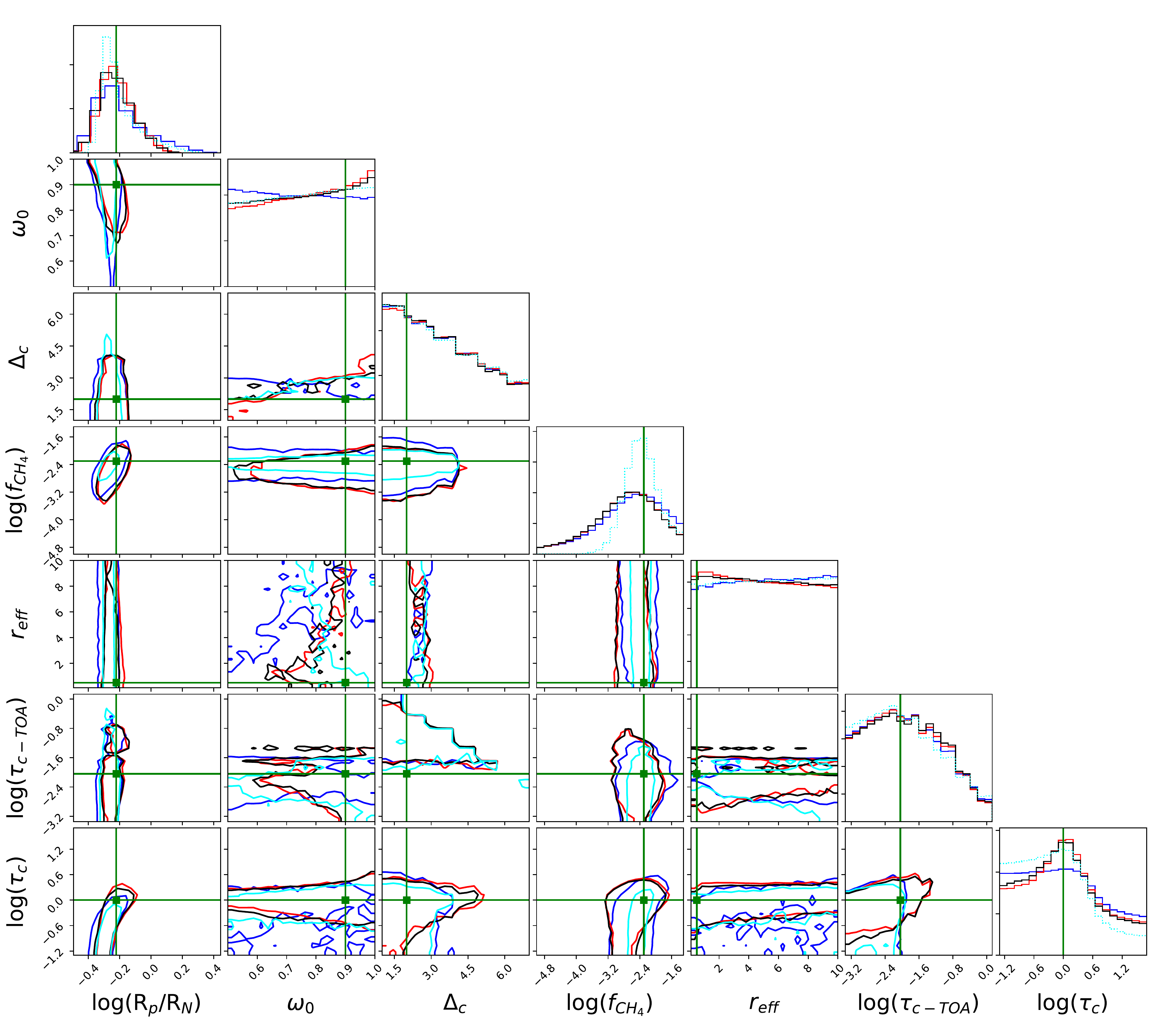}
      \caption{\label{fig:results_combined_thincloud}
      As Fig. \ref{fig:results_combined_nocloud}, but for the thin-cloud scenario.}
   \end{figure*}

\subsection{Thick-cloud scenario, combination of multi-phase observations}
\begin{figure*}[t]
   \centering
   \includegraphics[width=18.cm]{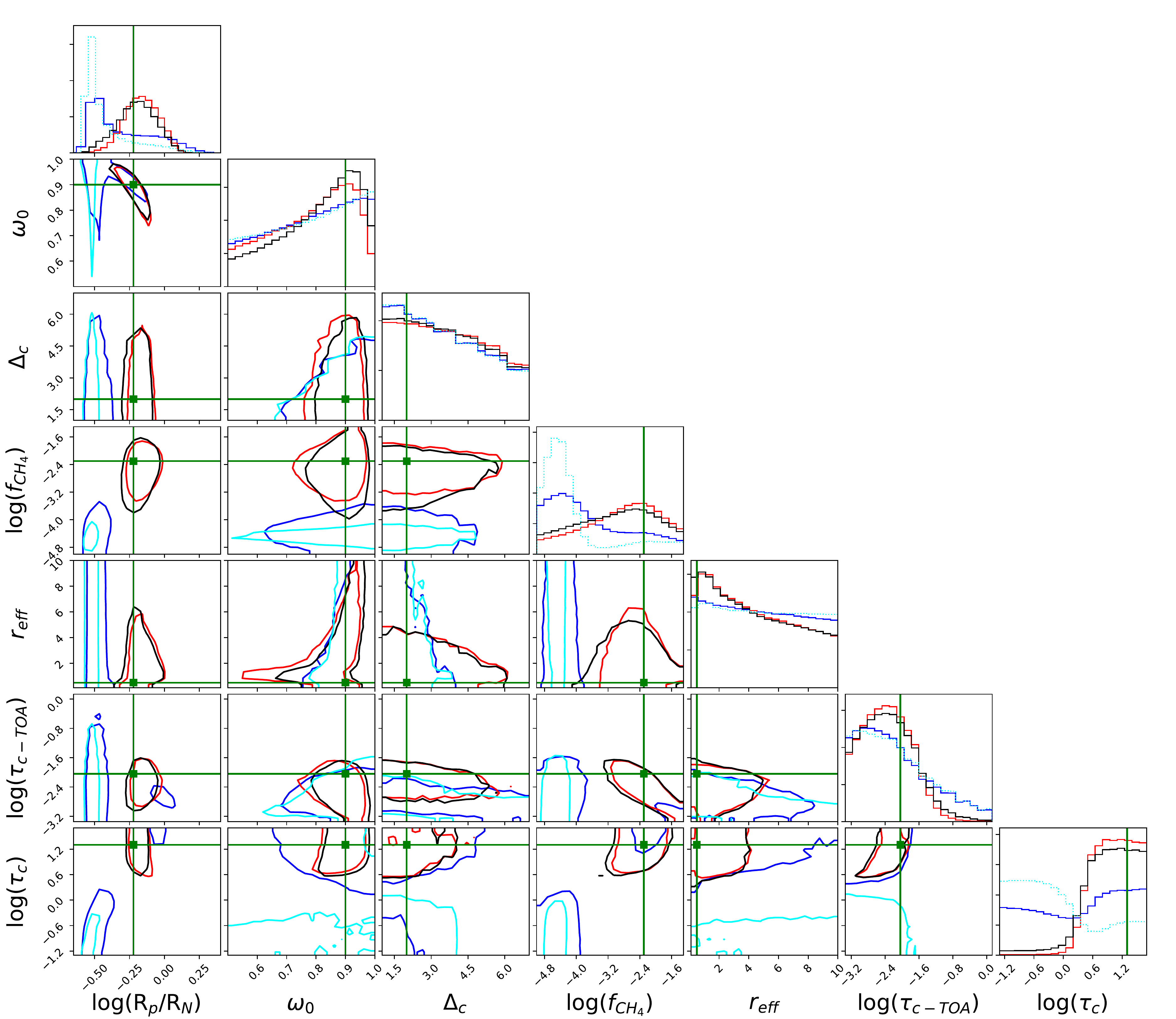}
      \caption{\label{fig:results_combined_thickcloud}
      As Fig. \ref{fig:results_combined_nocloud}, but for the thick-cloud scenario.
      }
   \end{figure*}

\end{appendix}

\end{document}